\documentclass[journal]{IEEEtran}
\IEEEoverridecommandlockouts
\usepackage{amsmath,amssymb,amsfonts}
\usepackage{algorithmic}
\usepackage{graphicx}
\usepackage{textcomp}
\usepackage{xcolor}
\usepackage{multicol}
\usepackage{xurl}
\usepackage[switch,columnwise]{lineno}
\usepackage[normalem]{ulem}
\usepackage{color}
\usepackage{subcaption}
\usepackage{graphicx}  
\usepackage{soul}
\usepackage{physics}
\usepackage{booktabs}

\usepackage[ruled,vlined,linesnumbered]{algorithm2e}
\usepackage{comment}
\usepackage[numbers,sort&compress,square]{natbib}
\usepackage{enumerate}
\usepackage{titlesec}
\titlespacing*{\section}
{0pt}{1 mm plus 1mm minus 1mm}{1 mm plus 1mm minus 1mm}
\titlespacing*{\subsection}
{0pt}{1 mm plus 1mm minus 1mm}{1 mm plus 1mm minus 1mm}

\usepackage[font=footnotesize]{caption}
\usepackage[font=footnotesize]{subcaption}
\usepackage[numbers,sort&compress,square]{natbib}
\usepackage{pdfpages}

\DeclareCaptionFormat{mysubcaption}{\footnotesize#1#2#3\par}
\usepackage[bookmarks=false, colorlinks=true, linkcolor=blue, citecolor=blue, urlcolor=blue]{hyperref}
\usepackage[normalem]{ulem}
\usepackage{float}
\usepackage{scalerel}
\usepackage{tikz}
\usetikzlibrary{svg.path}

\definecolor{orcidlogocol}{HTML}{A6CE39}
\tikzset{
  orcidlogo/.pic={
    \fill[orcidlogocol] svg{M256,128c0,70.7-57.3,128-128,128C57.3,256,0,198.7,0,128C0,57.3,57.3,0,128,0C198.7,0,256,57.3,256,128z};
    \fill[white] svg{M86.3,186.2H70.9V79.1h15.4v48.4V186.2z}
                 svg{M108.9,79.1h41.6c39.6,0,57,28.3,57,53.6c0,27.5-21.5,53.6-56.8,53.6h-41.8V79.1z M124.3,172.4h24.5c34.9,0,42.9-26.5,42.9-39.7c0-21.5-13.7-39.7-43.7-39.7h-23.7V172.4z}
                 svg{M88.7,56.8c0,5.5-4.5,10.1-10.1,10.1c-5.6,0-10.1-4.6-10.1-10.1c0-5.6,4.5-10.1,10.1-10.1C84.2,46.7,88.7,51.3,88.7,56.8z};
  }
}

\newcommand\orcidicon[1]{\href{https://orcid.org/#1}{\mbox{\scalerel*{
\begin{tikzpicture}[yscale=-1,transform shape]
\pic{orcidlogo};
\end{tikzpicture}
}{|}}}}

\def\BibTeX{{\rm B\kern-.05em{\sc i\kern-.025em b}\kern-.08em
    T\kern-.1667em\lower.7ex\hbox{E}\kern-.125emX}}
\begin{document}

\newenvironment{ldescription}[1]
  {\begin{list}{}%
   {\renewcommand\makelabel[1]{##1\hfill}%
   \setlength\itemsep{0pt}%
   \settowidth\labelwidth{\makelabel{#1}}%
   \setlength\leftmargin{\labelwidth}
   \addtolength\leftmargin{\labelsep}}}
  {\end{list}}

\title{Multi-Objective Transmission Expansion: An Offshore Wind Power Integration Case Study}

\author{Saroj Khanal,~\IEEEmembership{Graduate Student Member,~IEEE,} Christoph Graf, Zhirui Liang,~\IEEEmembership{Graduate Student Member,~IEEE,} Yury Dvorkin,~\IEEEmembership{Member,~IEEE,} and Burçin Ünel 
\thanks{S. Khanal and Z. Liang are with the Department
of Electrical and Computer Engineering, Johns Hopkins University, Baltimore,
MD 21218 USA (e-mail: skhanal8@jhu.edu, zliang31@jhu.edu)}
\thanks{Y. Dvorkin is with the Departments of Electrical and Computer Engineering Department and Civil and Systems Engineering, Johns Hopkins University, Baltimore,
MD 21218 USA (e-mail: ydvorki1@jhu.edu)}
\thanks{C. Graf and B. Ünel are with the Institute for Policy Integrity, New York University School of Law, New York, NY 10012 USA
(e-mail: christoph.graf@nyu.edu, burcin.unel@nyu.edu)}
\thanks{The authors thank Anita and Josh Bekenstein for their support of this work. Any opinions, findings, and conclusions expressed here are those of the individual authors. (\textit{Corresponding author}: S. Khanal)}
\thanks{Manuscript received November 17, 2023; revised March 1, 2024; accepted 2 April 2024.}
}

\maketitle

\begin{abstract}
Despite ambitious offshore wind targets in the U.S. and globally, offshore grid planning guidance remains notably scarce, contrasting with well-established frameworks for onshore grids. This gap, alongside the increasing penetration of offshore wind and other clean-energy resources in onshore grids, highlights the urgent need for a coordinated planning framework. Our paper describes a multi-objective, multistage generation, storage and transmission expansion planning model to facilitate efficient and resilient large-scale adoption of offshore wind power. Recognizing regulatory emphasis and, in some cases, requirements to consider externalities, this model explicitly accounts for negative externalities: greenhouse gas emissions and local emission-induced air pollution. Utilizing an 8-zone ISO-NE test system and a 9-zone PJM test system, we explore grid expansion sensitivities such as impacts of optimizing Points of Interconnection (POIs) versus fixed POIs, negative externalities, and consideration of extreme operational scenarios resulting from offshore wind integration. Our results indicate that accounting for negative externalities necessitates greater upfront investment in clean generation and storage (balanced by lower expected operational costs). Optimizing POIs could significantly reshape offshore topology or POIs, and lower total cost. Finally, accounting for extreme operational scenarios typically results in greater operational costs and sometimes may alter onshore line investment.
\end{abstract}


\section*{Nomenclature}
\newcommand{\longestitem}{$\mathrm{RR}^+_{i}, \mathrm{RR}^-_{i}$}
\subsection*{Indices and Sets}

\begin{ldescription}{\longestitem}
\item [$y \in \mathcal{Y}$] Years over the planning horizon.
\item [$e \in \mathcal{E}$] Representative days or scenarios.
\item [$h \in \mathcal{H}$] Hours on a representative day.
\item [$s \in \mathcal{S}$] Nodes or zones.
\item [$\mathcal{S}^1/ \mathcal{S}^0$] Onshore/Offshore nodes.
\item [$i \in \mathcal{G}$] Generators.
\item [$\mathcal{G}^D/\mathcal{G^I}$] Dispatchable/Intermittent (non-dispatchable) generators.
\item [$\mathcal{G}^N/ \mathcal{G}^E$] New/Existing generators.
\item [$k \in \mathcal{K}$] Generation technology types.
\item [$l \in \mathcal{L}$] Transmission lines.
\item [$\mathcal{L^{AC}}/\mathcal{L^{DC}}$] AC/DC transmission lines.
\item [$\mathcal{L^N}/\mathcal{L^E}$] New/Existing transmission lines.
\item [$\mathcal{L}^1/\mathcal{L}^0$] Onshore/Offshore transmission lines.
\item [$f(l) / t(l)$] Indices of sending/receiving nodes line $l$.
\item [$c \in \mathcal{C}$] Line types.
\item [$b \in \mathcal{B}^d$] Flexible demand block.
\item [$x \in \mathcal{X}$] Externalities.
\item [$j \in \mathcal{J}$] Regions for policy constraints.
\end{ldescription}

\subsection*{Parameters}
\begin{ldescription}{\longestitem}
\item [$\omega^{EC}$] Aggregate weighting parameter of discounted monetized externality costs.
\item [$\mathrm{N}^{(\cdot)}$] Useful life of asset $(\cdot)$ in years.
\item [$\mathrm{r}$] Discount rate.
\item [$\mathrm{GC}_k$] Cost per unit capacity of a generation technology type $k$.
\item [$\mathrm{LC}_c$] Cost per unit capacity-length of capacity of a transmission line of type $c$.
\item [$\mathrm{SC}^p$] Cost per unit power capacity of energy storage.
\item [$\mathrm{SC}^e$] Cost per unit energy capacity of battery storage.
\item [$\tau$] Number of days in a year.
\item [$\omega_e$] Weight of representative days.
\item [$\mathrm{RPS}_j$] Renewable Portfolio Standard in region $j$
\item [$\mathrm{PEN}^\times$] Policy mandate non-compliance penalty.
\item [$\mathrm{PEN}^-$] Under-generation penalty.
\item [$\mathrm{PEN}^+$] Over-generation penalty.
\item [$\mathrm{FC}_{i}$] Fixed annual operational cost per unit capacity of generator $i$.
\item [$\mathrm{VC}_{i}$] Variable operations cost of generator $i$.
\item [$\mathrm{WP}_b$] Willingness to pay for demand block $b$.
\item [$\mathrm{CE}_{i,x}$] Damage costs caused by externality $x$ per unit energy production by generator $i$.
\item [$\underline{\mathrm{P}}^g_i/ \overline{\mathrm{P}}_i^g$] Minimum/Maximum power limits of generator $i$.
\item [$\mathrm{RR}_{i}$] Ramp rate of generator $i$.
\item [$\mathrm{\Delta h}$] Temporal resolution of model.
\item [$\mathrm{\delta h}$] Duration (fraction of $\mathrm{\Delta h}$) within which reserves should be supplied.
\item [$\mathrm{M}_{i,s,k}$] Mapping of generator $i$, of type $k$, to node $s$.
\item [$\mathrm{M}_{i,s}$] Mapping generator $i$ to node $s$.
\item [$\mathrm{M}_{l,s}$] Mapping line $l$ to node $s$.
\item [$\mathrm{M}_{l,c}$] Mapping line $l$ to type $c$.
\item [$\mathrm{M}_{j,s}$] Mapping region $j$ to node $s$.
\item [$\mathrm{M}^{0}_{s,y}$] Mapping offshore node $s$ to online year $y$.
\item [$\phi$] Fraction of flexible load relative to the total load.
\item [$\alpha^+$] Acceptable fraction of renewable curtailment out of total renewable generation.
\item [$\mathrm{D}_{y, s, e, h}$] Forecasted load in year $y$, at node $s$, on representative day $e$, during hour $h$.
\item [$\eta^{s,ch}/\eta^{s,dis}$] Charging/Discharging efficiency of energy storage resources.
\item [$\mathrm{DoD}^s$] Allowable depth of discharge of energy storage resources.
\item [$\kappa^{s}$] Annual degradation factor of energy storage.
\item [$\mathrm{H}^{s}$] Energy storage duration.
\item [$\overline{\mathrm{F}}_{l,c}$] Flow limit (capacity) of transmission line $l$ of type $c$.
\item [$\mathrm{B}_{l}$] Susceptance of line $l$.
\item [$\mathrm{\mathbf{M}}$] Large positive number.
\item [$\mathcal{R}^{\star}_{y,s,e,h}$] Operating reserve requirement in year $y$, at node $s$, on representative day $e$, during hour $h$.
\end{ldescription}

\subsection*{Binary variables}
\begin{ldescription}{\longestitem}
\item [$i^{l}_{l,c,y}$] Investment (or construction start) of line $l$, of type $c$ in year $y$.
\item [$z^{l}_{l,c,y}$] Availability of line $l$, of type $c$ in year $y$.
\end{ldescription}

\subsection*{Continuous variables}
\begin{ldescription}{\longestitem}
\item [$OC^{(\cdot)}_{y}$] Discounted annual operational cost of asset ${(\cdot)}$ in year $y$.
\item [$IC^{(\cdot)}_{y}$] Discounted annual investment cost of asset ${(\cdot)}$ in year $y$.
\item [$EC_{y}$] Discounted annual damage costs caused by externalities in year $y$.
\item [$P^g_{y,s,k}$] New generation capacity in year $y$ at $s$ of type $k$.
\item [$p^g_{y,i,e,h}$] Total generation of generator $i$ in year $y$, on representative day $e$, during hour $h$.
\item [$p^{s,ch/dis}_{y, s, e, h}$] Charging/discharging power of energy storage resources at $s$ in year $y$, on representative day $e$, during hour $h$.
\item [$r^g_{y,i,e,h}$] Reserve provided by generator $i$ in year $y$, on representative day $e$, during hour $h$.
\item [$r^s_{y,s,e,h}$] Reserve provided by energy storage in year $y$, at node $s$, on representative day $e$, during hour $h$.
\item [$\psi^{+}_{y,s,e,h}$] Renewable energy curtailment in year $y$, at node $s$, on representative day $e$, during hour $h$.
\item [$\psi^{-}_{y,s,e,h}$] Unserved load in year $y$, at node $s$, on representative day $e$, during hour $h$.
\item [$\Delta d_{y,s,e,h}$] Flexible demand in year $y$, at node $s$, on representative day $e$, during hour $h$ (free variable).
\item [$\rho^\times_{y,j}$] Policy target noncompliance in year $y$ in region $j$.
\item [$f_{y,l,e,h}$] Power flow through line $l$ in year $y$, on representative day $e$, during hour $h$.
\item [$\theta_{y,s,e,h}$] Voltage angle at node $s$ in year $y$, on representative day $e$, during hour $h$ (free variable).
\end{ldescription}

\section{Introduction}

\IEEEPARstart{S}{treamlining} transmission expansion is required for the decarbonization of the U.S. electric grid \cite{jenkins2021mission, brinkman2021north, brown2021value}. This challenge is pertinent to increasing renewable penetration, the need to integrate offshore wind resources, and the expected load growth due to electrification of heating and transportation. Given that transmission investments last for decades, investing at the right place and time to ensure a cost-effective clean energy transition, while not sacrificing reliability, requires a planning framework that can take into account not only the cost and technology drivers but also policy drivers such as greenhouse gases, air pollution, and resilience to extreme weather events. At the same time,  interregional transmission lines cross state lines, the cost of these lines and the debates if the resource preference of one region requires additional transmission upgrades in another region is a significant barrier to transmission development \cite{Morehouse2021}. In the U.S., transmission planners seek to align centralized transmission network planning decisions with decentralized generation investments \cite{conejo2016investment}. Furthermore, despite Federal Energy Regulatory Commission (FERC) Order 1000, which highlights the importance of regional and interregional planning, the majority of transmission planning processes in the U.S. are driven by local reliability needs and separated from generation planning \cite{pfeifenbergerbenefit}. 

The main focus of current transmission planning, in practice, has been to minimize “economic” costs for solving a local reliability need. Even if planners look at regional needs, most U.S. transmission planners rely on generation, storage and transmission expansion planning (GS\&TEP) models, which include investment and operation cost estimates and often ignore any costs or benefits related to externalities such as greenhouse gas emissions and local air pollution from power generation that impose costs on society (including future generations). Similarly, there are societal benefits to a more reliable and resilient electric power system,  particularly with an increasing frequency of climate change--induced extreme weather events.


To overcome deficiencies of existing planning processes, FERC proposed to require transmission providers to conduct long-term regional transmission planning on a sufficiently forward-looking basis to meet the needs driven by changes in the resource mix and demand \cite{FERC2022}. FERC lists externalities  to include in the analysis, and asks transmission providers to develop selection criteria “to maximize benefits to consumers over time without over-building transmission facilities.” Similarly, National Grid GB has begun considering objectives other than just investment and operational costs when evaluating integration of offshore wind farms \cite{nationalgrid2022}. Another policy-relevant question for transmission planning is whether to minimize economic costs, or conversely, if active demand side participation is included, maximize economic welfare. Despite policy relevance, there is limited guidance in the economic or engineering literature on how to incorporate these factors or even whether these factors make a difference in the optimal investment topology, timing, or costs of both the offshore and onshore transmission systems.

This paper uses a multi-objective modeling framework to answer these questions. We assume the perspective of the social planner performing joint centralized transmission and decentralized generation expansion planning, including offshore wind. Guided by regulatory and policy emphasis on incorporating externalities in offshore wind power development, our GS\&TEP model uses multi-objective optimization to specifically address two critical negative externalities from the electric power sector highlighted in  onshore expansion planning studies: greenhouse gas emissions and local air pollution. Our results show that incorporating the externalities, along with the consideration of extreme operational scenarios, results in markedly different outcomes. The proposed multi-objective transmission planning framework equips electric power regulators with a tool that can help overcome balancing multiple policy objectives and address externalities in the face of the changing landscape in the generation mix, demand-side participation, and, most importantly, offshore wind power resources. Furthermore, our results show that considering these additional factors do not lead to significant changes in necessary onshore line investments, which may help alleviate tensions in the current policy debates. 

A rich literature exists on bottom-up, engineering-economic planning models and tools for the electric power sector, which consider both investment in and operation of generation, storage, and transmission assets, and utilization of electric power while capturing their economic and environmental impacts \cite{gacitua2018comprehensive, conejo2016investment}. Munoz et al. [17] describe an adaptive transmission and generation planning accounting for regulatory and market uncertainties. Qiu et al. [18] develop a co-planning framework for transmission and energy storage, where investment decisions are made in multiple stages and operational uncertainty is captured through representative days and addressed through a reserve allocation rule given by \cite{papavasiliou2011reserve}. Similar to other commodity markets, the existence of externalities in the electric power sector is well recognized \cite{costa2023review}, including the challenges associated with internalizing them \cite{Gies2017}. There have been significant efforts to internalize these through energy planning models. Although numerous social and environmental externalities of energy are acknowledged and deemed important, the focus has primarily been on global pollutants, such as CO$_2$ emissions, due to the climate crisis, and local pollutants that affect human health through air pollution. These planning models studies can be broadly categorized into optimization \cite{munoz2013engineering,qiu2016stochastic,rafaj2007internalisation,chen2018advances,rodgers2019assessing,quiroga2019power,chen2019multi,chiu2020future,lv2020generation,pereira2020power,gbadamosi2020multi,fitiwi2020enhanced,sani2021decarbonization,verastegui2021optimization} and multi-criteria decision analysis (MCDA) \cite{ribeiro2013evaluating,santos2017scenarios}. The optimization-based method has gained more attraction in planning studies for many reasons, including their more realistic simulation of how the actual short-term electricity market operates \cite{munoz2013engineering}. Methodologically, in optimization models, externalities can be found to be internalized as constraints or included in the objective as costs (negative externalities) or benefits (positive externalities), or combinations of them. Although these well-established frameworks for onshore grid planning exist, they cannot be readily used for coordinated grid planning, especially due to technical and non-technical reasons for offshore transmission compared to onshore transmission. 

Various studies underscore diverse challenges and benefits associated with offshore wind power integration. Studies by system operators, such as the Offshore Coordination Project by National Grid GB, the Electricity System Operator for Great Britain, \cite{musial2021offshore}, illustrate how an integrated approach can potentially reduce costs and enhance resiliency benefits. Additionally, consulting studies, such as \cite{pfeifenbergerbenefit}, underline benefits of offshore meshed networks. However, the limitations of the studies in \cite{pfeifenbergerbenefit, musial2021offshore} are in (i) lacking coordination with the onshore power grid and (ii) prescribed, rather optimized choices for offshore network configuration. Similarly,  Musial et al. \cite{musial2022offshore} stress the importance of interregional coordination for offshore transmission planning to achieve cost-effective and low-impact solutions. There also have been modeling studies, \cite{liu2022optimization, meng2019offshore,taylor2023wind, mehrtash2019graph,jin2019cable}, examining transmission options for the integration of offshore wind power, comparing AC and DC transmission options and cable routing. However, studies in \cite{musial2022offshore, liu2022optimization, meng2019offshore,taylor2023wind, mehrtash2019graph,jin2019cable} do not capture economic and/or non-economic benefits of large-scale offshore wind power integration that can be attained if co-designed with the onshore power system and if externalities are considered. With the rise in cost competitiveness and locational advantages of offshore wind power resources, this paper argues that these non-economic factors must be incorporated within transmission expansion planning to capture the full societal value of these resources and inform decision-makers on trade-offs between economic and non-economic factors.

In the face of a changing resource mix and demand driven mostly by clean energy and climate policies, the proposed model enables coordination and management of both onshore and offshore transmission needs. This paper layers on top of existing onshore grid planning methods, filling a gap in the need for a comprehensive planning framework that includes offshore resources and accounts for critical (negative) externalities, to help shape the existing planning and cost allocation process. We evaluate our proposed model in test systems designed to simulate ISO-NE and PJM systems, employing two sets of operational data sets with two different model details. This helps illuminate how model results are affected by difference in geographic scope, model complexity, and operational details while identifying overarching patterns.

The remainder of this paper is organized as follows. Section \ref{sec:model}  builds the proposed model. Sections \ref{sec:case_study}  and \ref{sec:case_study_pjm} detail ISO-NE and PJM case studies, covering data curation, results and discussions. Finally, Section \ref{sec:conclusion} concludes the paper.

\section{Model} \label{sec:model}
\begin{figure}
\centering
\includegraphics[width=0.5\textwidth]{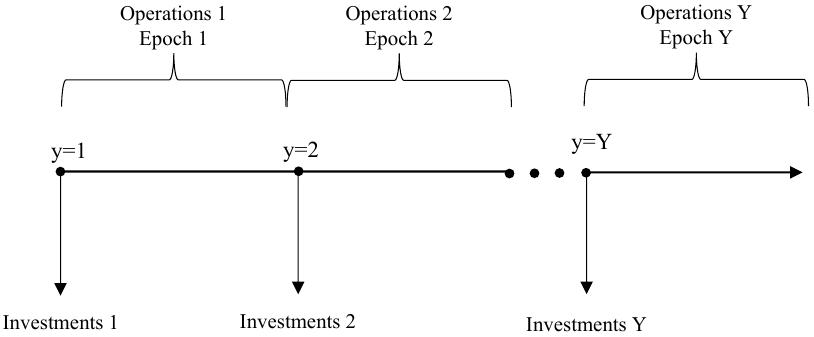}
\caption{Stages for Expansion/Investment Decisions and Ensuing Operations. (\textit{Notes:} Each expansion decision stage and ensuing operations can be subject to long- and short-term uncertainty.)}
\label{fig:timeline}
\end{figure}

Fig.~\ref{fig:timeline} illustrates the representation of multi-stage investments and operations in our model. Investments are allowed at the beginning of each epoch and system operations are modeled within each epoch. Each stage and the ensuing operations can have long- and short-term uncertainty factors. We capture the operational uncertainty by a set of {operational} scenarios, consisting of representative normal and extreme days. By employing different operational scenarios for each epoch, representative days encapsulate both long-term and short-term uncertainty factors, while optimizing investment and operational decisions. Unless otherwise noted in the nomenclature as free variables, all decision variables in our model are constrained to be non-negative.

\subsection{Objective Functions}
We enhance the GS\&TEP model formulation of \cite{qiu2016stochastic} by explicitly incorporating the social cost of environmental externalities into the objective function in Eq.~\eqref{main_min}.
\begin{align}
\label{main_min}
\min z = \sum_{y \in \mathcal{Y}} OC_y + IC_{y}^g + IC_{y}^{l} + IC_{y}^{s} + \omega^{EC} \, EC_{y},
\end{align}
where investment costs in generation ($IC_{y}^g$), transmission line ($IC_{y}^{l}$) and energy storage  ($IC_{y}^{s}$) are operational scenario-independent, while variables related to system operations, e.g., costs from externalities ($EC_y$) and operations ($OC_y$), are scenario-dependent. The weighting parameter, $\omega^{EC}$, can be used to weigh the aggregated ``soft" cost of externalities against the ``hard" economic costs. Setting $\omega^{EC}=0$ reduces the objective function to the traditional planning paradigm, accounting only for economic costs, as in \cite{qiu2016stochastic}. When considering externalities, we typically set it to one, and for sensitivity analyses, we alter the marginal cost estimates of each externality individually. This approach enables us to better address and account for the distinctiveness of each externality's valuation.

We adjust the annualized investment costs of each technology to account for their varying lifespans by using the capital recovery factor (CRF) and an annual discount rate ($\gamma$). This allows a consistent comparison between annualized investment costs across technologies. 
\begin{align}
&\operatorname{CRF}^{(\cdot)}\mathrm{ = \frac{r(1+\mathrm{r})^{{N}^{(\cdot)}}}{(1+\mathrm{r})^{{N}^{(\cdot)}}-1}} \\
&\gamma_n^{(\cdot)}= \frac{\mathbf{1}_n^{(.)}}{(1+\mathrm{r})^{n-1}}
\end{align}
where, $\mathbf{1}_n^{(.)}$ is an indicator function, as defined in Eq.~\eqref{indicator_fcn_availability}, denoting the availability of asset $(\cdot)$.
\begin{align}
\label{indicator_fcn_availability}
& \mathbf{1}_n^{(.)}= \begin{cases}1 & n \in \{1,\cdots,\mathrm{N}^{{(.)}} \} \\
0 & n \notin \{1,\cdots,\mathrm{N}^{(.)} \}\end{cases}.   
\end{align}

\subsubsection{Investment Cost}
Eqs.~\eqref{gen_inv}--\eqref{bess_inv} compute the generation, transmission and storage discounted annual investment costs.
\begin{align}
\label{gen_inv}
IC^{g}_{y} &= \operatorname{CRF}^g \sum_{n=1}^y \gamma_{y-n+1}^g \sum_{k \in \mathcal{K}} \sum_{s \in \mathcal{S}} \mathrm{GC}_k \, {P}^g_{n,s,k} \quad \forall \, y\\
\label{tx_inv}
IC^{l}_{y} &= \operatorname{CRF}^l \sum_{n=1}^y \gamma_{y-n+1}^l \sum_{l \in \mathcal{L}} \mathrm{LC}_c \, I^{l}_{l,c,n} \quad \forall \, y\\
\label{bess_inv}
IC_y^{s} & = \operatorname{CRF}^s \sum_{n=1}^y \gamma_{y-n+1}^l \sum_{s \in \mathcal{S}} \left(\mathrm{SC}^p \, P_{s, n}^{s} + \mathrm{SC}^e \, E^s_{s, n}\right) \quad \forall \, y.
\end{align}
\subsubsection{Operation Cost}
Eq.~\eqref{annual_op_cost} calculates the discounted annual operating cost for each year in the future. It includes the cost of operation of existing and new resources, payment to demand response, renewable energy curtailment cost, unserved load penalty, and penalty for non-compliance with policy targets.
\begin{align}
\label{annual_op_cost}
& OC_{y} = \frac{1}{(1+\mathrm{r})^{y-1}} \Biggl\{\mathrm{PEN}^\times \rho^\times_{y,j} + \sum_{i\in \mathcal{G}} \mathrm{FC}^g_{i} \, \overline{\mathrm{P}}_i^g + \\ \nonumber
& \sum_{e \in \mathcal{E}} \tau \omega_e \Biggl(\sum_{i\in \mathcal{G}} \sum_{h \in \mathcal{H}} \mathrm{VC}^g_{i} \, p^g_{y,i,e,h} + \sum_{s \in \mathcal{S}} \sum_{h \in \mathcal{H}} OC^{p}_{y,s,e,h} \\ \nonumber
& + \sum_{s \in \mathcal{S}} \sum_{h \in \mathcal{H}} OC^{d}_{y,s,e,h}\Biggr)\Biggr\} \quad \forall \, y
\end{align}

The penalty term $OC^{p}_{y,s,e,h}$ in Eq.~\eqref{annual_op_cost}, for both over-supply and under-supply penalties, is given by:
\begin{align}
OC^{p}_{y,s,e,h} = \mathrm{PEN}^+ \psi^+_{y, s, e, h} + \mathrm{PEN}^- \psi^-_{y, s, e, h} \quad \forall \, y,s,e,h. \label{eq:op_pen}
\end{align}

And, the demand (flexibility) cost ($OC^{d}_{y,s,e,h}$), where various demand blocks ($b$) are valued at different levels of willingness to pay for energy ($\mathrm{WP}_b$), is: 
\begin{align}
OC^{d}_{y,s,e,h} = \sum_{b \in \mathcal{B}^d} \mathrm{WP}_b \, \left |\Delta {d_{y,s,e,h,b}} \right| \quad \forall \,y,s,e,h. \label{demflex_1}
\end{align}

$\mathrm{WP}_b$ can range from the operating cost of the most expensive generator to the value of the lost load.

\subsubsection{Cost of Externalities}
Holistic energy planning models must consider the negative externalities such as environmental and social damage caused by the electric power sector \cite{nguyen2016quantifying, rafaj2007internalisation}. We refer to these negative externalities as `cost.' Given the costs of externalities per unit of energy production, the total cost of externalities in year $y$ is:
\begin{align}
\label{eq:ec}
EC_{y} &= \frac{1}{(1+\mathrm{r})^{y-1}} \sum_{e \in \mathcal{E}} \sum_{x \in \mathcal{X}} \tau \, \omega_e \sum_{i\in \mathcal{I}} \sum_{h \in \mathcal{H}} \mathrm{CE}_{i,x} \, p^g_{y,i,e,h} \quad \forall \, y. 
\end{align}

The term $EC{y}$ in the objective function (Eq.~\eqref{main_min}) represents the total cost of various externalities. The cost of externalities resembles the operation costs as they are caused by operating fossil-fueled generators. Since we assume that these costs are directly quantified in monetary terms as `soft' costs—inherently quantifying the `cost' relative importance compared to other hard costs—there is no need for additional weighting parameters in the rest of the objective function. Therefore, adjusting the sensitivity of the cost of externalities to the model outcomes can be more effectively and practically achieved by individually and differently modifying the cost estimates of each externality, instead of aggregating all externalities and assigning a single (aggregated) weight. As a major contributor of CO$_2$ emissions (the most common greenhouse gas causing global warming) and other emissions (VOC, NO$_x$, NH$_3$, SO$x$) causing local air pollution, it is \textit{important} to internalize these damages in electricity planning models to meet decarbonization targets with least cost and impact. However, estimating externality costs is not an easy task, but a crucial one as it significantly alters planning decisions \cite{Gies2017}. In this paper, we focus on internalizing the cost of two of the most important negative externalities considered in electricity planning models \cite{costa2023review}: 1) the cost of global pollutants (CO$_2$ emissions) and 2) human health costs caused by local air pollution from emissions of volatile organic compounds (VOC), nitrogen oxides (NO$_x$), ammonia (NH$_3$), and sulfur oxides (SO$x$). For the former, we use the estimated social cost of carbon by U.S. EPA (see \cite{EPA2022finalized}). For the latter, we estimate the marginal cost of damages caused by local air pollution caused by emissions using the Intervention Model for Air Pollution (InMAP) \cite{TeHi17} and a statistical life metric \cite{EPA2023MortalityRisk}, as detailed in Section \ref{sec:case_study}-3. Using these estimates we compute the total damage cost, which is a summand in $EC{y}$. Similarly,  other negative (or positive) externalities can be incorporated as a summand in $EC{y}$ in Eq.~\eqref{eq:ec}, while cost estimation techniques can vary depending on the types of externalities considered and the methodologies used.

\subsection{Operational Constraints}
\subsubsection{Generator Limits}
Eqs.~\eqref{gen_constraints}--\eqref{reserve_ramp_constraints} implement the constraints on capacity and ramping limits on generation and reserve provision. Eq.~\eqref{AP_gen} makes new generation capacity available.

\begin{align}
\label{gen_constraints}
& \underline{\mathrm{P}}^g_i \leq p^g_{y, i, e, h} \leq \overline{\mathrm{P}}^g_{i} \quad \forall \, y,i \in \mathcal{G^D} \cap \mathcal{G^E},e,h \\
\label{gen_ramp_constraints}
& p^g_{y, i, e, h} + r^g_{y, i, e, h} - p^g_{y, i, e, h-1} \leq \mathrm{RR}_i \quad \forall \, y,i \in \mathcal{G^D},e,h \\
& -\mathrm{RR}_i \leq p^g_{y, i, e, h} - p^g_{y, i, e, h-1} - r^g_{y, i, e, h-1} \, \forall \, y,i \in \mathcal{G^D},e,h \\
\label{reserve_ramp_constraints}
& r^g_{y, i, e, h} \leq \mathrm{RR}_i \, \mathrm{\delta h} \quad \forall \, y,i \in \mathcal{G^D},e,h 
\end{align}
\begin{align}
\label{AP_gen}
&{p}^{g}_{y, s, e, h} \, \mathrm{M}_{i,s,k}  \leq \sum_{n=1}^y \mathbf{1}_{y-n+1}^g \, {P}^{g}_{n,s,k} \quad \forall \, y, s, e, h, k, i \in \mathcal{G^N} \cap \mathcal{G^D}. 
\end{align}

\subsubsection{Demand Flexibility}
Eq.~\eqref{demflex_bnd} bounds the demand flexibility as a fraction of the total demand. Eq.~\eqref{int_demflex} ensures that flexible demand can be shifted within a day.
\begin{align}
\left| \sum_{b \in \mathcal{B^D}} \Delta {d_{y,s,e,h,b}} \right| \leq \phi \, \mathrm{D}_{y,s,e,h} \quad \forall \,y, s, e, h
\label{demflex_bnd} \\
\sum_{h \in \mathcal{H}} \sum_{b \in \mathcal{B^D}} \Delta {d_{y,s,e,h,b}} = 0\quad \forall \,y, s, e.
\label{int_demflex}
\end{align}

\subsubsection{Power Balance}
Eqs.~\eqref{sys_power_bal}--\eqref{eq:balance_last} enforces the nodal power balance.
\begin{align}
\label{sys_power_bal}
&\sum_{i \in \mathcal{G}} p^{g}_{y, i, e, h} \, \mathrm{M}_{i, s} -\sum_{l \in \mathcal{L}} f_{y, l, e, h} \, \mathrm{M}_{l, s} \nonumber \\
&= \mathrm{D}_{y,s,e,h} + \Delta d_{y,s,e,h} + \psi^+_{y, s, e, h} - \psi^-_{y, s, e, h} \nonumber \\ 
&\quad + p^{s,ch}_{y, s, e, h} - p^{s,dis}_{y, s, e, h}  \quad \forall \, y, s, e, h \\
\label{slack_lim}
& \psi^-_{y, s, e, h} \leq \mathrm{D}_{y,s,e,h} + \Delta d_{y,s,e,h} \quad \forall \, y, s, e, h \\
&\psi^+_{y, s, e, h} \leq \alpha^+ \sum_{i \in \mathcal{G^I}} p^{g}_{y, i, e, h} \, \mathrm{M}_{i, s} \quad \forall \, y, s, e, h. \label{eq:balance_last}
\end{align}

\subsubsection{Reserve Requirement}
Eq.~\eqref{res_req} imposes reserve requirements for system operations. 
\begin{align}
\label{res_req} 
& \sum_{i \in \mathcal{G^D}} r^g_{y, i, e, h} + \sum_{s \in \mathcal{S}} r^s_{y, s, e, h} \geq \mathcal{R}^{\star}_{y,s,e,h} \quad \forall\, y, s, e, h.
\end{align}
Similar to the operating reserve allocation method (3\%+5\% rule, \cite{papavasiliou2011reserve}) used to address the uncertainty of wind power integration, Eq.~\eqref{res_req} requests a certain amount of operating reserve ($\mathcal{R}^{\star}_{y,s,e,h}$) to compensate for anticipated real-time fluctuations in wind and solar generation, as well as in loads.

\subsubsection{Energy Storage}
While our energy storage model is broadly applicable to other types of energy storage, we have tailored it for grid-scale battery storage. In line with \cite{qiu2016stochastic,dvorkin2017co}, the battery energy storage system (BESS) planning and operation include tracking the state of charge (SoC) as per Eq.~\eqref{bess_soc}, enforcing capacity limits as in Eqs.~\eqref{bess_power_limits}--\eqref{bess_energy_limits}, and evaluating available energy and power capacities over time, as outlined in Eqs.~\eqref{AE_bess} and \eqref{AP_bess}, considering a constant annual degradation factor ($\kappa^s$). However, using a constant annual degradation factor may be an oversimplification of operational factors such as cycling, temperature, SoC, and depth of discharge (DoD). In operational models, binary variables are commonly used to prevent simultaneous charging and discharging in batteries (e.g., see \cite{arroyo2020use}). However, given our focus on planning rather than on detailed operations, we relax them to maintain model tractability at the expense of focusing on the other model complexities. It is worth noting that we did not encounter any instances of simultaneous charging and discharging in our case studies. Additionally, unlike \cite{qiu2016stochastic}, we include efficiencies directly into the SoC equation, Eq.~\eqref{bess_soc}, instead of the power balance equation, Eq.~\eqref{sys_power_bal} for the consistency of interpretation of discharging power ($p^{s,dis}_{y, s, e, h}$) and charging power ($p^{s,ch}_{y, s, e, h}$). 
\begin{align}
\label{bess_soc}
& SoC^s_{y, s, e, h}= SoC^s_{y, s, e, h-1} + \eta^{s,ch} \, p^{s,ch}_{y, s, e, h} \mathrm{\Delta h} \nonumber \\
&  - \frac{p^{s,dis}_{y, s, e, h}}{\eta^{s,dis}} \mathrm{\Delta h}  \quad \forall\, y, s, e, h \\
& \underline{E}^s_{s,y} = \mathrm{DoD}^s \, \overline{E}^s_{s,y} \\
& SoC^s_{y, s, e, \min (\mathcal{H})}, \, SoC^s_{y, s, e, \max (\mathcal{H})} = \mathrm{SoC}^{s,0} \\
\label{bess_power_limits}
& p^{s,ch}_{y, s, e, h}, p^{s,dis}_{y, s, e, h} \leq \overline{P}^s_{s,y}  \quad \forall\, y, s, e, h \\
& r^s_{y, s, e, h} + p^{s,dis}_{y, s, e, h} - p^{s,ch}_{y, s, e, h} \leq \overline{P}^s_{s,y} \\
& (r^s_{y, s, e, h} - p^{s,ch}_{y, s, e, h}) \, \mathrm{\delta h}\leq \eta^{s,dis} \, (SoC^s_{y, s, e, h} - \underline{E}^s_{s,y}) \\
\label{bess_energy_limits}
& \underline{E}^s_{s,y} \leq SoC^s_{y, s, e, h} \leq \overline{E}^s_{s,y}  \quad \forall\, y, s, e, h\\
\label{AE_bess}
& \overline{E}^s_{s,y} =\sum_{n=1}^y \mathbf{1}^s_{y-n+1} (1-\kappa^s)^{y-n} \,E^s_{s, n} \quad \forall\, y, s \\
\label{AP_bess}
& \overline{P}^s_{s,y} =\sum_{n=1}^y \mathbf{1}^s_{y-n+1} \, P^s_{s, n} \quad \forall\, y, s \\
\label{bess_size_hour}
& \mathrm{H}^s \, P^s_{s, y} = E^s_{s, y}  \quad \forall\, y, s.
\end{align}

\subsubsection{Transmission Constraints}
We model a typical DC power flow for transmission planning, \cite{munoz2013engineering, qiu2016stochastic, conejo2016investment}, and extend it to incorporate DC lines by only constraining flows \cite{hobbs2008improved}. Eq.~\eqref{dc_opf} represents the DC power flows for existing lines, while Eq.~\eqref{line_flow} imposes the flow limits, $\forall \, y, l, e, h$.
\begin{align}
\label{dc_opf}
f_{y, l, e, h} = & \mathrm{B}_{l} \left ( \theta_{y, f(l), e, h}-\theta_{y, t(l), e, h} \right ) \\
\label{line_flow}
&|f_{y,l, e, h}| \leq \overline{\mathrm{F}}_{l}, \qquad|\theta_{y, s, e, h}| \leq \pi .
\end{align}

Eq.~\eqref{count_line_construction_new} records the year when a line is constructed, which adds to the investment cost in Eq.~\eqref{tx_inv}. Eq.~\eqref{delay_1} enforces a delay of $\Delta y$ years between the investment decision and the availability of an expanded line. Eq.~ \eqref{pf_tx_1} and \eqref{pf_tx_2} determine the flows on new/upgraded transmission lines. Eq.~\eqref{offshore_export} ensures at least one line is available by the time any offshore node is online (or electrically charged) and connected to the onshore grid. Eq.~\eqref{only_once_line} ensures that no line is upgraded or built twice throughout the planning horizon.

\begin{align}
\label{count_line_construction_new}
& z^{l}_{l, c, y} = \sum_{n=1}^{y-1} i^{l}_{l, c, n} \quad \forall \, l \in \mathcal{L^N}, c, y \\
\label{delay_1}
& z^{l}_{l,c, y} \geq z^{l}_{l,c, y-\Delta y}  \quad \forall \, l \in \mathcal{L^N}, c\\
\label{only_once_line}
& \sum_{l \in \mathcal{L^N}} \sum_{c \in \mathcal{C}} \sum_{y \in \mathcal{Y}} i^{l}_{l,c,y} \leq 1 \\
\label{pf_tx_1}
& \left | f_{y, l, e, h} - \mathrm{B}_l \left ( \theta_{y, t(l), e, h}-\theta_{y, f(l), e, h} \right ) \right| \, \mathrm{M}_{l,c} \nonumber\\ 
& \leq \mathbf{M} \left ( 1-z^{l}_{l, c, y} \right )\quad \forall \, l \in \mathcal{L^{AC}} \cap \mathcal{L^N}, c, y \\
\label{pf_tx_2}
& | f_{y, l, e, h}| \leq \overline{\mathrm{F}}_{l, c} \, z^{l}_{l, c, y} \quad \forall \, l \in \mathcal{L^N}, c \\
\label{offshore_export}
& \sum_{l \in \mathcal{L}^N} \sum_{c \in \mathcal{C}}z^{l}_{l, c, y} \, \mathrm{M}^{0}_{s,y}  \geq 1  \quad \forall \, s \in \mathcal{S}^0.
\end{align}

\subsection{Policy Constraints}
Eq.~\eqref{RPS} implements the Renewable Portfolio Standard (RPS) constraint for each region $j$. Since offshore regions are not physically located within the onshore regions where these standards are in place, we attribute the flow contribution to the specific onshore connection point where it integrates.
\begin{align}
    &\sum_{l \in \mathcal{L}^0 | s \in \mathcal{S}^1} \sum_{h \in \mathcal{H}} \sum_{e \in \mathcal{E}} f_{y,l,e,h} \, \mathrm{M}_{l,s} \, \mathrm{M}_{s,j} \nonumber \\
    &\quad + \sum_{h \in \mathcal{H}} \sum_{e \in \mathcal{E}} \sum_{i \in \mathcal{G^I}, s \in \mathcal{S}^1} p^g_{y,i,e,h} \, \mathrm{M}_{s,i} \, \mathrm{M}_{s,j} + \rho^\times_{y,j} \nonumber \\
    &\geq \mathrm{RPS}_{j} \sum_{s \in \mathcal{S}^1} \sum_{e \in \mathcal{E}} \sum_{h \in \mathcal{H}}\mathrm{M}_{s,j} \,\mathrm{D}_{y,s,e,h} \quad \forall y, j. 
\label{RPS}
\end{align}

\subsection{Model Implementation}
The model was implemented in Julia v1.9 and JuMP v1.15.0 and solved with Gurobi v10.0.2 on a Mac with an Apple M2 Max chip and 32 GB of memory. The MIP gap was set by default to 0.01\%. Runtimes for the two case studies presented in Sections \ref{sec:case_study} and \ref{sec:case_study_pjm} range from approximately 15 minutes (the simplest specification) to around 48 hours (the most complex specification). The code is available online through our GitHub repository at \url{https://github.com/SarojKhanal/MOTEP-OSW}.

\section{ISO-NE Case Study} \label{sec:case_study}
We deploy our model using an ISO-NE 8-Zone Test System \cite{krishnamurthy20158}, which has been enhanced and curated to include updated or additional data for generation fleet, load, renewables, and transmission parameters, thereby making it suitable for GS\&TEP studies.

\subsection{Data Setup}
We updated the generation mix using the Form EIA-860 from the U.S. Energy Information Administration (EIA) \cite{us3form}. Generators are aggregated at the power plant level and if a power plant consists of generators with different production technologies we treat them separately. 
Since 99.7\% of the region's electricity comes from natural gas, nuclear, and imported electricity, we ignore existing non-thermal generators as in \cite{krishnamurthy20158}. For renewables, we only consider the net injection of solar and wind. Operational characteristics such as minimum and maximum power levels and maximum capacity are from the EIA-860 form, and, if missing, replaced by standard technology values from \cite{krishnamurthy20158} and the International Renewable Energy Agency (IRENA) \cite{IRENA2019Flexibility}.
We use average variable operating costs (derived from the quadratic generating cost functions in \cite{krishnamurthy20158} assuming operation at 80\% output) for each technology.

We utilize actual hourly load data from the ISO-NE website, reported for the operations of the year 2022 at each load zone \cite{ISO_NE2023}. Additionally, in line with ISO-NE's expectations of an annual 2.3\% increase in electricity use, we adopted the same assumption for load growth over the planning horizon \cite{ISO_NE2023}.
Unlike the loads, ISO-NE reports wind and solar generation data only at the system level. Therefore, we distributed these data across all zones, based on the fraction of installed wind and solar capacities, as reported in the 2022 CELT (Capacity, Energy, Loads, and Transmission) Report from ISO-NE website \cite{ISO_NE2023}. For the generation profiles of the six offshore wind farms under consideration, we utilize the wind power data set curated by the National Renewable Energy Lab (NREL) and offered by the U.S. Department of Energy through their Open Energy Data Initiative \cite{DOEOEDI2023}, incorporating an assumed 10\% loss in energy due to wake effects.

\subsubsection{Operational Scenarios}
We model hourly operating conditions to capture short-term uncertainty across multiple years using different representative days (or operational scenarios) across different long-term epochs. However, we acknowledge that representative days may not fully capture the unconditional probability distributions of hourly load factors and wind and solar capacity factors. Therefore, we also include extreme scenarios that may in turn vary from one year to another. Scott et al. \cite{scott2019clustering} survey different clustering-based methods that are common (as highlighted in \cite{merrick2016representation}) to select representative days to capture the dynamics of intermittent supply and demand in electric power sector planning models. They discuss different ways to select time series (historical load, wind, and solar) as input, which can be broadly classified into net load time series and individual time series, with their pros and cons. Below we describe a method based on net load series that we use as a base set of operational scenarios.

\begin{figure}
    \centering
    \includegraphics[width=\linewidth]{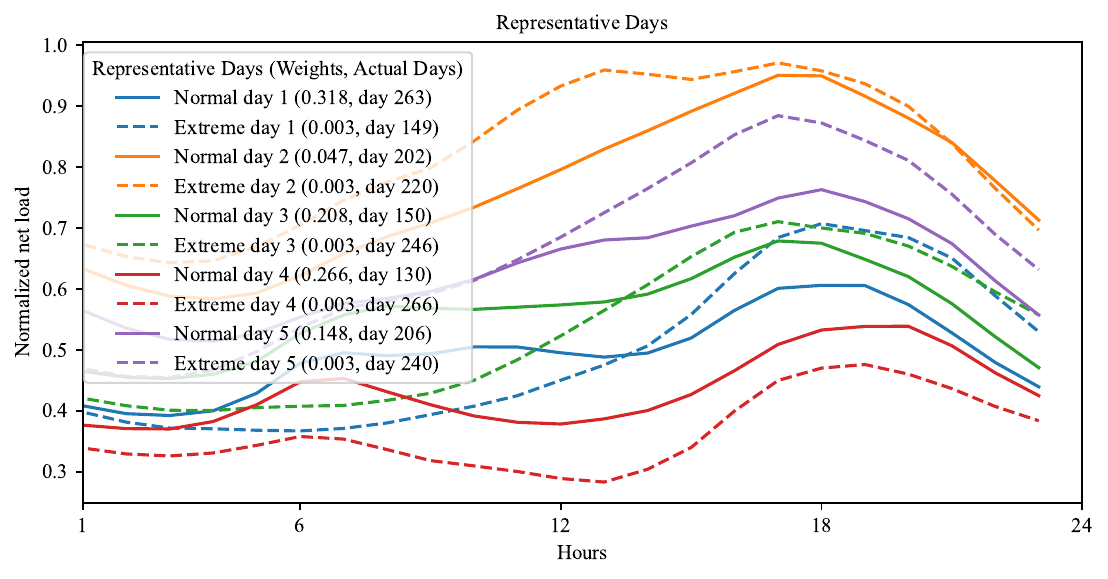}
    \caption{Hourly Normalized Net Load for Normal and Extreme Scenarios.}
    \label{fig:1_representative_days}
\end{figure}

 We extract representative days as operational scenarios from hourly net load data, i.e., hourly load minus onshore renewable generation. To obtain representative days, we use k-means clustering to identify the representative k-cluster centers from the annual time series of the net load. Since the k-means clustering algorithm has an averaging or smoothing tendency, for each cluster, we find the day closest to its center from the original time series to select a representative day, which we call a normal day, and also select the day that is the farthest and call it an extreme day (thus, we avoid smoothing extreme scenarios). We determine the weight of representative days based on the number of days that fall into the corresponding cluster, excluding the day for extreme days, resulting in extreme days having a one-day occurrence per cluster. Fig.~\ref{fig:1_representative_days} shows normalized net load profiles of representative normal days and extreme days, with their corresponding weights. It indicates that the probability of a normal representative day varies from 4.6 to 31.8\%, while the probability of an extreme representative day is significantly lower (e.g., 0.27\% or an equivalent of a one-day-in-a-year event). 


In addition, following the clustering methodology of \cite{scott2019clustering}, we also find representative days and their weights based on individual series to evaluate how that affects planning outcomes. We have not provided those details here in the interest of space and scope.

\subsubsection{Average Marginal Damage from Local Air Pollution}
We use the Intervention Model for Air Pollution (InMAP) \cite{TeHi17} to compute average marginal damages from local air pollution of existing power plants. InMAP uses air pollution source-receptor matrices to relate emissions at source location to concentration at receptor locations. These matrices then can be used to estimate locational damages from air pollution without simulations with computationally demanding air quality models. We use 2016 annual emissions of volatile organic compounds (VOC), Nitrogen oxides (NO$_x$), ammonia (NH$_3$), Sulfur oxides (SO$_x$), and fine particulate matter (PM$_{2.5}$) measured in short tons as inputs to InMAP. To map the Air Emissions Modeling data to our power plant database, we first match the power plant data to Clean Air Markets Program Data (CAMD) using the EPA-EIA-Crosswalk \cite{HuTa21}. To calculate the damages from air pollution, we follow InMAP's methodology \cite{TeHi17}. We first simulate the total PM$_{2.5}$ concentration from emissions at the power plant level. The total PM$_{2.5}$ concentration is the sum of primary PM$_{2.5}$ concentration, particulate NH$_4$ concentration, particulate SO$_4$ concentration, particulate NO$_3$ concentration, and secondary organic aerosol concentration all measured in ($\mu$g/m$^3$). We then use the estimated total PM$_{2.5}$ concentration to estimate the number of deaths using the Cox proportional hazards equation, along with information on population counts and baseline mortality rates. We assume that the overall mortality rate increases by 14\% for every 10 $\mu$g/m$^3$ increase in total PM$_{2.5}$ concentration, as shown in \cite{LeLa12}. Finally, to estimate the economic damage, we apply the value of a statistical life metric set to \$9 million \cite{EPA2023MortalityRisk}.

The above estimation procedure is repeated for each power plant in the sample and provides monetary estimates of economic damages from local pollution caused annually. To get an estimate for the marginal emissions for each power plant, we repeat the above estimation procedure adding additional emission from generating 1 extra \;MWh of electricity. The difference between these two estimates gives us an estimate of the average (annual) marginal damages from local air pollution for each power plant in the sample.

\begin{figure}
    \centering
    \includegraphics[width=0.25\textwidth]{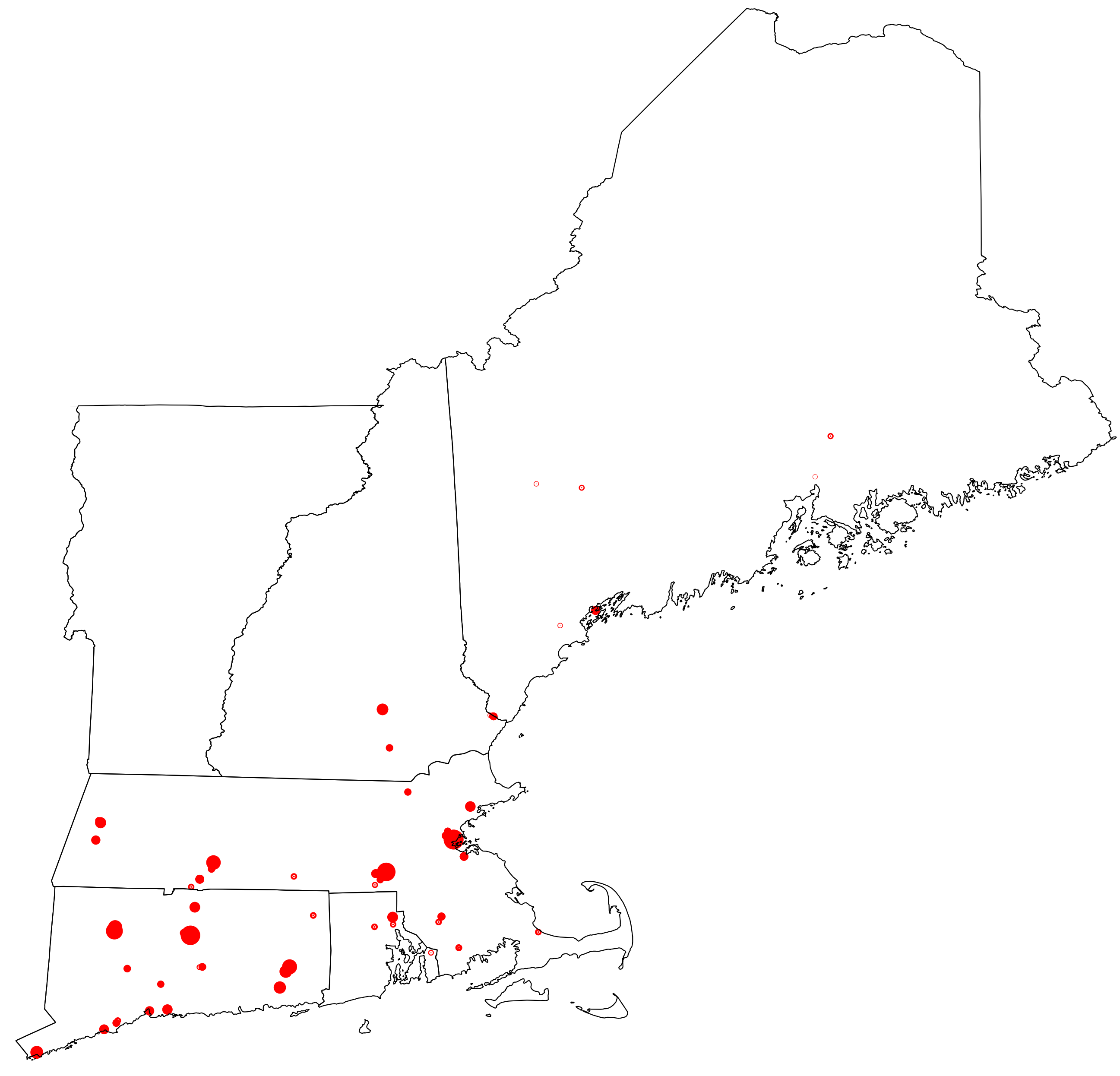}
        \caption{Average Marginal Damages from Local Air Pollution in ISO-NE. (\textit{Notes:} Size of the red dots represents the \$/MWh average marginal damages with a maximum value of 535.75\;\$/MWh and a minimum value of 0.31\;\$/MWh.)}
            \label{fig:1_ave_marg_damages_map}
\end{figure}

In Fig.~\ref{fig:1_ave_marg_damages_map}, we show the different estimates of the average (annual) marginal damages from local air pollution. The average values for each technology are: Gas CCGT: 15.42\;\$/MWh, Gas GT: 26.22\;\$/MWh, Gas Steam: 29.41\;\$/MWh, Coal: 60.38\;\$/MWh, and Oil:  133.70\;\$/MWh. We use those values for unmatched and newly built power plants. 

\subsubsection{Average Marginal Emissions}
We calculate the marginal emissions rates for existing generators in ISO-NE states by deriving the average emissions rates of CO$_2$, SO$_2$, and NO$_x$ across different technologies and states using the Power Sector Emissions Data  \cite{USEPA2023}.  We also directly price CO$_2$ emissions, as it is a global pollutant, and do not price local pollutants, e.g., SO$_2$ and NO$_x$, to avoid double counting with air quality damage costs.

\subsubsection{Technology Investment Options}
For generation investment options, we consider fossil-fuel-fired generators, solar, and wind, while retiring existing generators exogenously. For the sake of consistency with available data, we include Natural Gas Combustion Turbine (NG-CT) and Natural Gas Combined Cycle Carbon Capture and Sequestration (NG-CC-CCS).

We also consider investments in land-based wind and solar photovoltaic (PV) resources. We use normalized reduced profiles of existing renewable profiles to be multiplied by installed capacity for the generation contribution of those non-dispatchable resources. Although the model can handle general capacity expansion in offshore nodes, we exogenously consider build-outs of offshore wind resources on the commercial online year from \cite{musial2022offshore} that have offtake agreements in ISO-NE footprint (see Table~\ref{table:osw projects}). It should be noted that for the Revolution Wind Farm (REV) with its 704 MW capacity, 304 MW is offtaken by Connecticut (CT), while the rest by Rhode Island (RI). To adequately capture this distribution, we create two separate nodes (one for each state).  However, when the model is permitted to optimize the point of interconnection (POI), we treat REV as a single node like other offshore projects. We do not account for the investment cost of the already obligated exogenous offshore wind. 

\begin{table}
  \centering
  \caption{Offshore Wind Projects in the ISO-NE region \cite{musial2022offshore}.}
  \resizebox{\columnwidth}{!}{
    \begin{tabular}{lrrl}
    \toprule
      Project Name (Offshore Node) & Online Year & Capacity (MW) & Candidate POI \\
      \midrule
      Revolution (REV) & 2024 & 704 & CT/RI \\
      Vineyard 1 (VINE) & 2024 & 800 & SEMA \\
      Park City (PKCTY) & 2025 & 800 & SEMA \\
      Commonwealth (COMW) & 2027 & 1,232 & SEMA \\
      Mayflower 1 (MFLR1) & 2025 & 804 & SEMA \\
      Mayflower 2 (MFLR2) & 2025 & 400 & SEMA \\
    \bottomrule
    \end{tabular}
  }
  \label{table:osw projects}
\end{table}

For transmission network investment options, we allow building new lines to offshore nodes and onshore line upgrades. We particularly focus on optimal inter-farm configurations of offshore wind farms, points of offshore interconnection to onshore nodes, and onshore grid upgrades. Each offshore farm is considered a separate offshore node, requiring an investment in at least one candidate line to ensure grid integration of the offshore wind farm by the commercial online date of the offshore project. We incorporate three discrete cable choices for the offshore grid: one HVAC (high-voltage alternating current) line of 400 MW capacity and two HVDC (high-voltage direct current) cables of sizes 1,400 MW and 2,200 MW, reflecting the current standard sizes in ongoing projects in the U.S. For onshore line expansion, we only consider grid reinforcement by doubling the existing capacity. 
And, if onshore upgrade decisions are deemed desirable, line upgrade decisions could potentially double the existing capacity. Additionally, given the prevailing uncertainties surrounding the eventual connection points of these offshore projects, we permit a greater number of interconnection points on land than what the offtake agreements specify. We optimize inter-farm line configurations and export lines.

\subsubsection{Policy Assumptions}

We impose Renewable Portfolio Standards (RPS) for the states in Table~\ref{table:rps} with either strict or soft mandates (accompanied by a non-compliance penalty). Note that MA has a clean energy target instead of RPS, but this study treats clean energy targets as RPS. For this case study, we assume strict compliance with RPS mandates. 

\begin{table}
  \centering
  \footnotesize
  \caption{RPS by States \cite{spglobal_renewable}.}
  \begin{tabular}{lrr}
    \toprule
    State & Target Year & RPS (\%) \\
    \midrule
    ME & 2030 & 80.0 \\
    NH & 2025 & 25.2 \\
    VT & 2032 & 75.0 \\
    MA & 2030 & 80.0 \\
    CT & 2030 & 48.0 \\
    RI & 2035 & 38.5 \\
    \bottomrule
  \end{tabular}
  \label{table:rps}
\end{table} 

\subsubsection{Model Parameters and Specifications}
In Table~\ref{table:model_params}, additional model parameters are presented. Rather than solving for an annual resolution, we adopt an epoch approach, using a 5-year duration within a 20-year planning horizon, starting from 2022. Given the length of each epoch, we do not impose delays on the availability of resources following investment. Consequently, operational parameters are derived from the final year of each epoch, whereas investment-related variables pertain to the epoch's initial year. Costs, generation, and air quality damages are calculated per epoch, i.e., annual variables are scaled by the duration of each epoch. Furthermore, new incremental investment in transmission, generation, and storage capacity in each epoch becomes available at the beginning of the epoch and remains unchanged throughout the planning horizon.
\begin{table}
  \centering
  \footnotesize
  \caption{Model Parameters.}
  \begin{tabular}{cc}
    \toprule
    Parameter & Value \\
    \midrule
    $y$ & $\{2023, \ldots, 2042 \}$ \\
    $e$ & 5 / 10 [with extreme days]\\
    $h$ & 24 \\
    $r$ & 5\% \\
    $\mathrm{PEN}^-, \mathrm{PEN}^+$ & 5,000 [\$/MWh], 0 [\$/MWh] \\ 
    $\phi$ & 10\% \\ 
    $\mathrm{WP}_b$ & \{383, 575, 1,149, 5,000 ($=\mathrm{PEN}^-$)\} [\$/MWh] \\
    $\kappa^s$ & 6\% \\ 
    $\mathrm{H}^s$ & 4 h \\ 
    $\eta^{s,ch}, \eta^{s,dis}$ & 86\%, 86\% \\ 
    $\mathrm{DoD}^s$ & 0.2 \\
    $\alpha^+$ & 0.5 \\
    $\mathrm{\delta h}$ & 1/6 (= $\mathrm{\delta h}$/$\Delta h$ =  10 min/60 min) \\
    \bottomrule
  \end{tabular}
  \label{table:model_params}
\end{table}

We use projections for generation technology cost and performance from the NREL ATB 2022 dataset \cite{vimmerstedt2022annual} (see Table~\ref{table:tech_cost_epochs}), complemented by onshore transmission line data from \cite{ho2021regional}: 3,888.5 U.S. dollars per MW-mile for ISO-NE and 1,499.85 U.S. dollars per MW-mile for PJM. Additionally, offshore line data are derived from \cite{xiang2021comparison}. Specifically, \cite{xiang2016cost} provides cost analyses for offshore HVAC and HVDC transmission systems 0.6 GW and 1.4 GW considering varying distances from the coast. We employ linear approximations of line characteristics based on this information and perform coefficient-wise linear interpolation on regression formulas for the non-linearity of costs because of distance and capacity. To ensure relevance to the U.S. context and reflect the most recent cost range, we adjust these projections by scaling estimates from \cite{Catapult2024WindFarmCosts}, thereby updating the cost analysis originally presented in \cite{xiang2016cost}. The calibrated costs (in million U.S. dollars) for offshore transmission systems of length $l$ miles are as follows: for a 400 MW HVAC: $0.0229 \, l^2 +  1.5093 \, l + 40.13$, for a 1,400 MW HVDC: $2.6763 \, l + 448.58$, and for a 2,200 MW HVDC: $3.5421 \, l + 687.44$. Finally, all monetary values are adjusted to 2020 U.S. dollars for uniformity.
\begin{table}
\centering
\caption{Cost Components for Various Technologies across Epochs.}
\resizebox{\columnwidth}{!}{
    \begin{tabular}{l l l l l l}     
    \toprule
    Technology & Cost Component & Epoch 1 & Epoch 2 & Epoch 3 & Epoch 4 \\
    \midrule
    NG-CC-CCS & Investment (\$/MW) &  2,209,365.98  &  2,059,151.02  &  1,917,707.74  &  1,759,817.57 \\
      & Fixed operation (\$/MW-yr) &  62,000.00  &  61,000.00  &  60,000.00  &  60,000.00 \\
      & Variable operation (\$/MWh) &  27.29  &  29.46  &  28.59  &  28.54 \\
    \midrule
    NG-CT & Investment (\$/MW) &  853,045.52  &  791,643.79  &  766,425.22  &  746,688.95 \\
      & Fixed operation (\$/MW-yr) &  21,000.00  &  21,000.00  &  21,000.00  &  21,000.00 \\
      & Variable operation (\$/MWh) &  34.48  &  39.29  &  40.13  &  40.50 \\
    \midrule
    Wind & Investment (\$/MW) &  1,308,400.00  &  1,052,400.00  &  921,500.00  &  874,000.00 \\
     & Fixed operation (\$/MW-yr) &   40,165.00 	& 38,365.75 	& 36,905.13 	&  35,444.50  \\
     & Variable operation (\$/MWh) &  -    &  -    &  -    &  -   \\
    \midrule
    Solar PV & Investment (\$/MW) &  1,073,834.12  &  843,632.15  &  731,460.46  &  697,975.63 \\
     & Fixed operation (\$/MW-yr) &  16,983.15  &  15,020.43  &  14,521.95  &  14,029.36 \\
     & Variable operation (\$/MWh) &  -    &  -    &  -    &  -   \\
    \midrule
    Battery & Investment (\$/MWh) &  254,835.78  &  178,968.64  &  151,709.97  &  141,861.83 \\
     & Investment (\$/MW) &  236,365.10  &  252,067.20  &  254,746.03  &  238,220.32 \\
     & Fixed operation (\$/MW-yr) &  25,375.77  &  21,819.24  &  20,421.28  &  19,023.33 \\
     & Variable operation (\$/MWh) &  -    &  -    &  -    &  -   \\
    \bottomrule
    \end{tabular}
}
\label{table:tech_cost_epochs}
\end{table}

\subsubsection{Simplifications and Limitations}
The case study rests on several assumptions to keep the model computationally tractable. We use an eight-zone representation of the ISO-NE system and model transmission corridors, rather than specific lines. These publicly available data may be considerably inferior to the data accessible by transmission planners. We also do not include some non-technical constraints in the model, e.g., renewable capacity deployment constraints due to land use restrictions or public opposition to renewable energy projects. Furthermore, the extreme days considered are based on historic observations and do not necessarily reflect the increased likelihood or magnitude of climate change--induced extreme weather conditions. Similarly, we omit the effect of climate change and increasing (average) temperatures on loads,  thermal efficiency factors, and power line ratings. As mentioned earlier, we adapted cost coefficients for market operations from the test system as in \cite{krishnamurthy20158}. A benchmarking analysis of the operational costs for an annual representative year against actual ISO reports for 2022 shows a close alignment. Consequently, we have applied the same cost coefficients throughout the planning horizon, consistent with the EIA's base forecast, which indicates a stable trend for natural gas prices.

Our model also presupposes perfect foresight of the demand growth and technological cost/performance.  Furthermore, our model represents new generation decisions as aggregate capacity within a given area, rather than a standalone unit, which limits the ability to compute local air pollution. 


\subsection{Results and Discussions}
We compare capacity expansion (generation, storage, and transmission) decisions and operational decisions, along with their associated costs, including those of externalities, by varying the terms included in the objective function. The objective function of our baseline model includes only investment costs and expected operational cost ($\omega^{EC} = 0$). We denote this model specification as the single-objective (SO) model. We benchmark SO results against results from a multi-objective (MO) optimization model ($\omega^{EC} = 1$), where we also include monetized environmental externalities, such as air quality damages and carbon emission costs in the objective function. Model outcomes computed with five extreme days (scenarios) are denoted with the suffix X5. This sensitivity analysis exposes the GS\&TEP model with a more complete distribution of capacity factors of weather-dependent resources such as wind and solar, as well as load to better capture extreme events. In addition, we refine MO by optimizing the points of interconnection (POI) between cables connecting offshore wind hubs (farms) and onshore zones. We call this model specification multi-objective, optimized POI (MO OPOI) optimization model. All non-OPOI specifications feature candidate offshore export lines that are restricted to connections with fixed (or a predetermined set of) POIs on land (see candidate POI in Table~\ref{table:osw projects}). However, in the OPOI specification, this constraint is relaxed. For all offshore projects, the candidate POIs are expanded to include multiple locations: ME, NH, NEMA, CT, RI, and SEMA. Finally, the U.S. EPA recently finalized revising the SCC estimate from \$51 to \$190 per metric ton \cite{EPA2022finalized}. We therefore analyze the \emph{sensitivity} of our results with changes in the SCC estimate. Specifically, we compare the results using estimates of \$190 (MO EM190) per metric ton to our baseline multi-objective (MO) specification with \$51 per metric ton.

\begin{figure}
    \centering
    \begin{subfigure}{0.24\textwidth}
        \includegraphics[width=\textwidth]{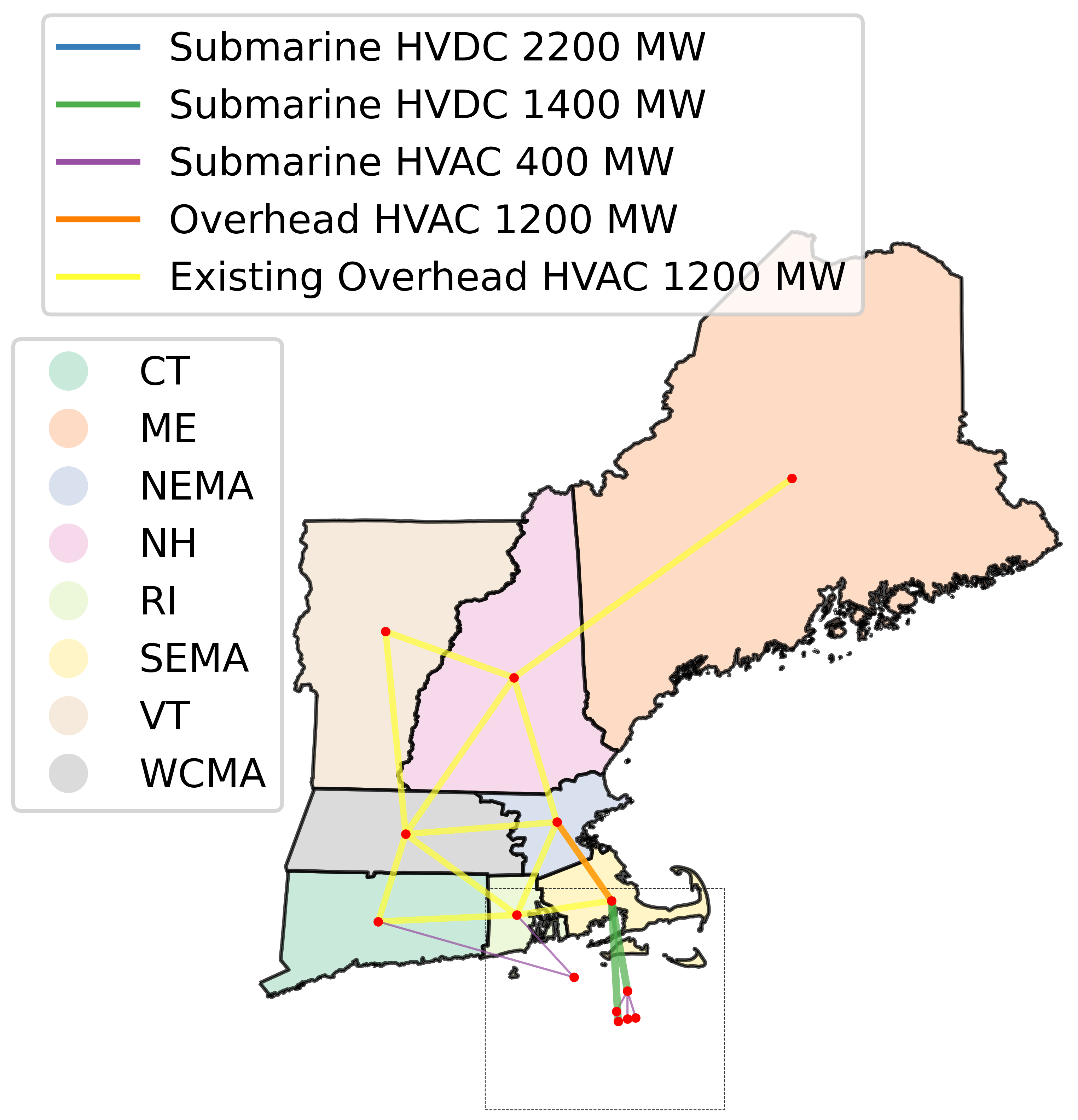}
        \caption{SO, SO X5}
        \label{fig:1_SO}
    \end{subfigure}
    \hfill
    \begin{subfigure}{0.24\textwidth}
        \includegraphics[width=\textwidth]{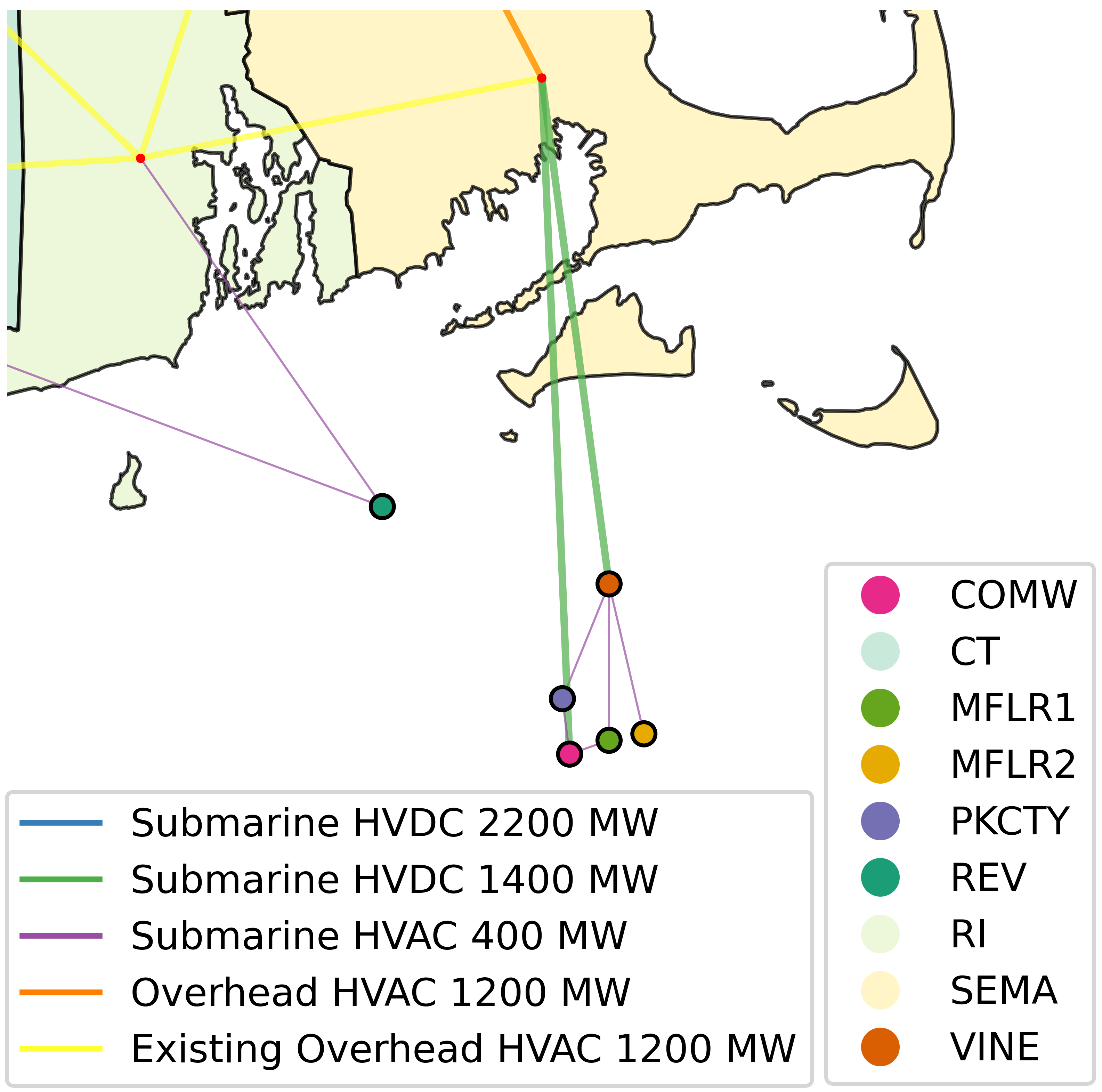}
        \caption{focused (a)}
        \label{fig:1_SO_focus}
    \end{subfigure}
    \newline
    \begin{subfigure}{0.24\textwidth}
        \includegraphics[width=\textwidth]{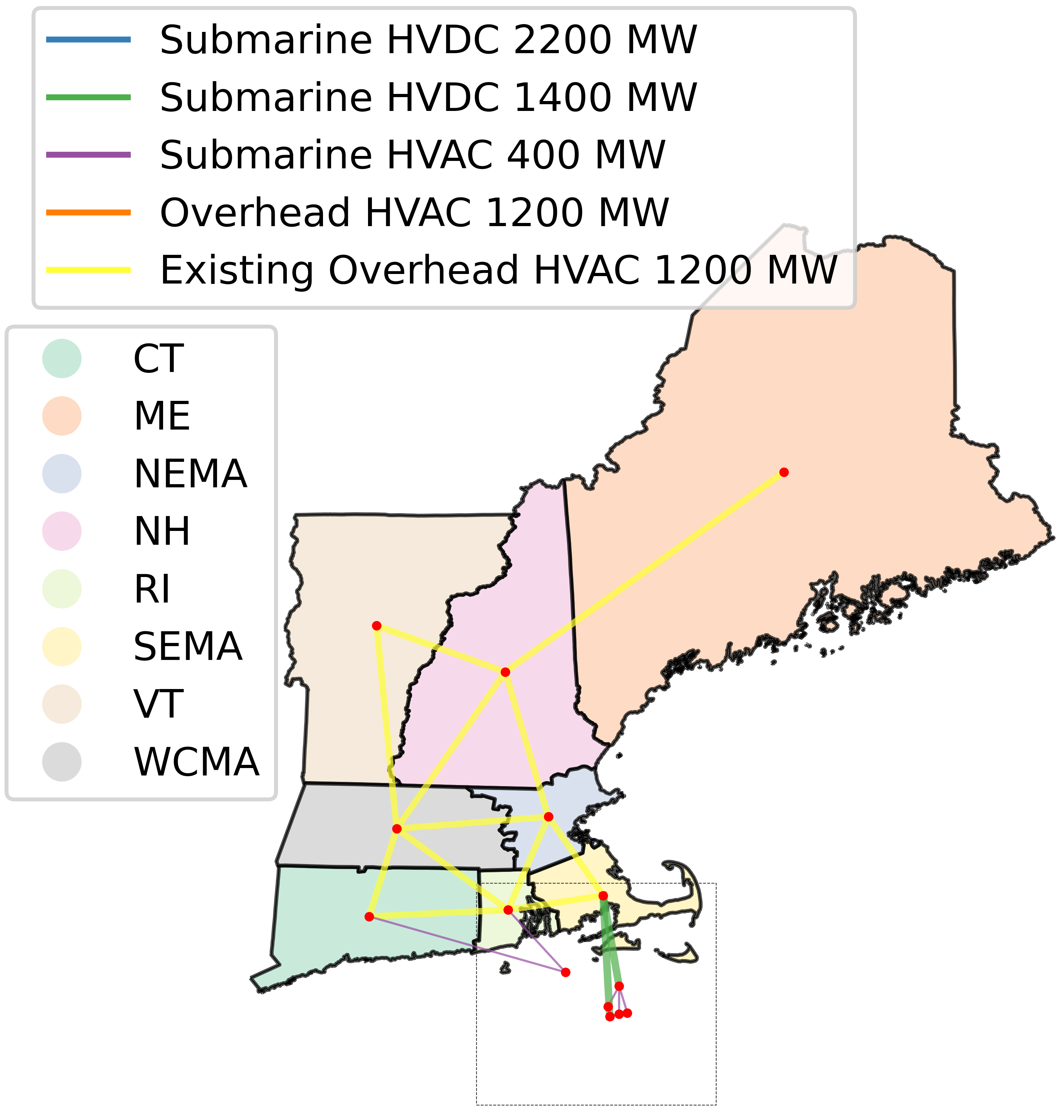}
        \caption{MO, MO X5}
        \label{fig:1_MO}
    \end{subfigure}
    \hfill
    \begin{subfigure}{0.24\textwidth}
        \includegraphics[width=\textwidth]{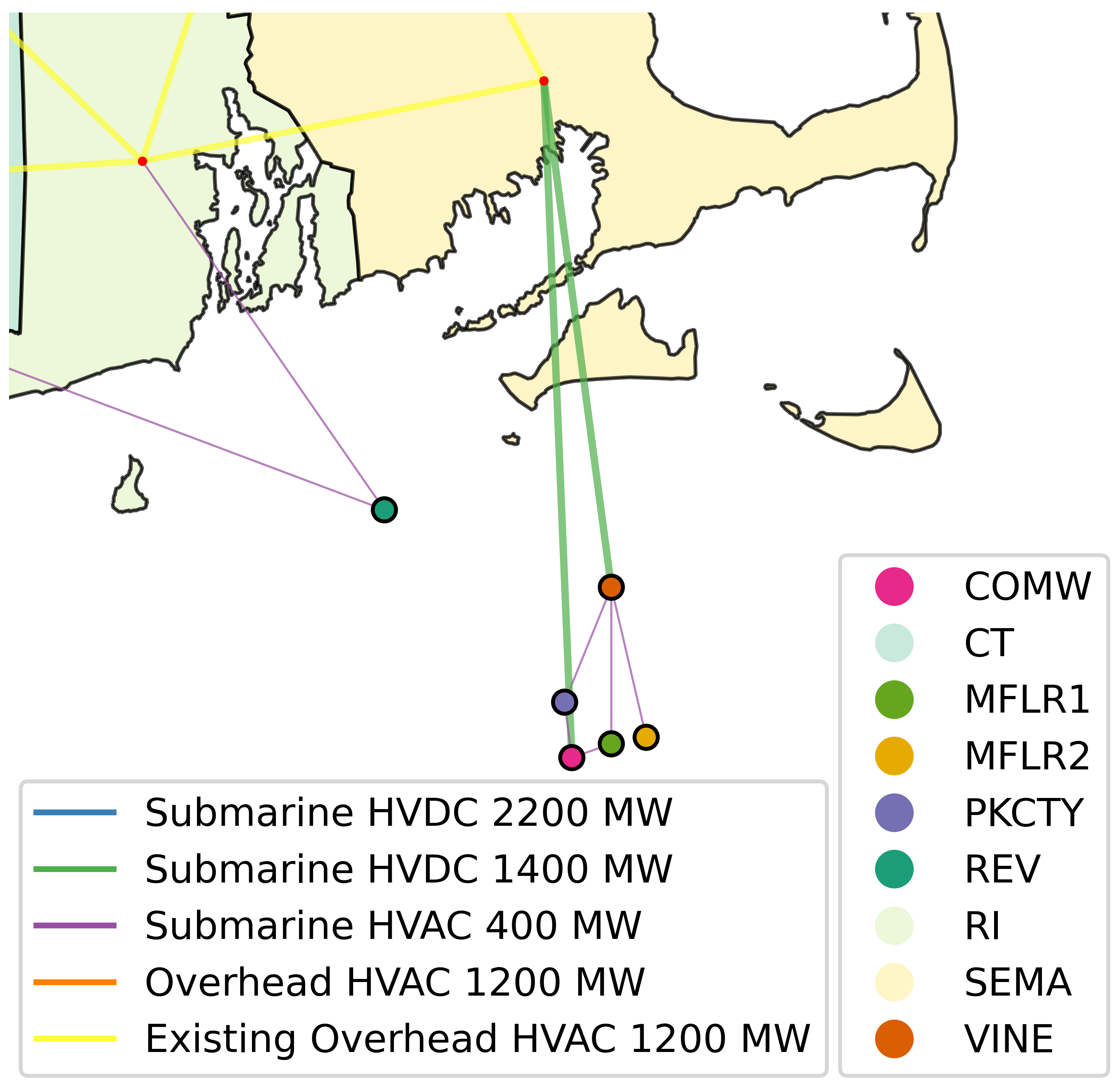}
        \caption{focused (c)}
        \label{fig:1_MO_focus}
    \end{subfigure}
    \caption{Optimal Onshore and Offshore Topology: Impacts of Accounting for Externalities and Extreme Days. (\textit{Notes:} The line between NEMA and SEMA in SO is upgraded in Epoch 3 whereas in SO X5 in Epoch 1, the default epoch for transmission buildouts if not mentioned explicitly.)} 
    \label{fig:1_topology_externalities}
\end{figure}

Fig.~\ref{fig:1_topology_externalities} summarizes the optimal transmission expansion decisions for SO, SO X5, MO and MO X5 specifications to capture the impact of externalities, extreme days, and both. We find that the SO specifications (SO and SO X5) require an upgrade of the onshore transmission line between SEMA and NEMA: for SO in Epoch 3 and for SO X5 in Epoch 1 (see Fig.~\ref{fig:1_topology_externalities} a and b), while the MO specifications (MO and MO X5) do not (see Fig.~\ref{fig:1_topology_externalities} c and d). The offshore topology remains the same for SO and MO  cases (see Fig.~\ref{fig:1_topology_externalities} b and d). Therefore, accounting for a more comprehensive set of economic costs will decrease the need to upgrade onshore transmission.

\begin{figure}
    \centering
    \begin{subfigure}{0.24\textwidth}
        \includegraphics[width=\textwidth]{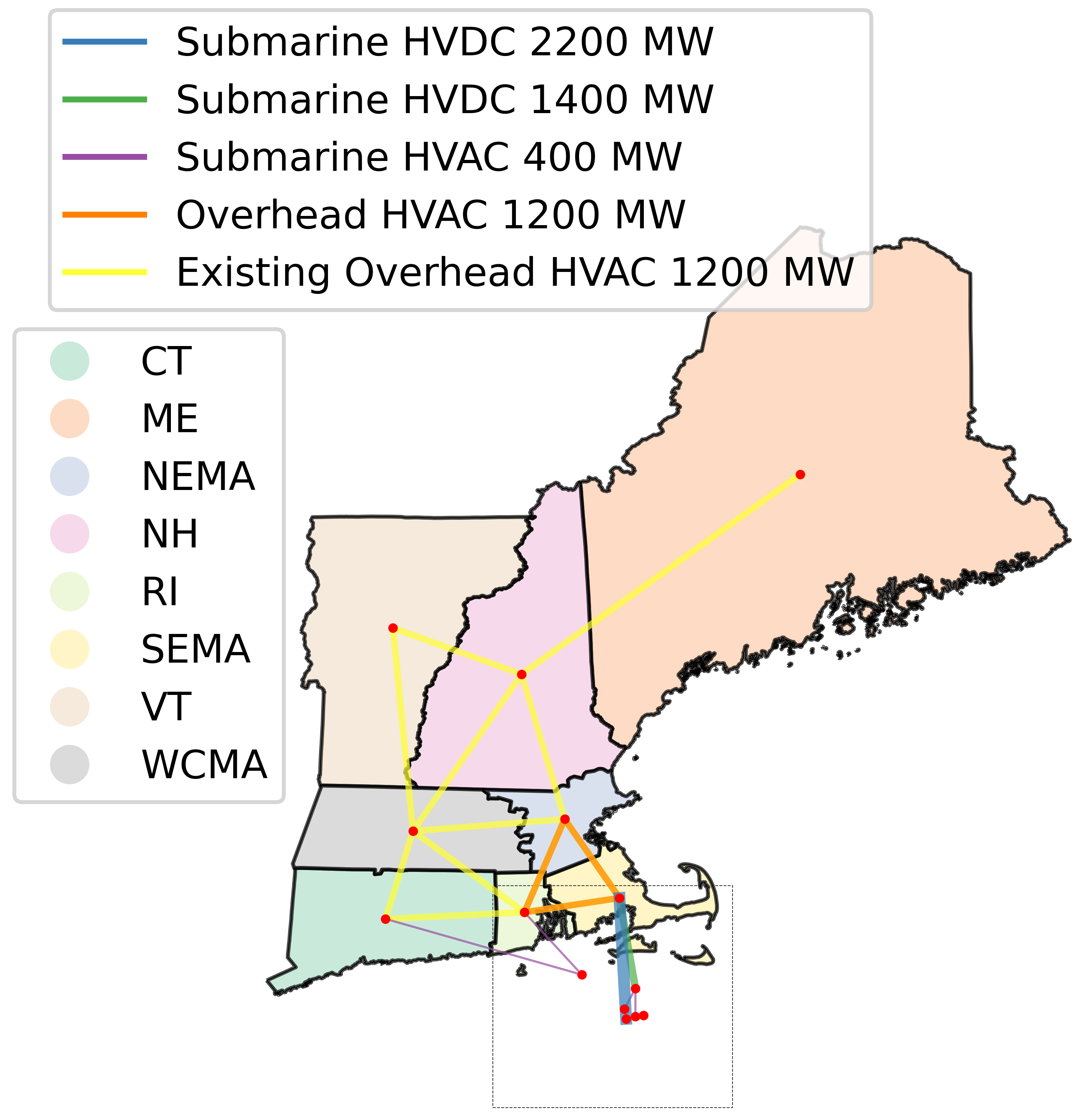}
        \caption{SO, SO X5, MO}
        \label{fig:1_NS_SO}
    \end{subfigure}
    \hfill
    \begin{subfigure}{0.24\textwidth}
        \includegraphics[width=\textwidth]{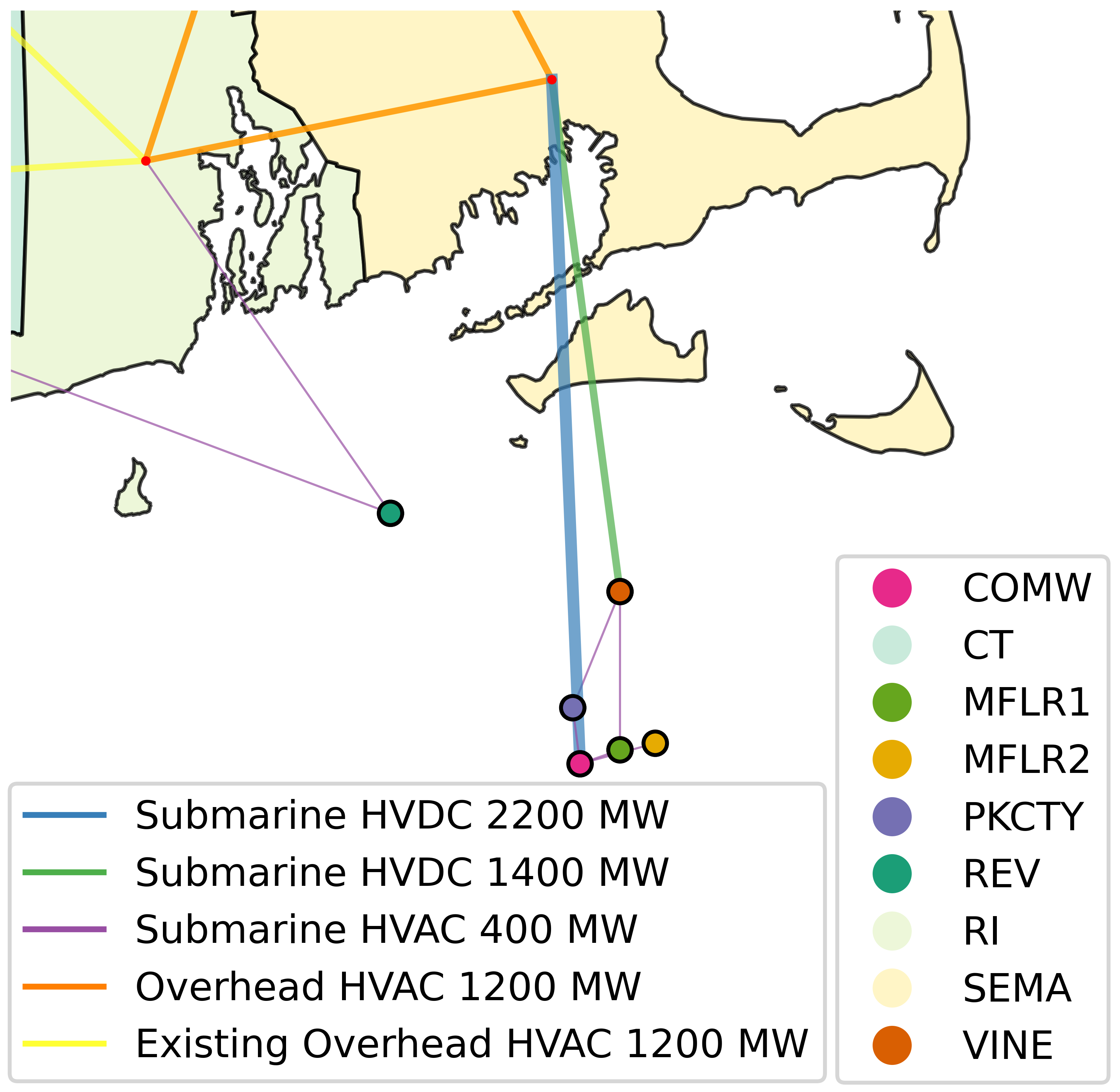}
        \caption{focused (a)}
        \label{fig:1_NS_SO_focus}
    \end{subfigure}
    \newline
        \centering
    \begin{subfigure}{0.24\textwidth}
        \includegraphics[width=\textwidth]{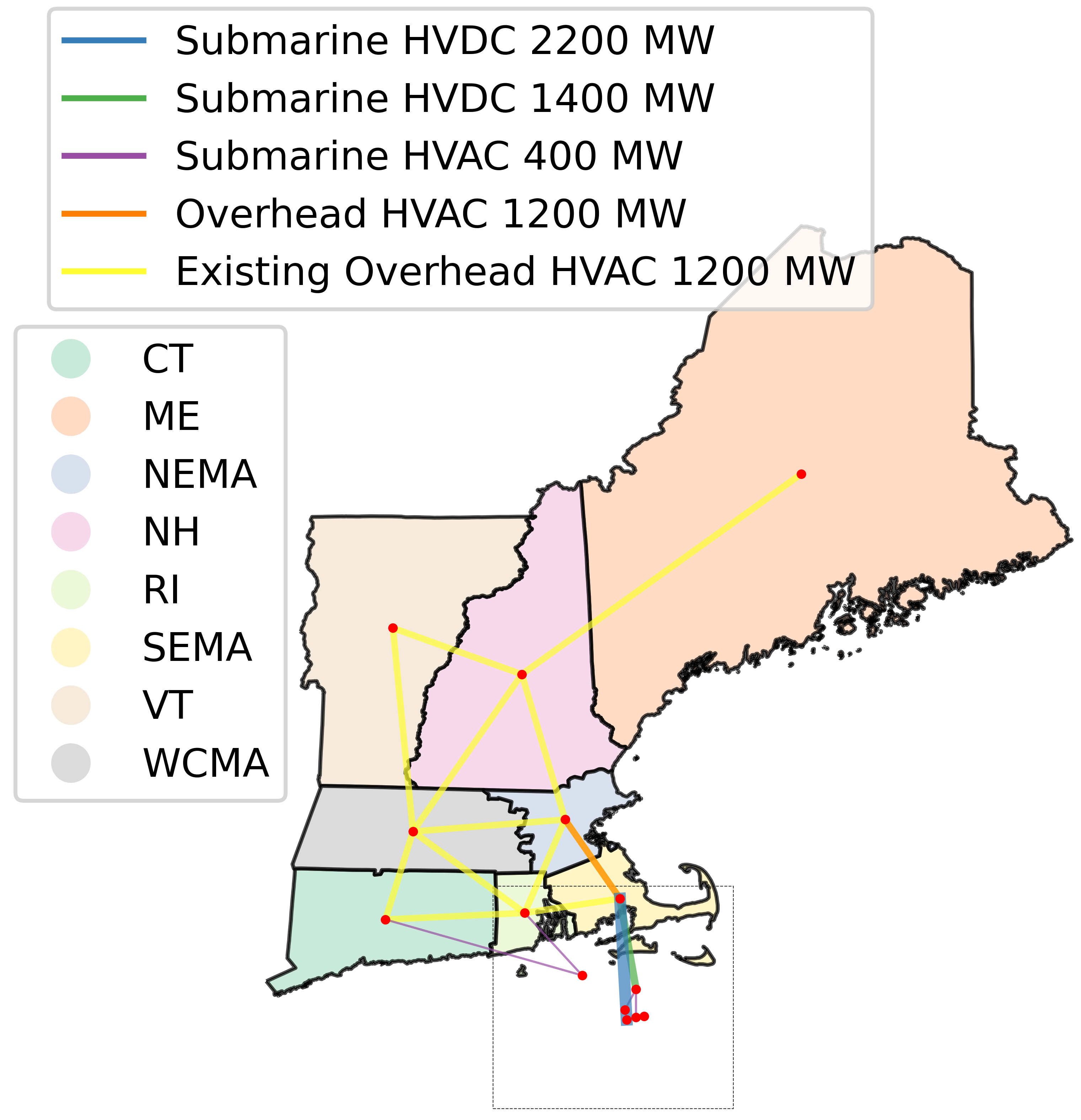}
        \caption{MO X5}
        \label{fig:1_NS_MO_X5}
    \end{subfigure}
    \hfill
    \begin{subfigure}{0.24\textwidth}
        \includegraphics[width=\textwidth]{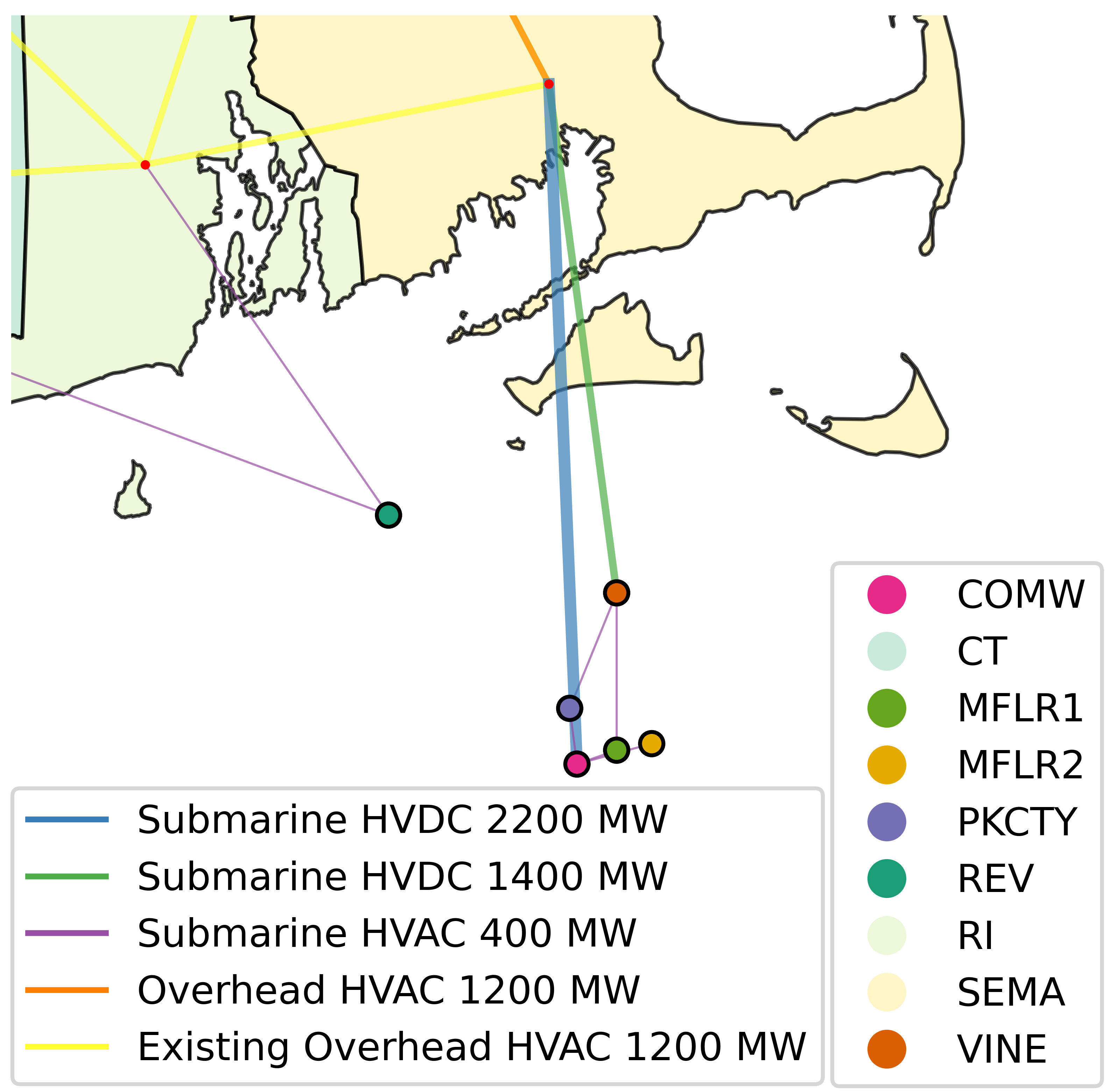}
        \caption{focused (e)}
        \label{fig:1_NS_MO_X5_focus}
    \end{subfigure}
    \caption{Optimal Onshore and Offshore Topology: Impacts of Accounting for Externalities and Extreme Days Using Operational Scenarios Derived from \cite{scott2019clustering}. (\textit{Notes:} The line between NEMA and RI in SO is upgraded in Epoch 2, whereas in SO X5 and MO in Epoch 1.)} 
    \label{fig:1_ns_topology_externalities}
\end{figure}

Fig.~\ref{fig:1_ns_topology_externalities} summarizes again the optimal transmission expansion decisions for SO, SO X5, MO and MO X5 specifications, but now with the new set of operational scenarios following \cite{scott2019clustering}, to capture the impact of externalities, consideration of extreme days, and both. Here, we find that both SO and MO specifications, including SO with extreme days (SO X5), require the same onshore transmission line upgrades between SEMA and NEMA, SEMA and RI, and NEMA and RI (see Fig.~\ref{fig:1_ns_topology_externalities} a and b), while MO with extreme days (MO X5) requires fewer onshore line upgrades (see Fig.~\ref{fig:1_ns_topology_externalities} c and d). As in Fig.~\ref{fig:1_topology_externalities}, offshore topologies remain the same across all four specifications (see Fig.~\ref{fig:1_ns_topology_externalities}). Therefore, accounting for a more comprehensive set of economic costs, as well as operational scenarios, \textit{generally} decrease the need to upgrade onshore transmission. 

\begin{figure}
    \begin{subfigure}{0.24\textwidth}
        \includegraphics[width=\textwidth]{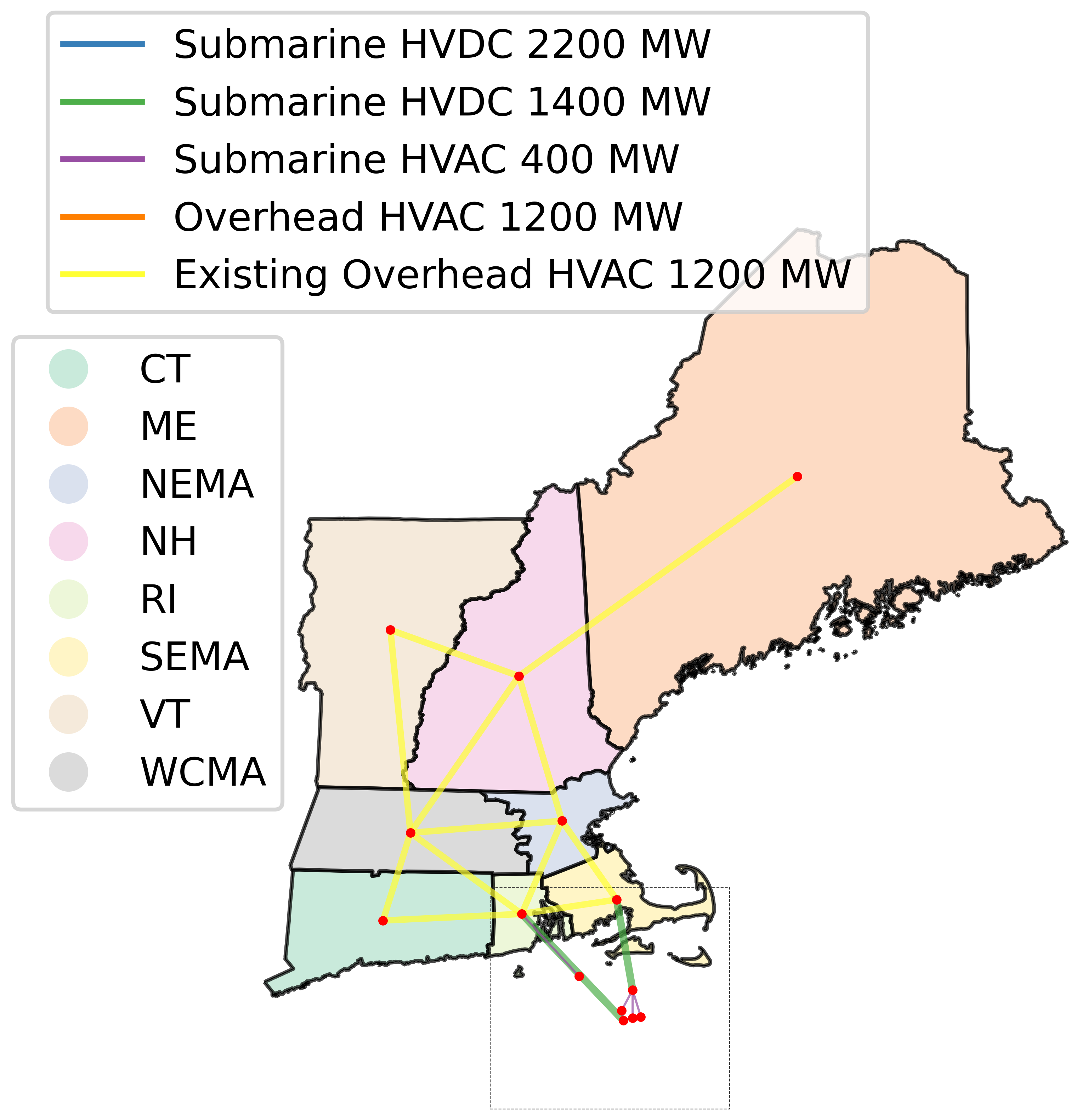}
        \caption{MO OPOI, MO OPOI X5}
        \label{fig:1_MO_OPOI}
    \end{subfigure}
    \hfill
    \begin{subfigure}{0.24\textwidth}
        \includegraphics[width=\textwidth]{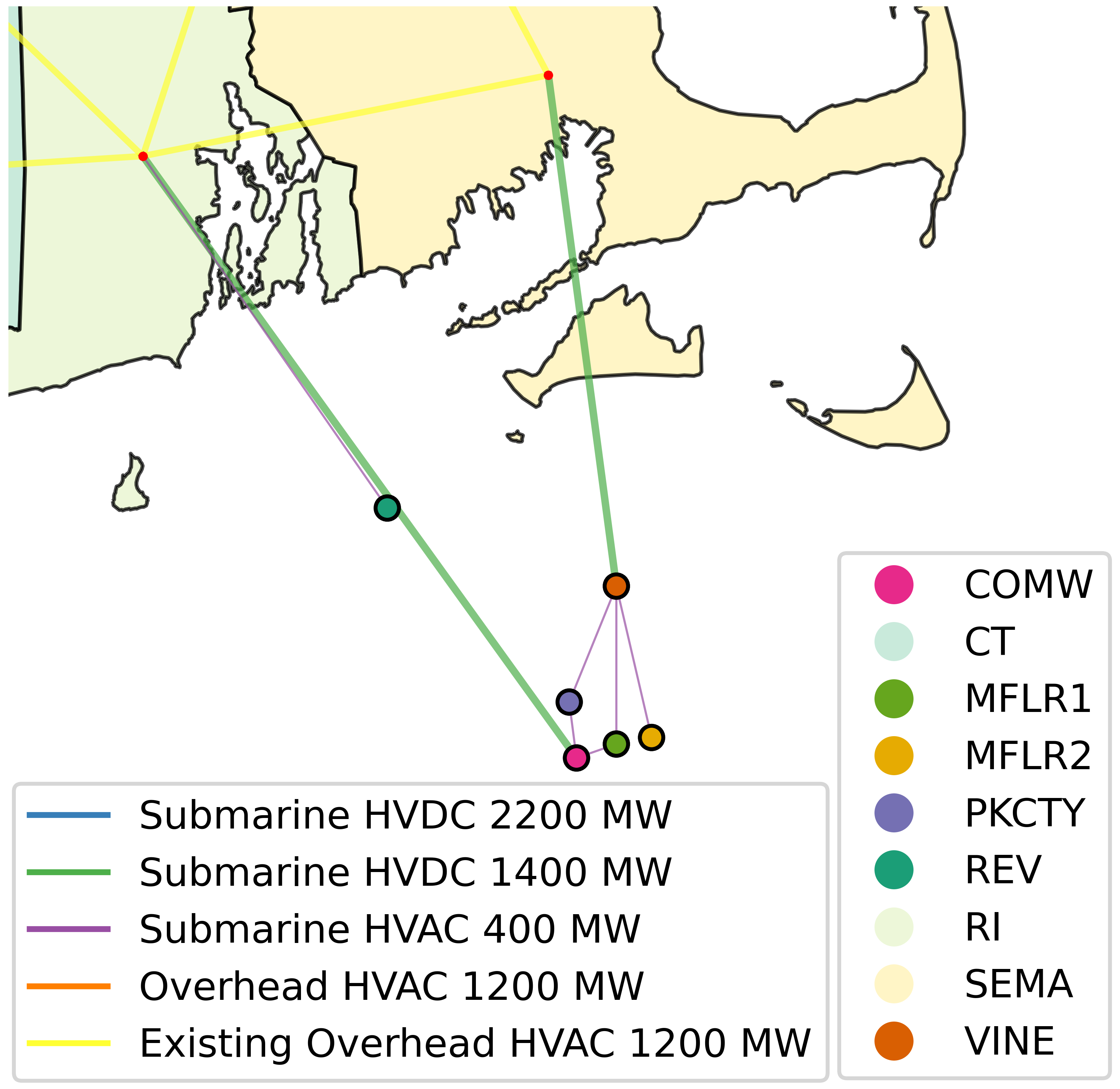}
        \caption{focused (a)}
        \label{fig:1_MO_OPOI_focus}
    \end{subfigure}
    \caption{Optimal Onshore and Offshore Topology: Impacts of Optimizing POI and Consideration of Extreme Days.} 
    \label{fig:1_topology_opoi}
\end{figure}

\begin{figure}
    \begin{subfigure}{0.24\textwidth}
        \includegraphics[width=\textwidth]{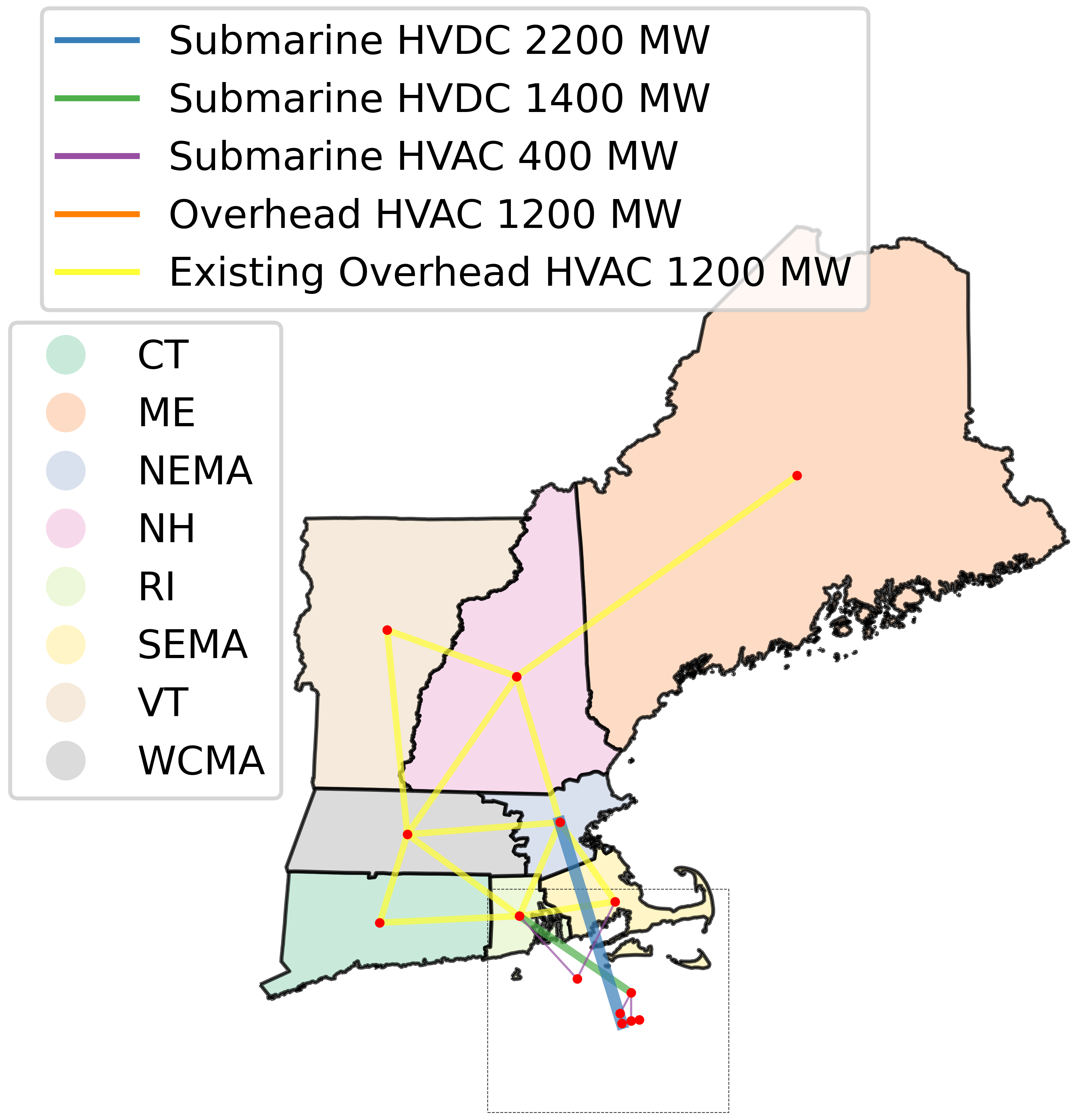}
        \caption{MO OPOI}
        \label{fig:1_NS_MO_OPOI}
    \end{subfigure}
    \hfill
    \begin{subfigure}{0.24\textwidth}
        \includegraphics[width=\textwidth]{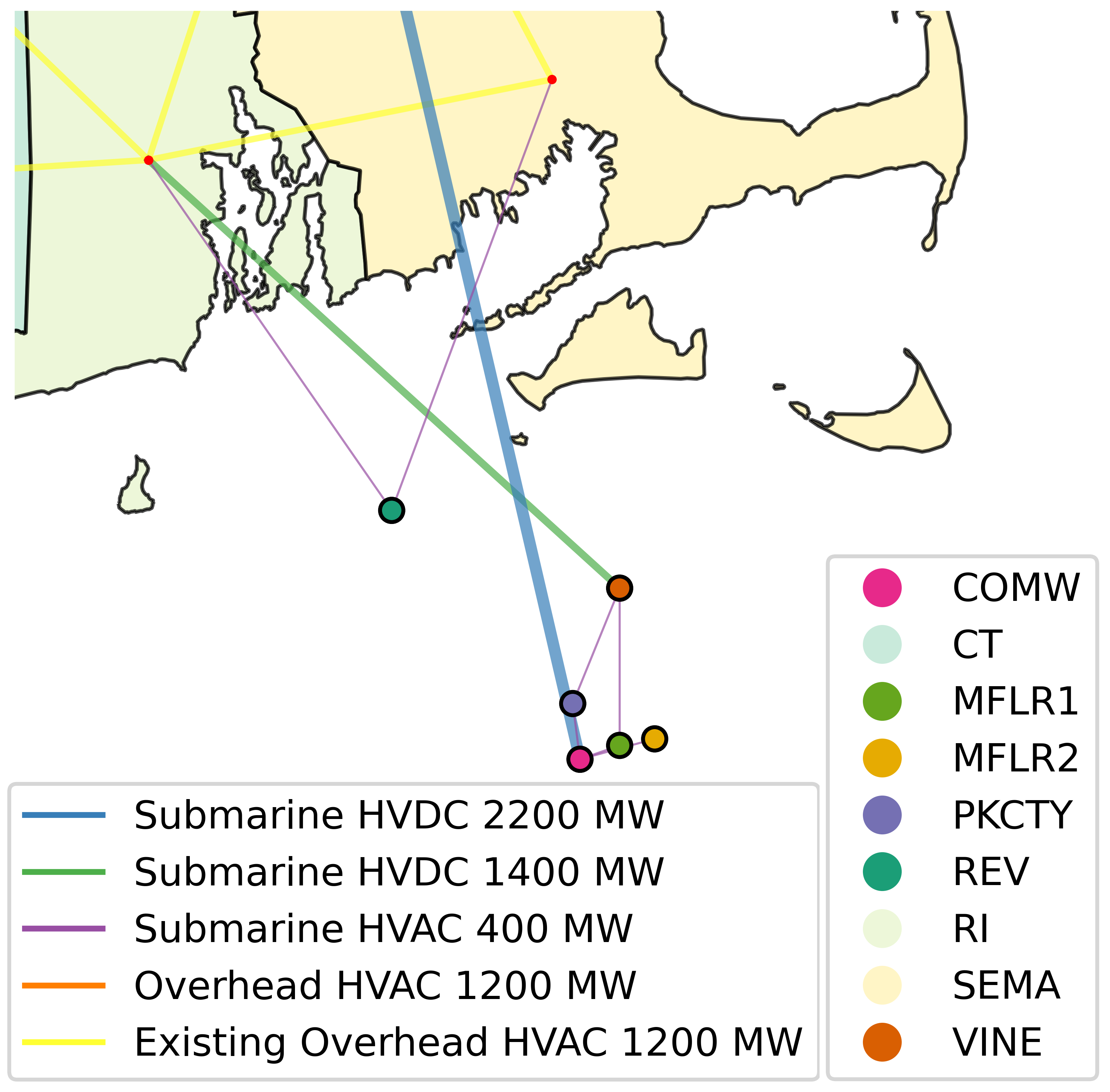}
        \caption{focused (a)}
        \label{fig:1_NS_MO_OPOI_focus}
    \end{subfigure}
        \newline
        \centering
    \begin{subfigure}{0.24\textwidth}
        \includegraphics[width=\textwidth]{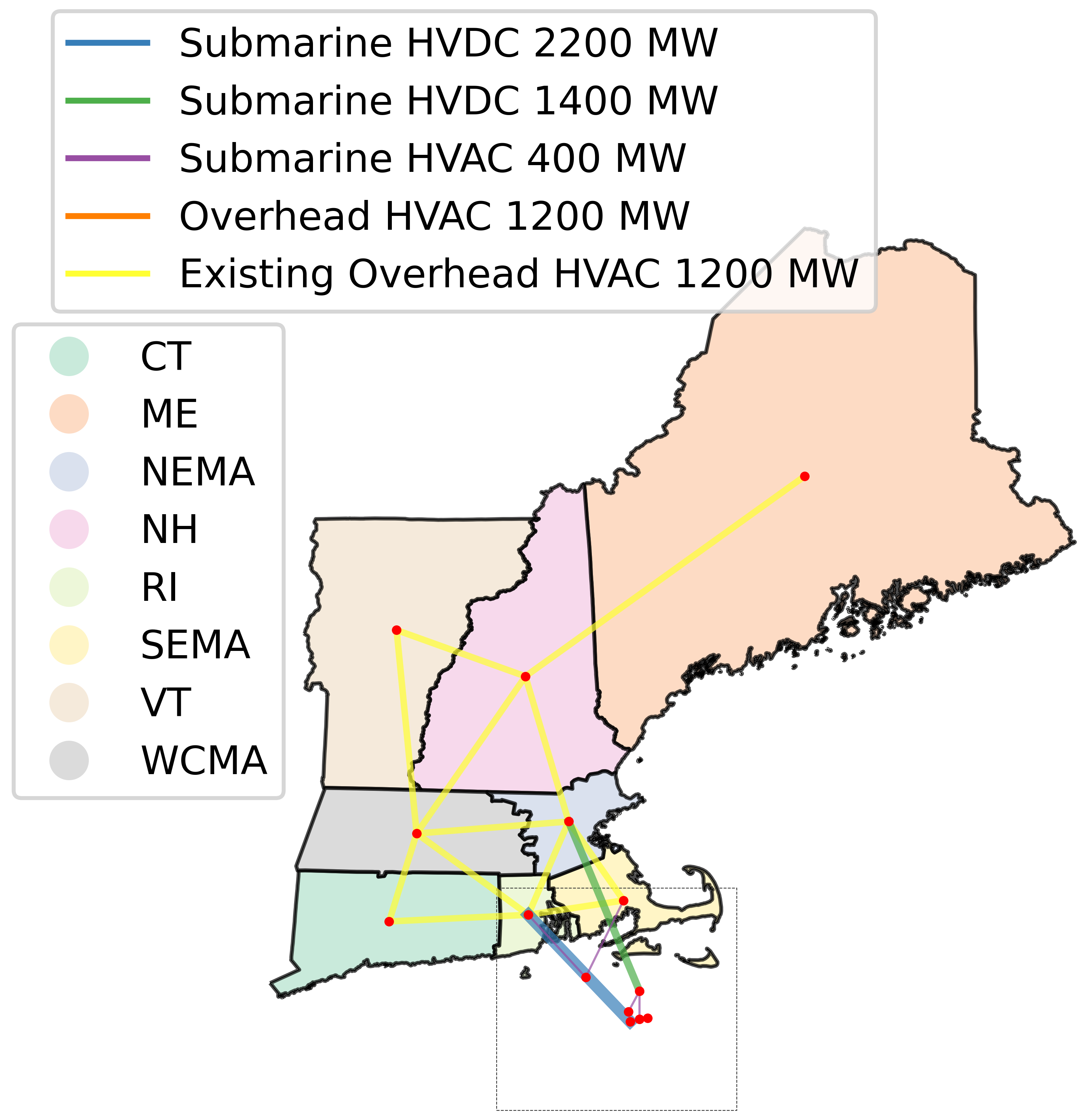}
        \caption{MO OPOI X5}
        \label{fig:1_NS_MO_OPOI_X5}
    \end{subfigure}
    \hfill
    \begin{subfigure}{0.24\textwidth}
        \includegraphics[width=\textwidth]{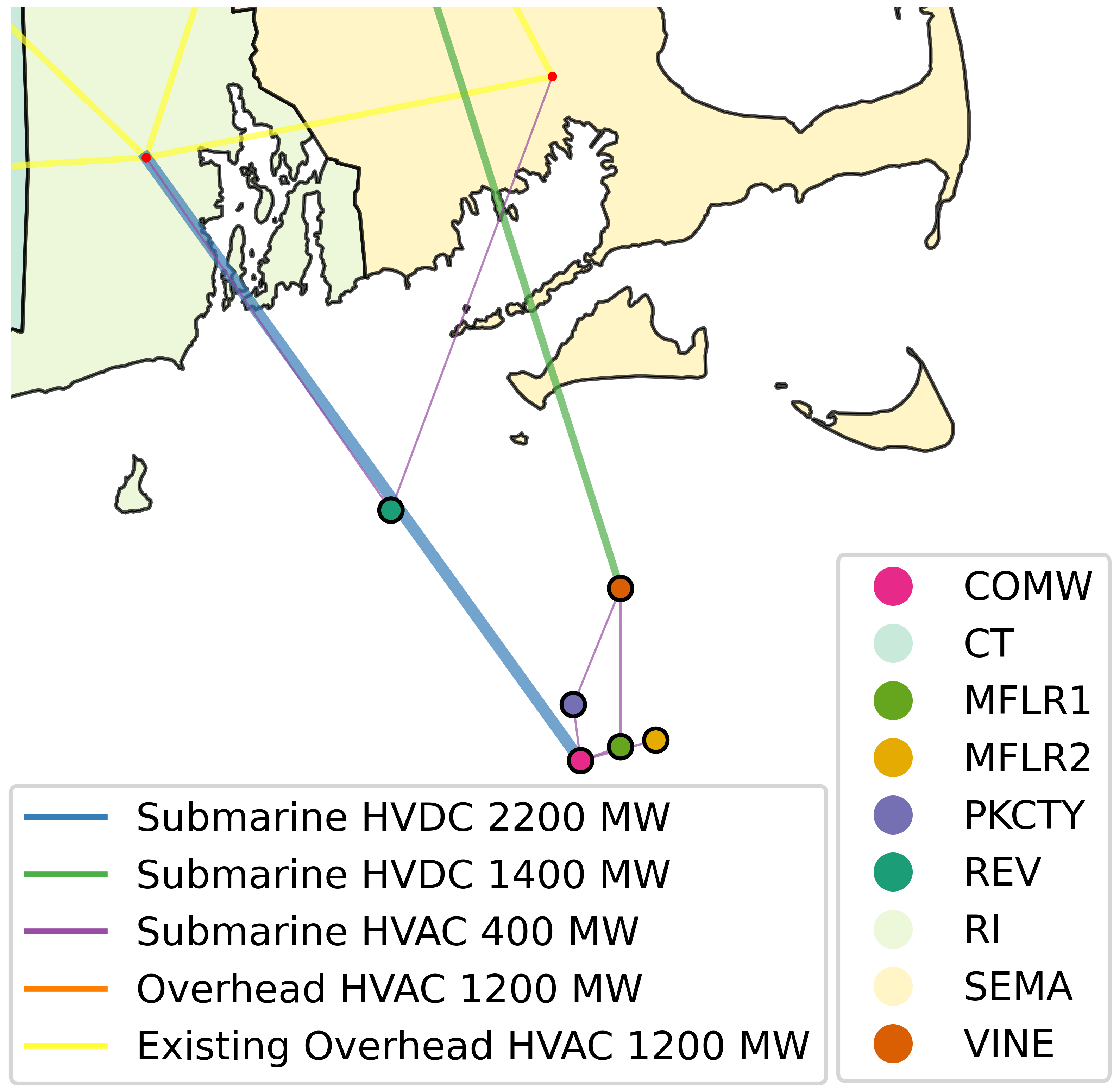}
        \caption{focused (c)}
        \label{fig:1_NS_MO_OPOI_X5_focus}
    \end{subfigure}
    \caption{Optimal Onshore and Offshore Topology: Impacts of Optimizing POI and Consideration of Extreme Days Using Operational Scenarios Derived from \cite{scott2019clustering}.} 
    \label{fig:1_ns_topology_opoi}
\end{figure}

Fig.~\ref{fig:1_topology_opoi} shows optimal transmission outcomes with optimizing POIs, both with normal days (MO OPOI) and extreme days (MO OPOI X5). In both specifications compared to fixed POI specifications, the optimal offshore topologies become more meshed with the onshore grid, favoring an extra HVDC line to RI. The optimal offshore topologies are similarly meshed, but with different POIs, e.g., NEMA instead of SEMA (see Fig.~\ref{fig:1_ns_topology_opoi}), when using the new set of operational scenarios from \cite{scott2019clustering}.

\begin{figure}
    \begin{subfigure}{0.24\textwidth}
        \includegraphics[width=\textwidth]{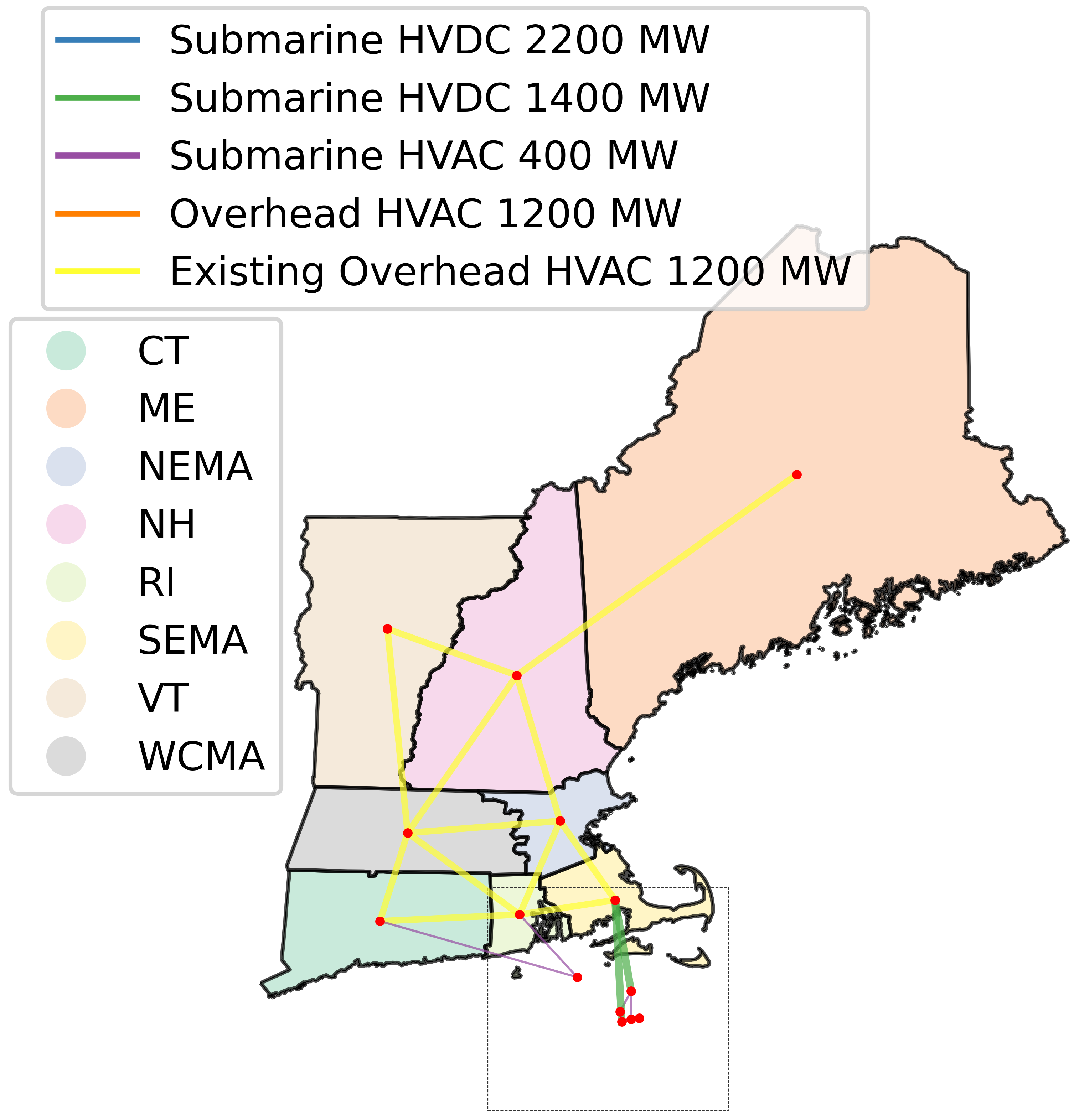}
        \caption{MO EM190}
        \label{fig:1_MO_EM190}
    \end{subfigure}
    \hfill
    \begin{subfigure}{0.24\textwidth}
        \includegraphics[width=\textwidth]{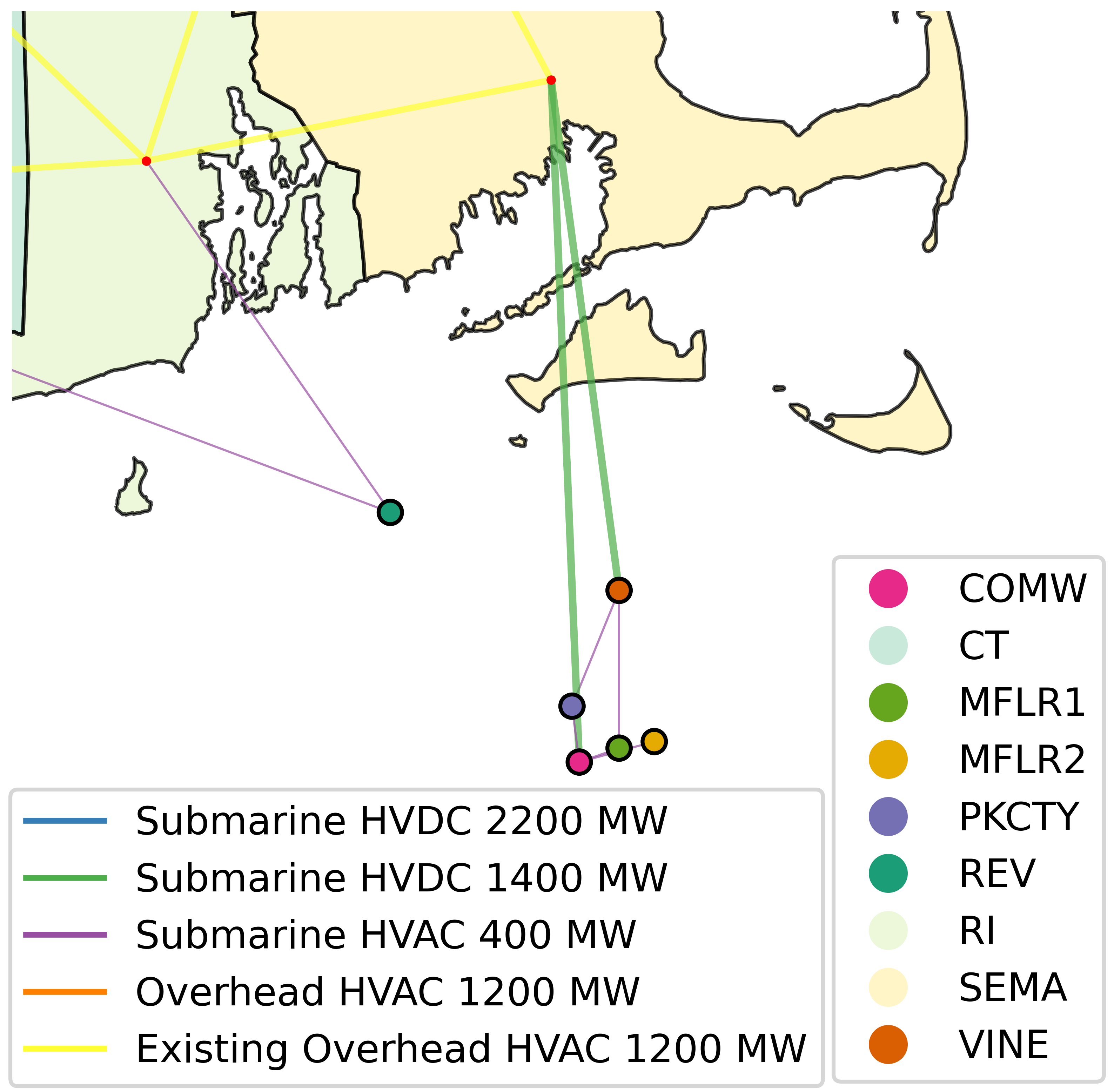}
        \caption{focused (a)}
        \label{fig:1_MO_EM190_focus}
    \end{subfigure}
        \newline
        \centering
    \begin{subfigure}{0.24\textwidth}
        \includegraphics[width=\textwidth]{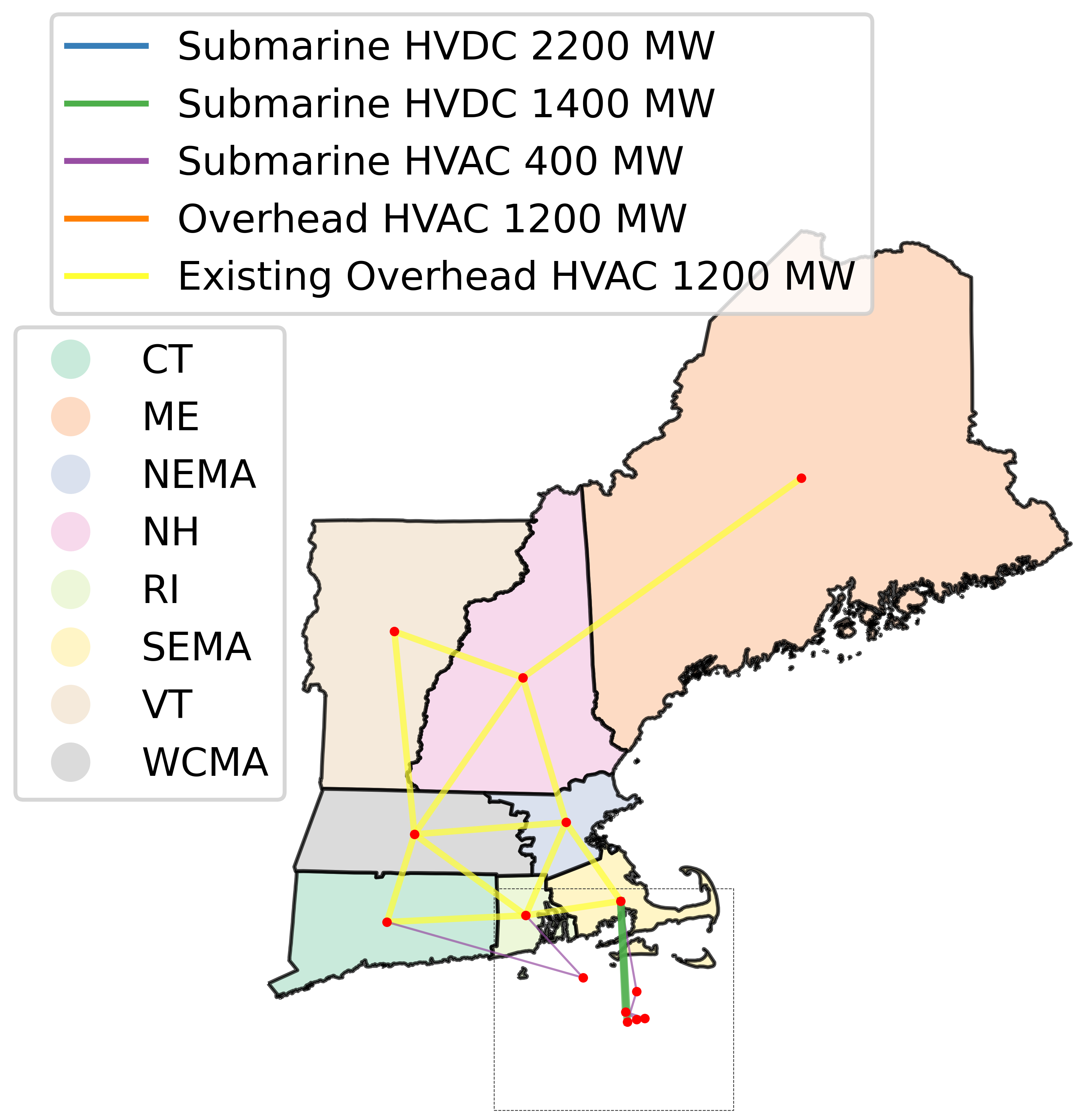}
        \caption{MO EM190 X5}
        \label{fig:1_MO_EM190_X5}
    \end{subfigure}
    \hfill
    \begin{subfigure}{0.24\textwidth}
        \includegraphics[width=\textwidth]{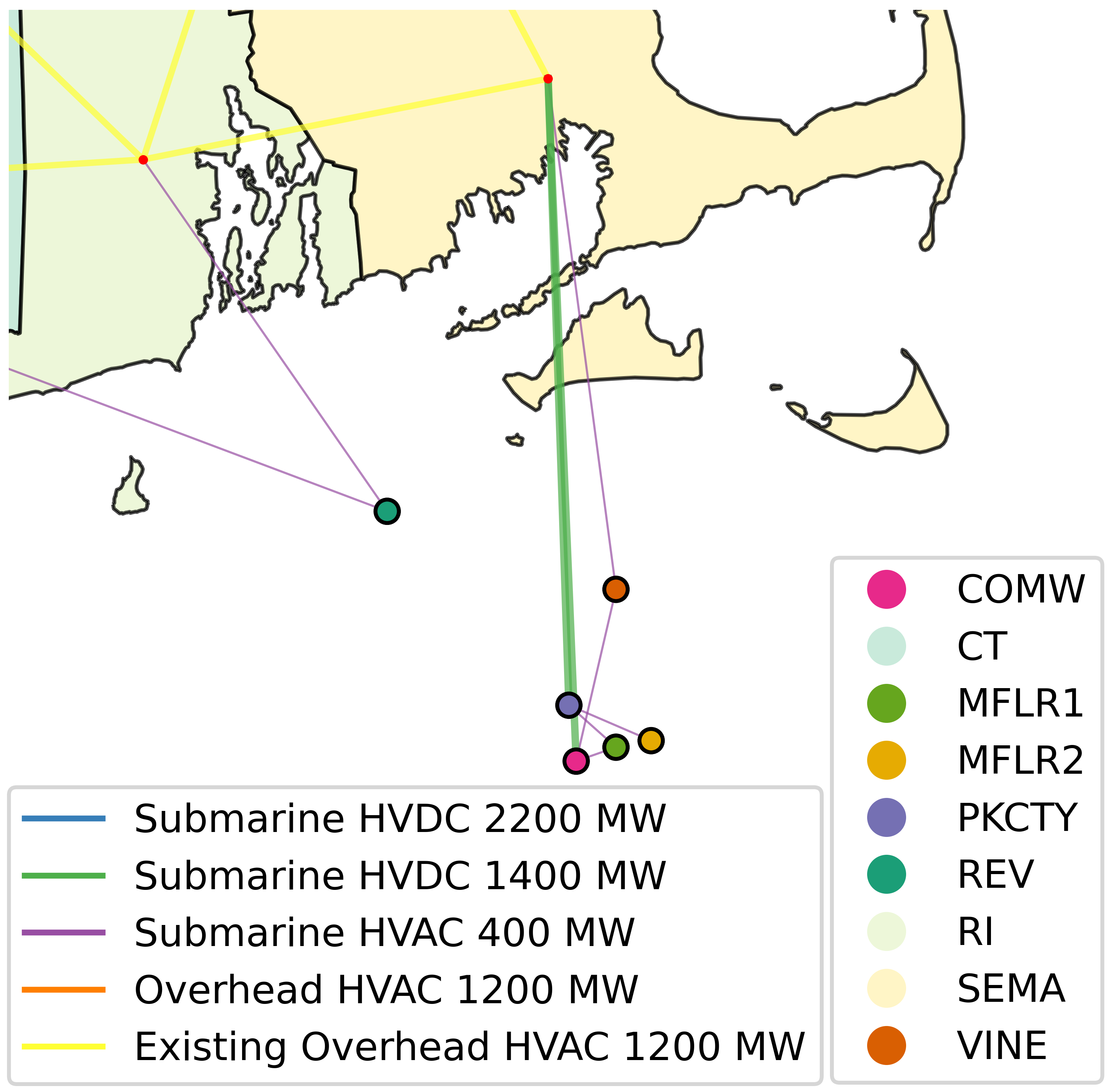}
        \caption{focused (c)}
        \label{fig:1_MO_EM190_X5_focus}
    \end{subfigure}
    \caption{Optimal Onshore and Offshore Topology: Impacts of Higher Social Cost of Carbon Estimate and Consideration of Extreme Days.} 
    \label{fig:1_topology_scc}
\end{figure}

\begin{figure}
    \begin{subfigure}{0.24\textwidth}
        \includegraphics[width=\textwidth]{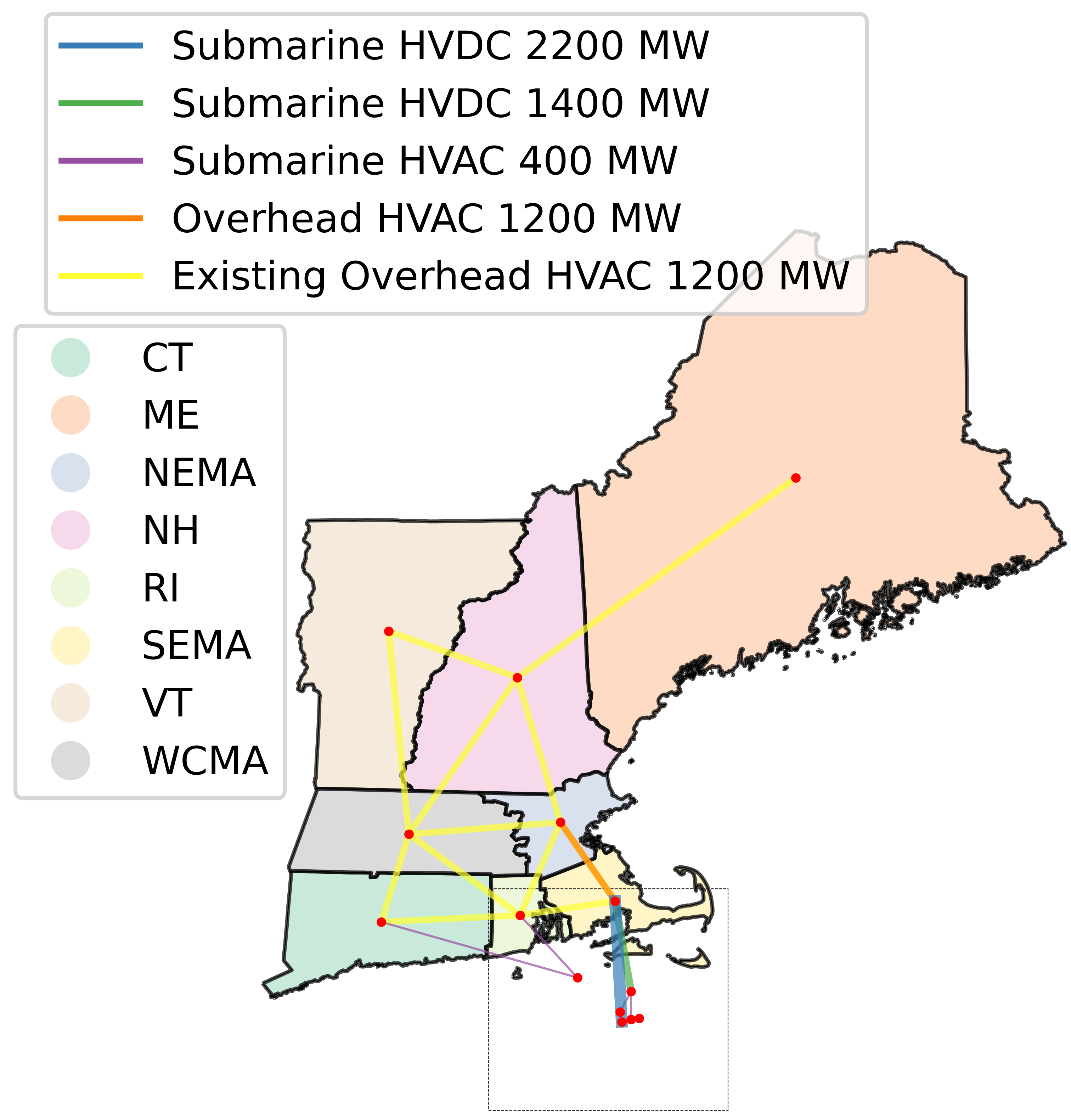}
        \caption{MO EM190, MO EM190 X5}
        \label{fig:1_NS_MO_EM190}
    \end{subfigure}
    \hfill
    \begin{subfigure}{0.24\textwidth}
        \includegraphics[width=\textwidth]{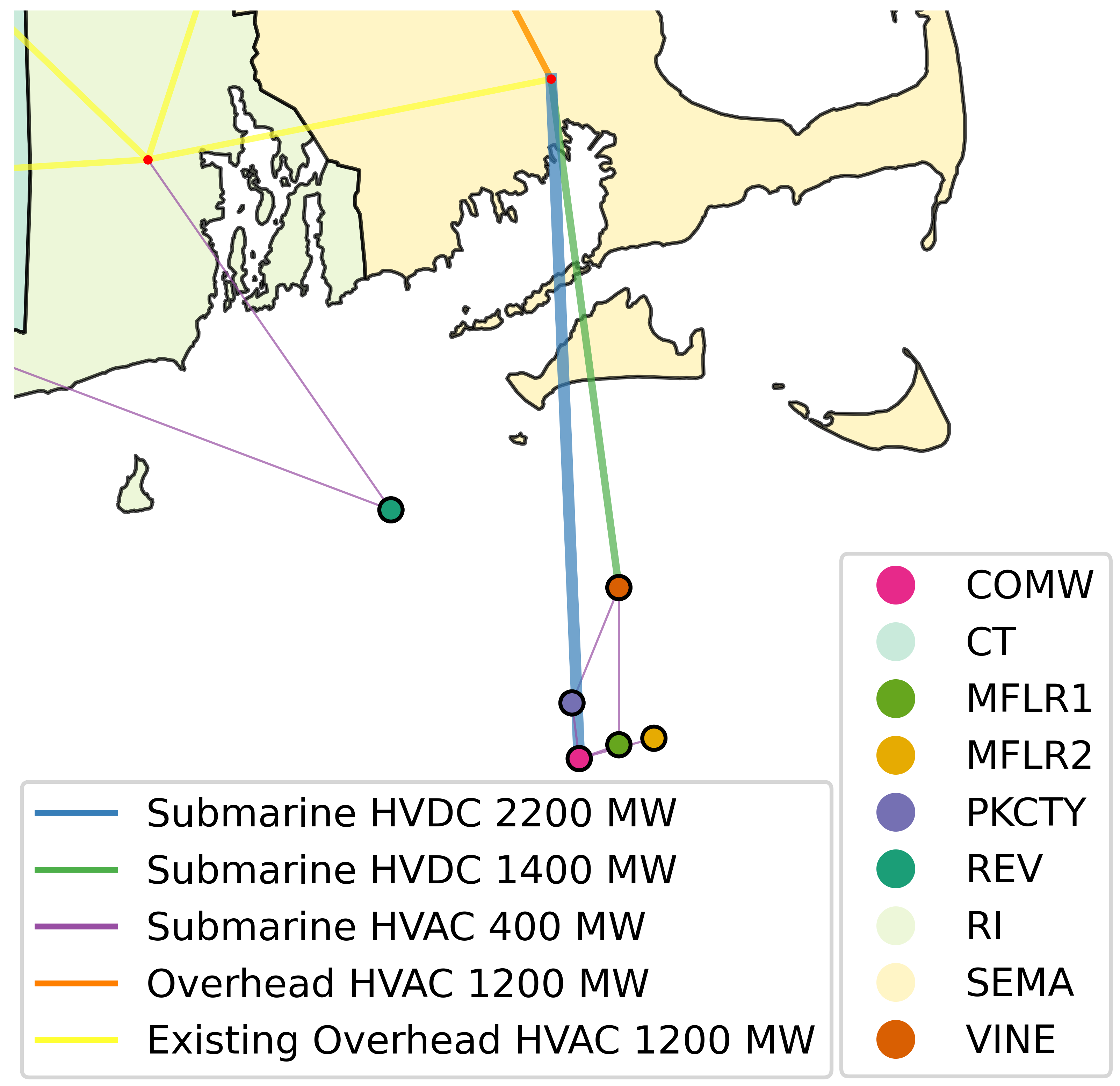}
        \caption{focused (a)}
        \label{fig:1_NS_MO_EM190_focus}
    \end{subfigure}
    \caption{Optimal Onshore and Offshore Topology: Impacts of Higher Social Cost of Carbon Estimate and Consideration of Extreme Days Using Operational Scenarios Derived from \cite{scott2019clustering}.} 
    \label{fig:1_ns_topology_scc}
\end{figure}

Fig.~\ref{fig:1_topology_scc} shows the optimal transmission decisions with a higher SCC estimate equal to \$190 per metric ton. Relative to MO with the baseline SCC of \$51 per metric ton (see Fig.~\ref{fig:1_topology_externalities} c and d), MO EM190 has fewer meshed connections in offshore topology, whereas MO EM190 X5 has more locally meshed connections and has altered the connection point of one HVDC line. However, using the new set of operational scenarios requires one more onshore transmission line upgrade (Fig.~\ref{fig:1_ns_topology_scc}). But compared to the corresponding MO cases (see Fig.~\ref{fig:1_ns_topology_externalities}), there are two fewer onshore line upgrades than the MO, and there is no change compared to MO X5, indicating that a higher SCC may lead to decreased investment in onshore lines.

\begin{figure}
    \centering
    \includegraphics[width=\linewidth]{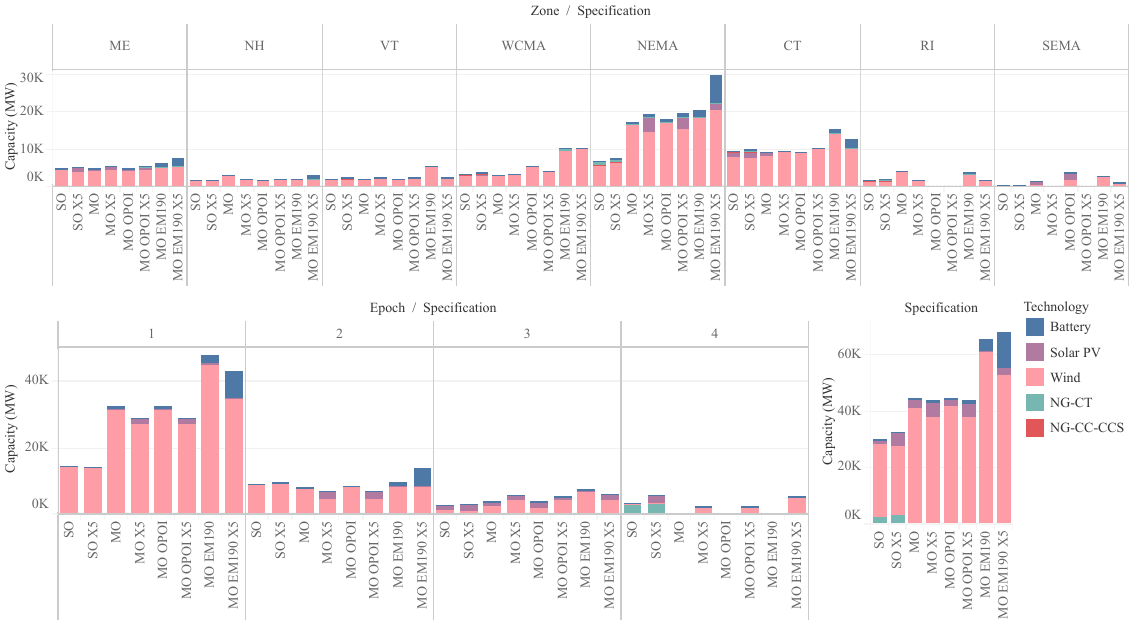}
    \caption{Optimal Generation and Storage Capacity Expansion Decisions with Varying Specifications. \emph{Top}: Across existing onshore zones and model specifications. \emph{Bottom left}: Across epochs and model specifications. \emph{Bottom right}: Total across model specifications. X5 denotes consideration of extreme scenarios.}
    \label{fig:1_capacity}
\end{figure}

\begin{figure}
    \centering
    \includegraphics[width=\linewidth]{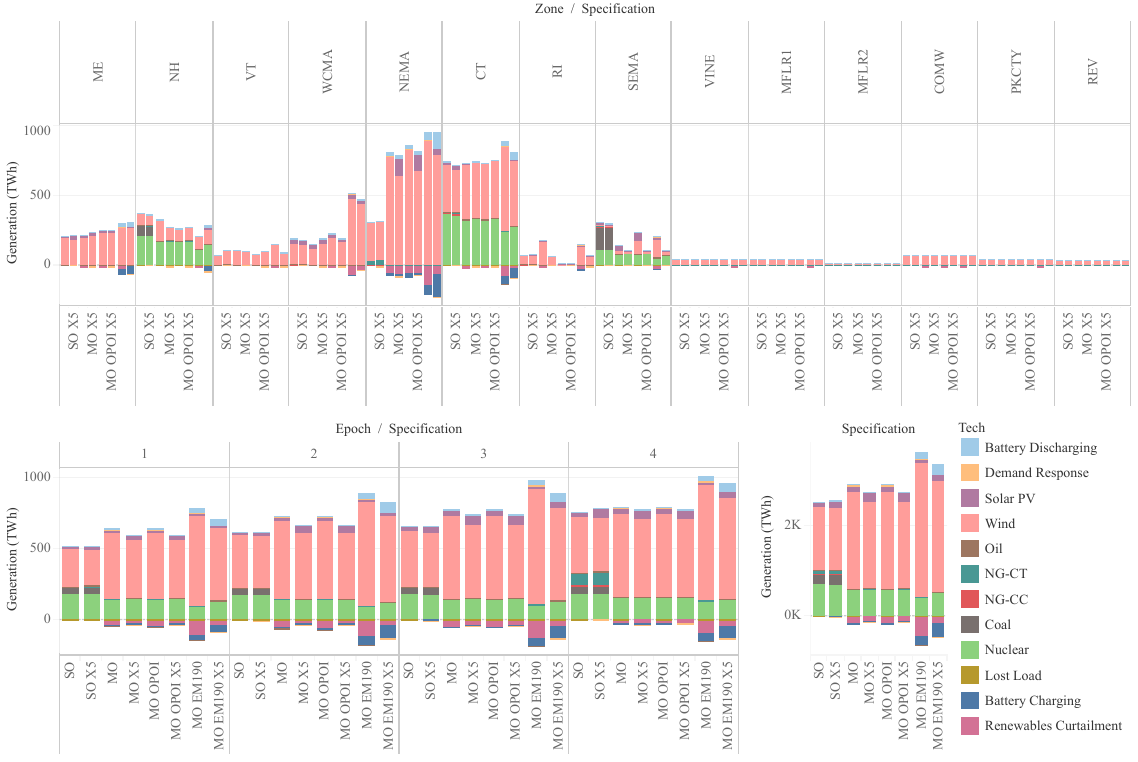}
    \caption{Optimal Generation Decisions with Varying Specifications. \emph{Top}: Across existing onshore zones and model specifications. \emph{Bottom left}: Across epochs and model specifications. \emph{Bottom right}: Total across model specifications. X5 denotes consideration of extreme scenarios.}
    \label{fig:1_generation}
\end{figure}

Fig.~\ref{fig:1_capacity} presents the generation and storage capacity expansion decisions. Most planned generation investment comes from onshore (or land-based) wind resources, some from solar PV, and far less from gas-fired generating and battery storage resources. Note that some states in the ISO-NE footprint have very ambitious RPS targets, which contribute to the onshore wind expansion decisions (and also to the exogenous offshore wind expansion plans). MO and MO OPOI lead to no or insignificant investment in gas-fired generating resources but to more investment in battery storage (SO: 0 MW, MO: 670.05 MW, and MO OPOI:  679.39 MW). Furthermore, land-based wind expansion increases drastically in the MO specifications (SO: 25.46 GW, MO: 41.40 GW, and MO OPOI: 41.43 GW). The increased share of onshore wind capacity will replace less efficient (and hence dirtier) existing generators, as evident from the generation mixes in Fig.~\ref{fig:1_generation}. Epoch-related generation mixes in Fig.~\ref{fig:1_generation} depict the total generation decisions over the entire epoch, i.e., for five years. Relative to the SO specification, accounting for environmental externalities significantly reduces the operation of coal-fired power plants. 

Regarding generation and storage capacity expansion decisions when accounting for extreme days (see Fig.~\ref{fig:1_capacity}), we find that more wind and gas-fired resources are added to the system in the SO specification (comparing SO X5 to SO). This pattern is different in the MO specification (comparing MO X5 to MO) where effectively less total capacity is added but the capacity mix of new generation and storage is more diversified by replacing some of the onshore wind additions with solar PV and storage.

\begin{figure}
    \centering
    \includegraphics[width=\linewidth]{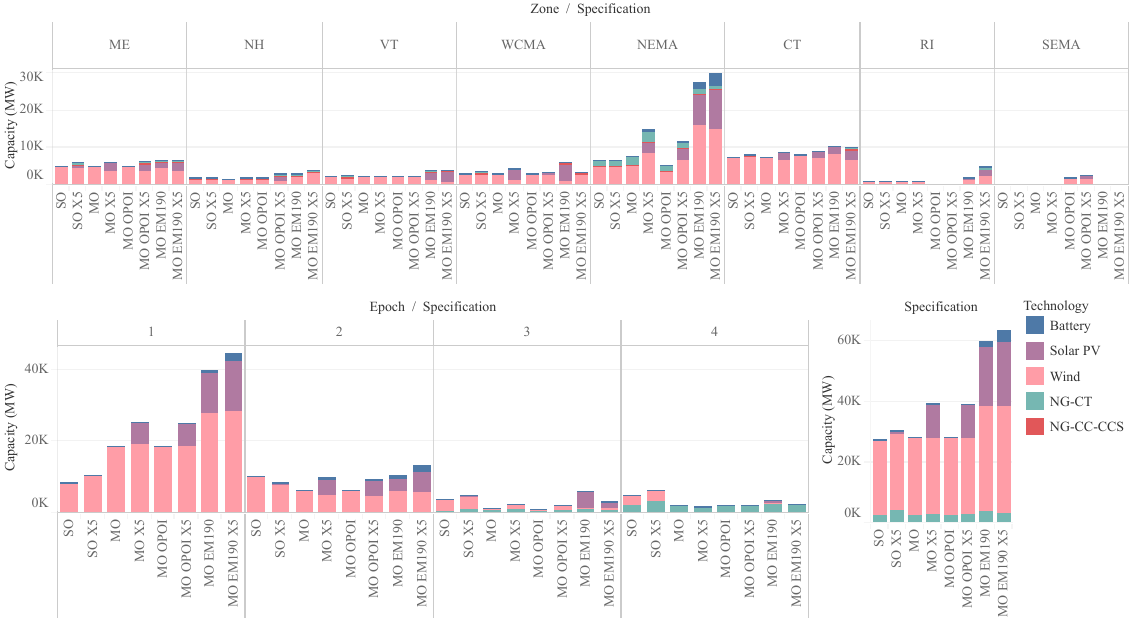}
    \caption{Optimal Generation and Storage Capacity Expansion Decisions with Varying Specifications Using Operational Scenarios Derived from \cite{scott2019clustering}. \emph{Top}: Across existing onshore zones and model specifications. \emph{Bottom left}: Across epochs and model specifications. \emph{Bottom right}: Total across model specifications. X5 denotes consideration of extreme scenarios.} 
    \label{fig:1_ns_capacity}
\end{figure}

\begin{figure}
    \centering
    \includegraphics[width=\linewidth]{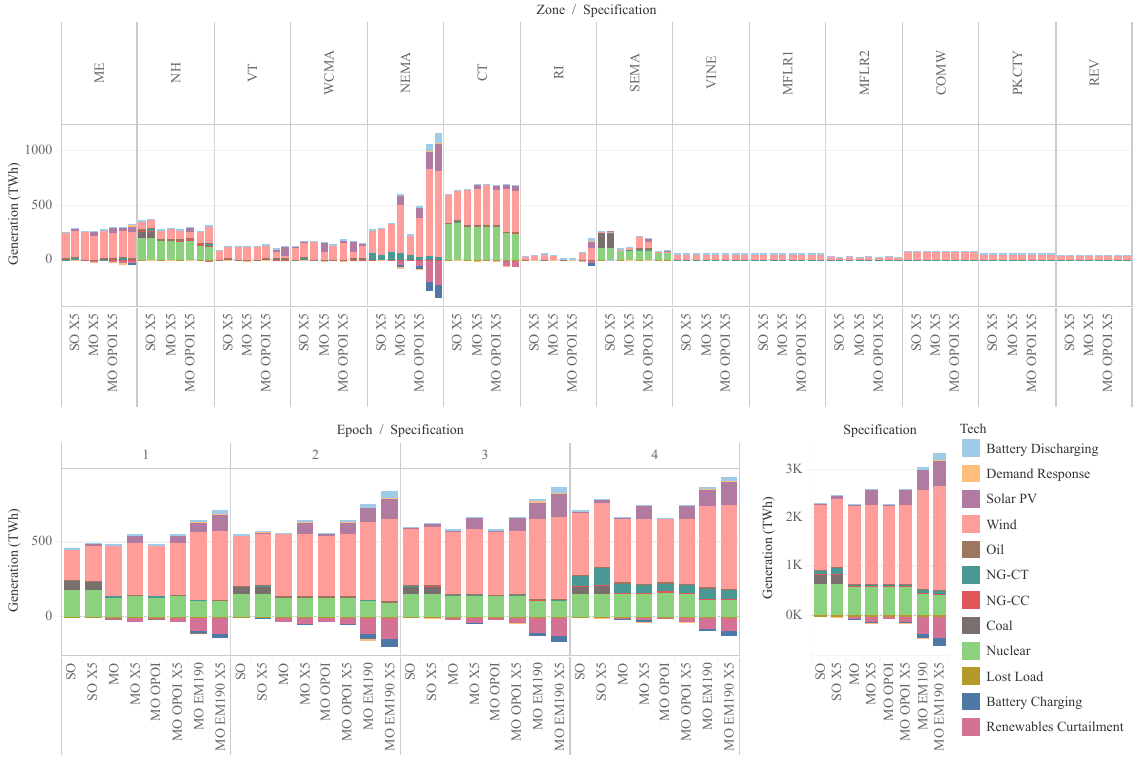}
    \caption{Optimal Generation Decisions with Varying Specifications Using Operational Scenarios Derived from \cite{scott2019clustering}. \emph{Top}: Across existing onshore zones and model specifications. \emph{Bottom left}: Across epochs and model specifications. \emph{Bottom right}: Total across model specifications. X5 denotes consideration of extreme scenarios.} 
    \label{fig:1_ns_generation}
\end{figure}

Fig.~\ref{fig:1_ns_capacity} shows again the same set of results as Fig.~\ref{fig:1_capacity}, but with the new set of operational scenarios. In cases without extreme scenarios, the optimal solution features similar levels of onshore wind investments (SO: 25.23 GW, MO: 25.68 GW, and MO OPOI: 25.61 GW) and no battery storage investment in SO, MO, and MO OPOI. However, with extreme days, there is battery storage investment (SO X5: 99.03 MW, MO X5: 811.15 MW, MO OPOI X5: 770.30 MW), and with MO X5, MO OPOI X5 cases favor investment in solar. And, both MO EM190 and MO EM190 X5 favor significantly higher investment in battery and solar. Finally, Fig.~\ref{fig:1_ns_generation} shows corresponding generation dispatch decisions.

\begin{figure}
    \centering
    \includegraphics[width=\linewidth]{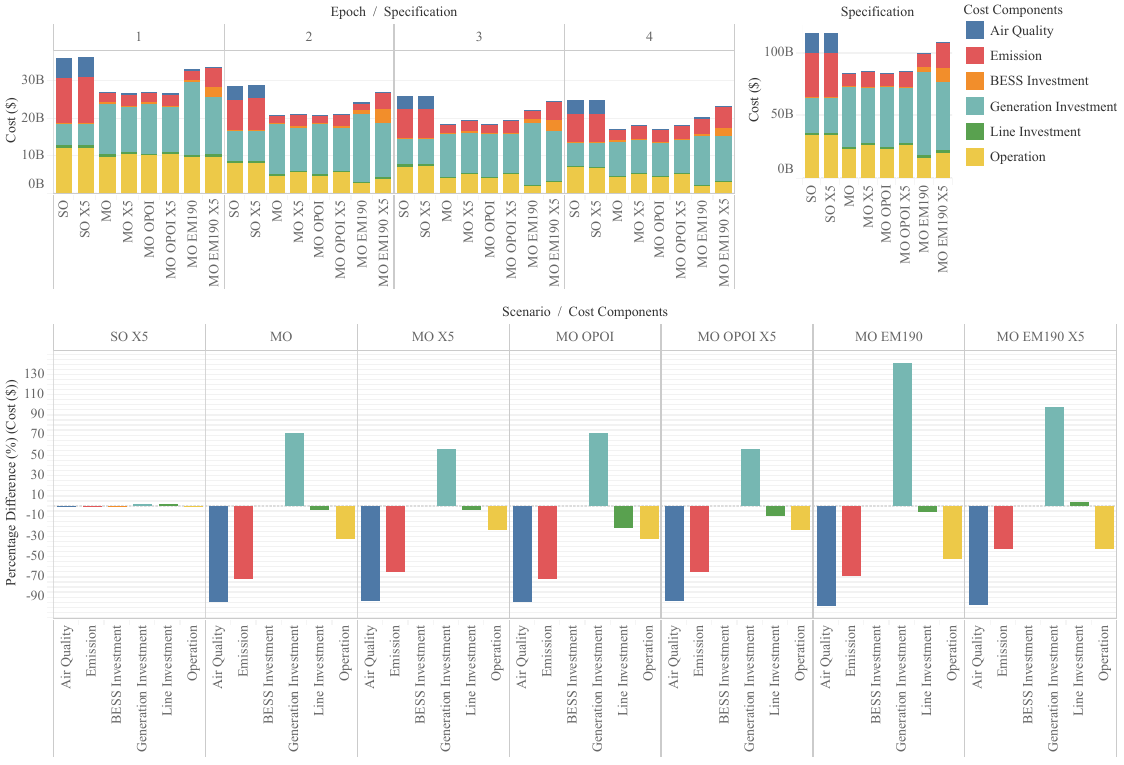}
    \caption{Optimal Costs with Varying Specifications. \emph{Top}: Costs across epochs and model specifications. \emph{Bottom left}: Cost comparisons across model specifications considering SO as the baseline. \emph{Bottom right}: Total costs across model specifications. X5 denotes consideration of extreme scenarios.}
    \label{fig:1_costs}
\end{figure}

Fig.~\ref{fig:1_costs} compares the varying sets of solutions with respect to their cost, considering SO as the base specification. Surprisingly, the total ``hard economic costs,'' i.e., investment costs and expected operating costs, are the same order of magnitude in all three specifications (SO: \$64.05 billion, MO: \$73.41 billion, and MO OPOI: \$73.17 billion). However, both MO cases end up with higher investment costs (SO: \$29.57 billion, MO: \$50.10 billion, and MO OPOI: \$49.88 billion) that are traded off with lower expected operational costs (SO: \$34.48 billion, MO: \$23.31 billion, and MO OPOI: \$23.29 billion). The higher (upfront) investment costs in MO and MO OPOI come with two benefits: lower overall expected operational costs, and a stark decrease in environmental externalities (SO: \$51.72 billion, MO: \$10.56 billion, MO OPOI: \$10.53 billion).

In the MO specifications (MO X5), environmental externality costs are slightly increasing as a response to better accounting for extreme days, but generation and investment costs do not starkly vary across all specifications (see Fig.~\ref{fig:1_costs}).

\begin{figure}
    \centering
    \includegraphics[width=\linewidth]{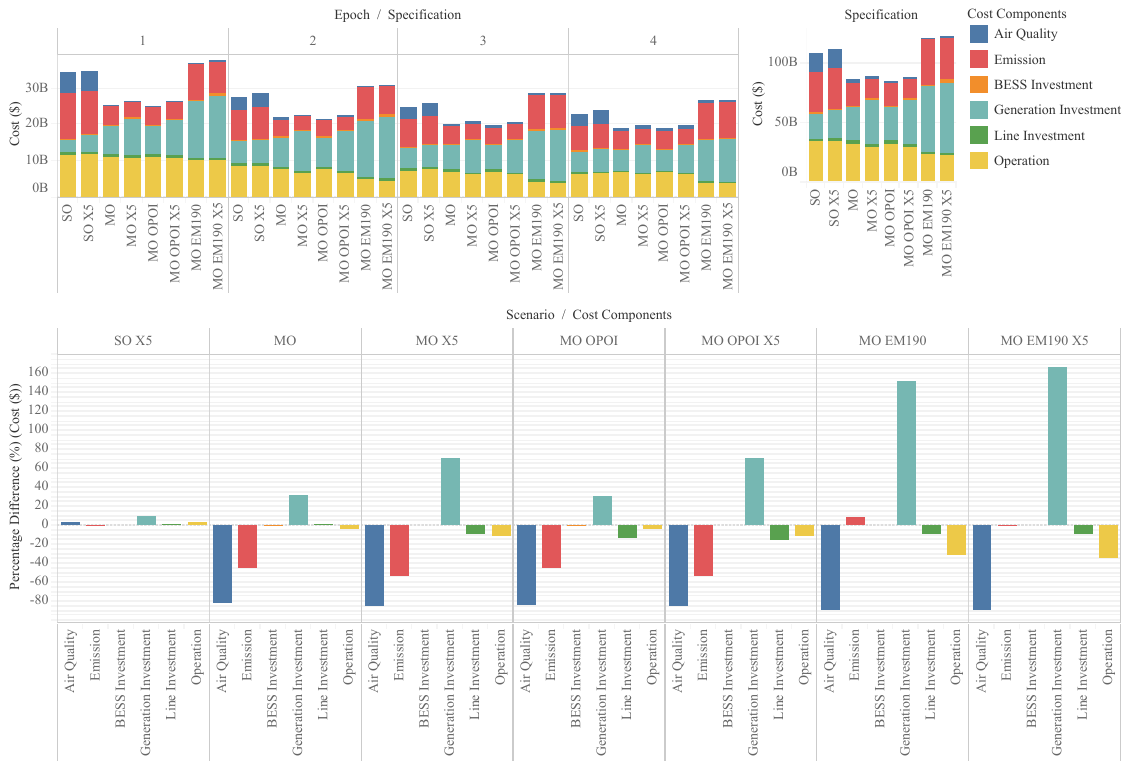}
    \caption{Optimal Costs with Varying Specifications Using Operational Scenarios Derived from \cite{scott2019clustering}. \emph{Top}: Costs across epochs and model specifications. \emph{Bottom left}: Cost comparisons across model specifications considering SO as the baseline. \emph{Bottom right}: Total costs across model specifications. X5 denotes consideration of extreme scenarios. }
    \label{fig:1_ns_costs}
\end{figure}

In Fig.~\ref{fig:1_capacity}, we show that more onshore wind, solar PV, and storage capacity is installed, and operated (see Fig.~\ref{fig:1_generation}) when we use a higher SCC. Consistent with the capacity expansion decisions, the increased capital expenditures are offset by lower expected operational costs, as can be seen in Fig.~\ref{fig:1_costs}. A higher SCC value drastically reduces carbon dioxide emissions over the total planning horizon (MO: 71.65\;MMT\ and MO EM190: 22.34\;MMT). However, because the reduced carbon dioxide emissions are valued at a higher SCC estimate, the carbon externality costs are not necessarily decreasing (MO: \$9.82 billion and MO EM190: \$10.93 billion). Furthermore, additional reductions in greenhouse gas emissions come with increased investment costs (MO: \$50.10 billion and MO EM190: \$71.96 billion), offset by reduced operating costs (MO: \$23.31 billion and MO EM190: \$16.42 billion). 

\begin{figure}
    \centering
    \includegraphics[width=\linewidth]{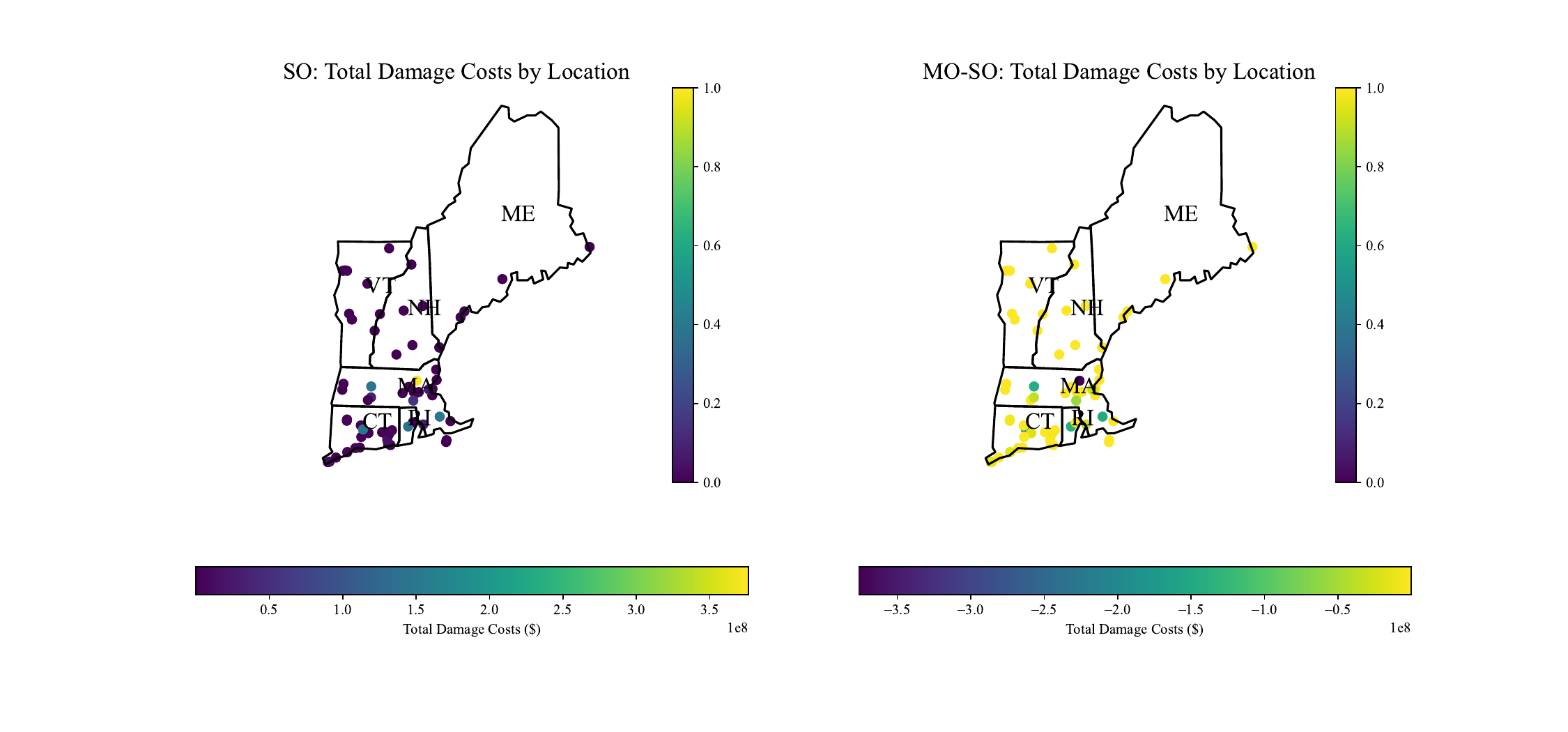}
    \caption{Distribution of Cost Impact Due to Air Quality Damage. \emph{Left}: SO, \emph{Right}: Difference between MO and SO. Horizontal bar: absolute scale, vertical bar: normalized scale. \emph{Notes}: two coal units in SEMA and NH  are removed for scaling.}
    \label{fig:1_SO_and_MO-SO_aq_costs}
\end{figure}

\begin{figure}
    \centering
    \includegraphics[width=\linewidth]{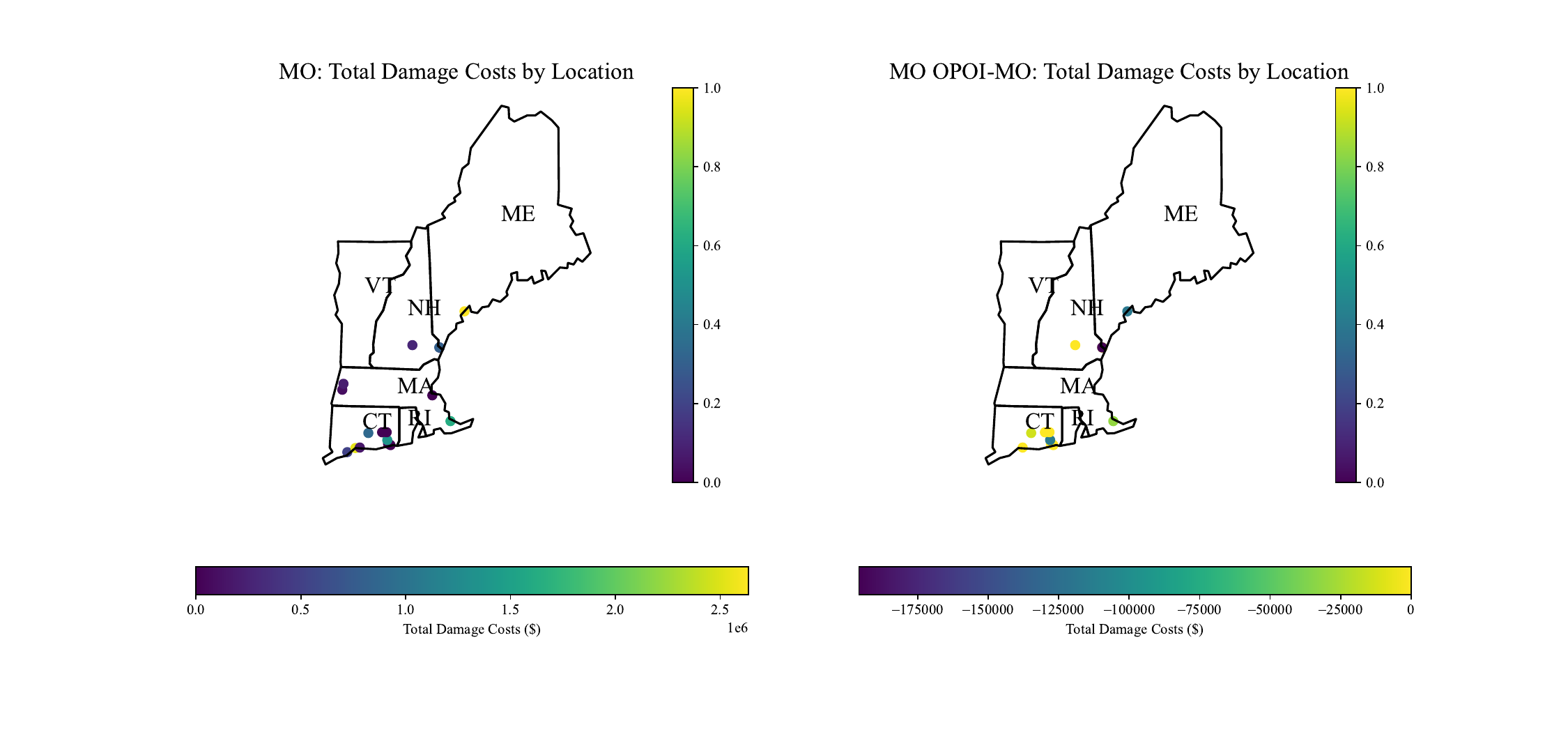}
    \caption{Distribution of Cost Impact Due to Air Quality Damage. \emph{Left}: MO, \emph{Right}: Difference between MO OPOI and MO. Horizontal bar: absolute scale, vertical bar: normalized scale. \emph{Notes}: two coal units in SEMA and NH  are removed for scaling.}
    \label{fig:1_MO_and_MO_OPOI-MO_aq_costs}
\end{figure}

Fig.~\ref{fig:1_SO_and_MO-SO_aq_costs} displays the total air quality damage estimates of the SO specification and the difference between MO and SO specifications, summed over the planning horizon. Fig.~\ref{fig:1_MO_and_MO_OPOI-MO_aq_costs} presents the total air quality damage estimates for the MO specification and the difference between the MO OPOI and MO specifications. Both accounting for externalities and optimizing points of interconnection lead to a reduction in air quality damages. However, it should be noted that accounting for externalities has a greater impact compared to optimizing interconnection points. Total air quality damage costs for all zones across all specifications are given in Table~\ref{table:aq_costs}.

\begin{table}[H]
\centering
\caption{Air Quality Damage Costs with Varying Specifications [\$MM].}
\resizebox{\columnwidth}{!}{
    \begin{tabular}{l l l l l l l l l}     
    \toprule
    Zone\textbackslash{}Specification & SO & SO X5 & MO & MO X5 & MO OPOI & MO OPOI X5 & MO EM190 & MO EM190 X5 \\
    \midrule
    ME & 6.91 & 11.26 & 2.63 & 7.61 & 2.52 & 7.45 & -   & 2.96 \\
    NH & 4,198.67 & 4,121.65 & 11.19 & 24.26 & 10.69 & 22.22 & -   & 3.61 \\
    VT & 13.76 & 146.70 & -   & 1.15 & -   & 0.99 & -   & 0.83 \\
    WCMA & 184.82 & 182.86 & 0.31 & 6.30 & 0.31 & 4.95 & -   & 2.86 \\
    NEMA & 376.22 & 411.39 & -   & -   & -   & -   & -   & -   \\
    CT & 261.29 & 261.39 & 5.49 & 21.03 & 5.36 & 18.84 & - &	 9.43 \\
    RI & 142.45 & 166.31 & -   & 0.05 & -   & 0.05 & -   & 0.03 \\
    SEMA & 6,050.95 & 5,993.22 & 10.76 & 32.14 & 10.73 & 34.36 & -   & 2.74 \\
    \bottomrule
    \end{tabular}
}
\label{table:aq_costs}
\end{table}

\section{PJM Case Study} \label{sec:case_study_pjm}
\begin{figure}
    \centering
    \includegraphics[width=0.50\textwidth]{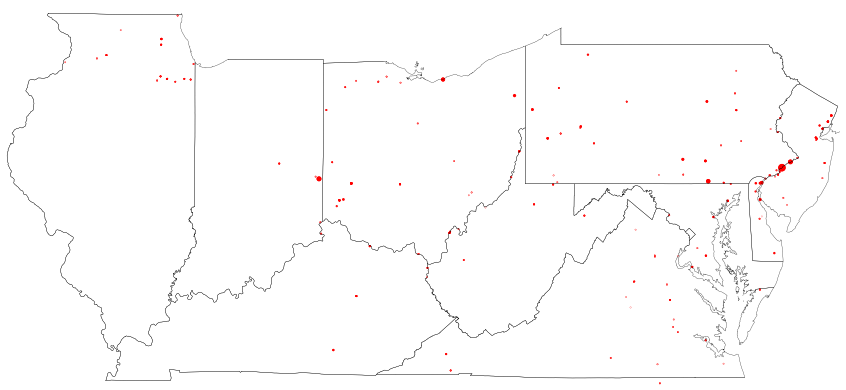}
        \caption{Average Marginal Damages from Local Air Pollution in PJM. (\textit{Notes:} Size of the red dots represents the \$/MWh average marginal damages with a maximum value of 7,237.12\;\$/MWh and a minimum value of 0.47\;\$/MWh.)}
            \label{fig:2_ave_marg_damages_map}
\end{figure}

This section describes our numerical results for the PJM test system. PJM data was sourced from the EPA Zonal Model \cite{EPA2024PowerSectorModeling}, following fossil-fueled generator retirement data from \cite{grubert2020fossil} that is roughly consistent with the PJM estimates \cite{pjm2023retirements}, which is different from the ISO-NE test system in \cite{krishnamurthy20158} used above. Therefore, we had to appropriate the dataset to fit the proposed model to allow for a systematic comparison. In particular, we have set the limit of transmission expansion capability to match the capacity of the existing corridor, consistent with the ISO-NE case study, where discrete sizes of the existing corridors can be selected. Additionally, the PJM region does not have well-defined state borders for Renewable Portfolio Standards (RPS) mandates, leading to potential over- or under-estimation in the buildouts of new renewables. Despite these limitations, analyzing the PJM system remains valuable for understanding the impact of coordinated onshore and offshore grid planning while accounting for critical negative externalities. This approach also provides insights into how other sensitivities compare to the ISO-NE Case Study. We employed a methodology similar to the one used for ISO-NE to prepare the input data for the PJM system and use the same model parameters given in Table~\ref{table:model_params}. Table~\ref{table:rps_states_zones_pjm} outlines the RPS requirements for each state within PJM. However, since parts of these states may not be within PJM, some zones might be constructing excess renewables relative to their share. Fig.~\ref{fig:2_ave_marg_damages_map} shows average marginal damages from local air pollution in PJM.

\begin{table}
  \centering
  \footnotesize
  \caption{RPS by States and Zones.}
  \begin{tabular}{lll}
    \toprule
    State (Zone) & Target Year & RPS (\%) \\
    \midrule
    MD (PJM\_AP, PJM\_SMAC) & 2030 & 50 \\
    OH (PJM\_ATSI, PJM\_West) & 2026 & 8.5 \\
    IL (PJM\_COMD) & 2040 & 50 \\
    VA (PJM\_Dom) & 2045 & 100 \\
    NJ (PJM\_EMAC) & 2030 & 50 \\
    PA (PJM\_PENE, PJM\_WMAC) & 2021 & 18 \\
    \bottomrule
  \end{tabular}
  \label{table:rps_states_zones_pjm}
\end{table}

\begin{figure}
    \centering
    \begin{subfigure}{0.36\textwidth}
        \includegraphics[width=\textwidth]{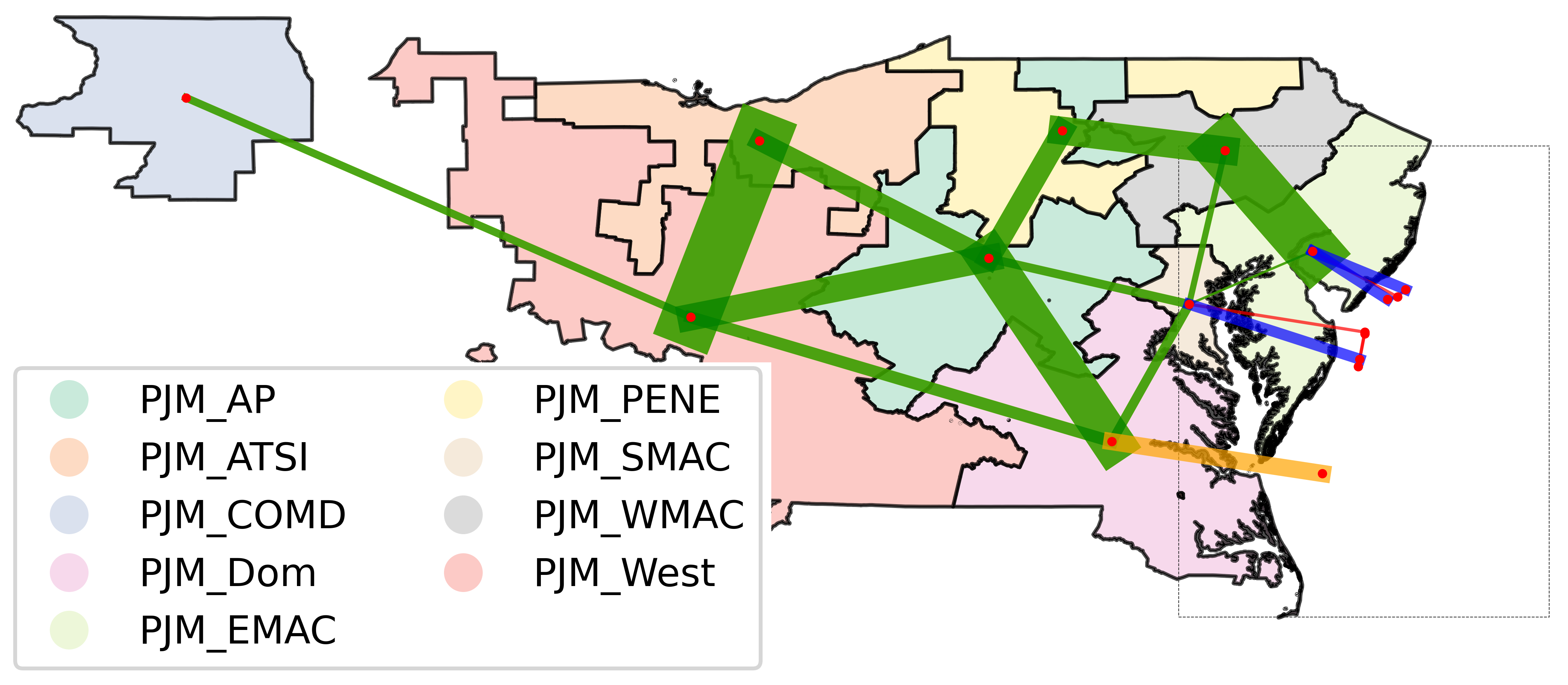}
        \caption{SO}
        \label{fig:2_SO}
    \end{subfigure}
    \hfill
    \begin{subfigure}{0.12\textwidth}
        \includegraphics[width=\textwidth]{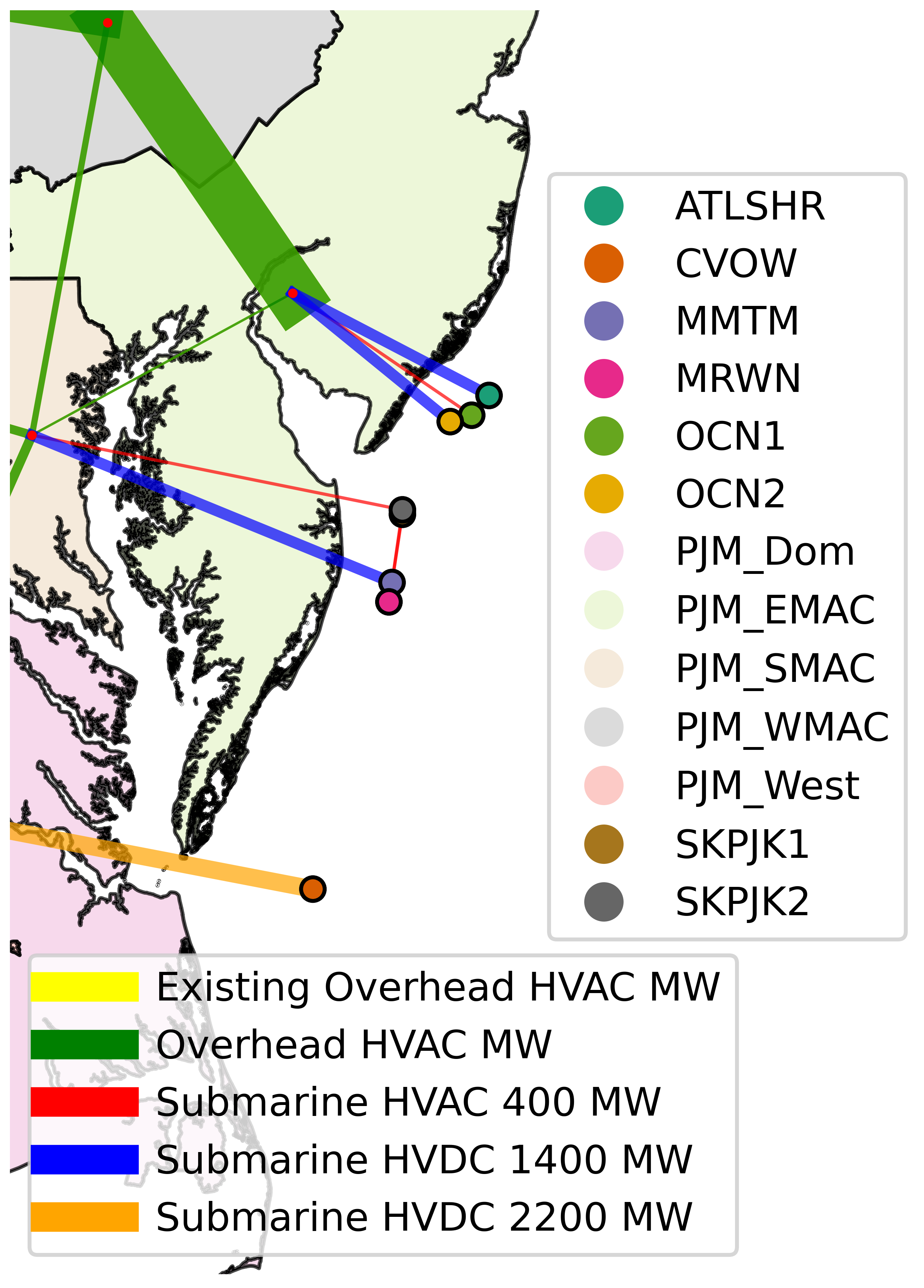}
        \caption{focused (a)}
        \label{fig:2_SO_focus}
    \end{subfigure}
    \newline
    \begin{subfigure}{0.36\textwidth}
        \includegraphics[width=\textwidth]{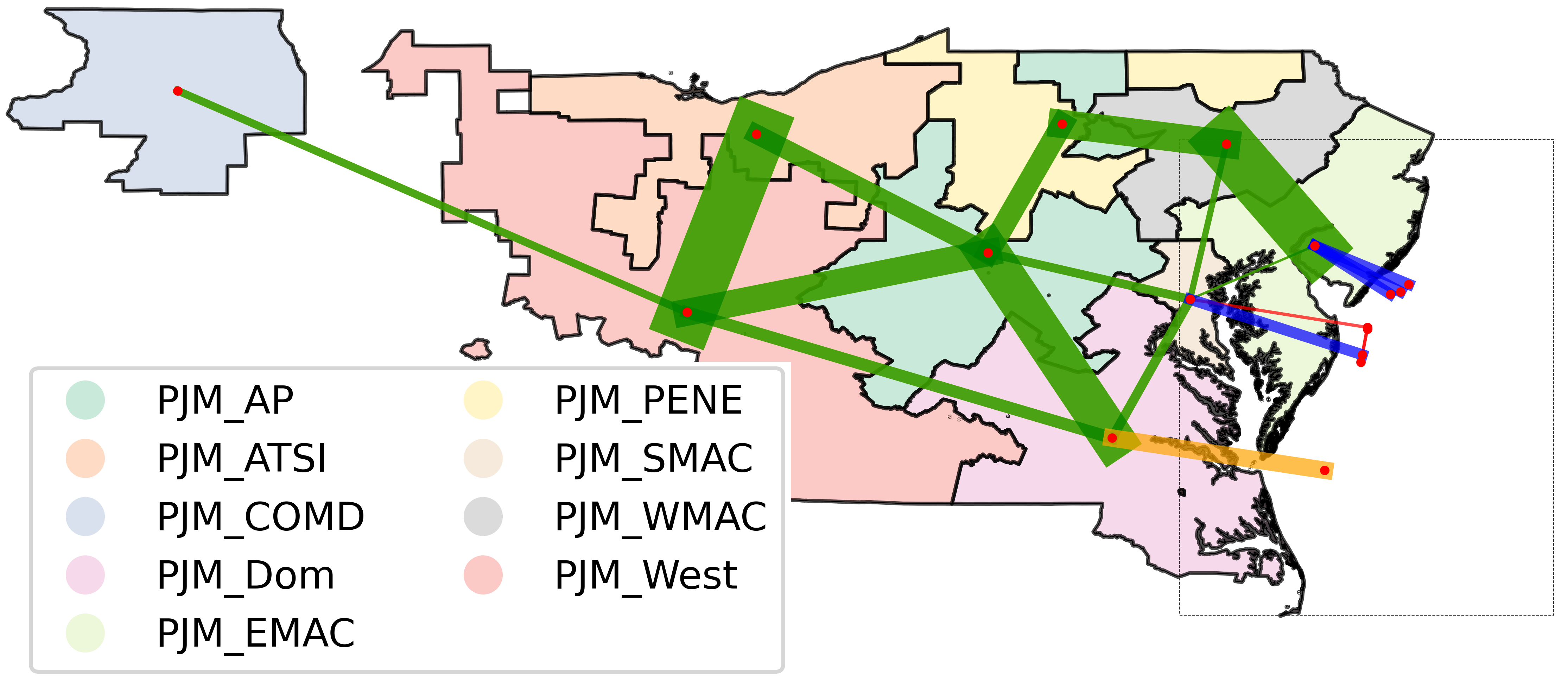}
        \caption{MO}
        \label{fig:2_MO}
    \end{subfigure}
    \hfill
    \begin{subfigure}{0.12\textwidth}
        \includegraphics[width=\textwidth]{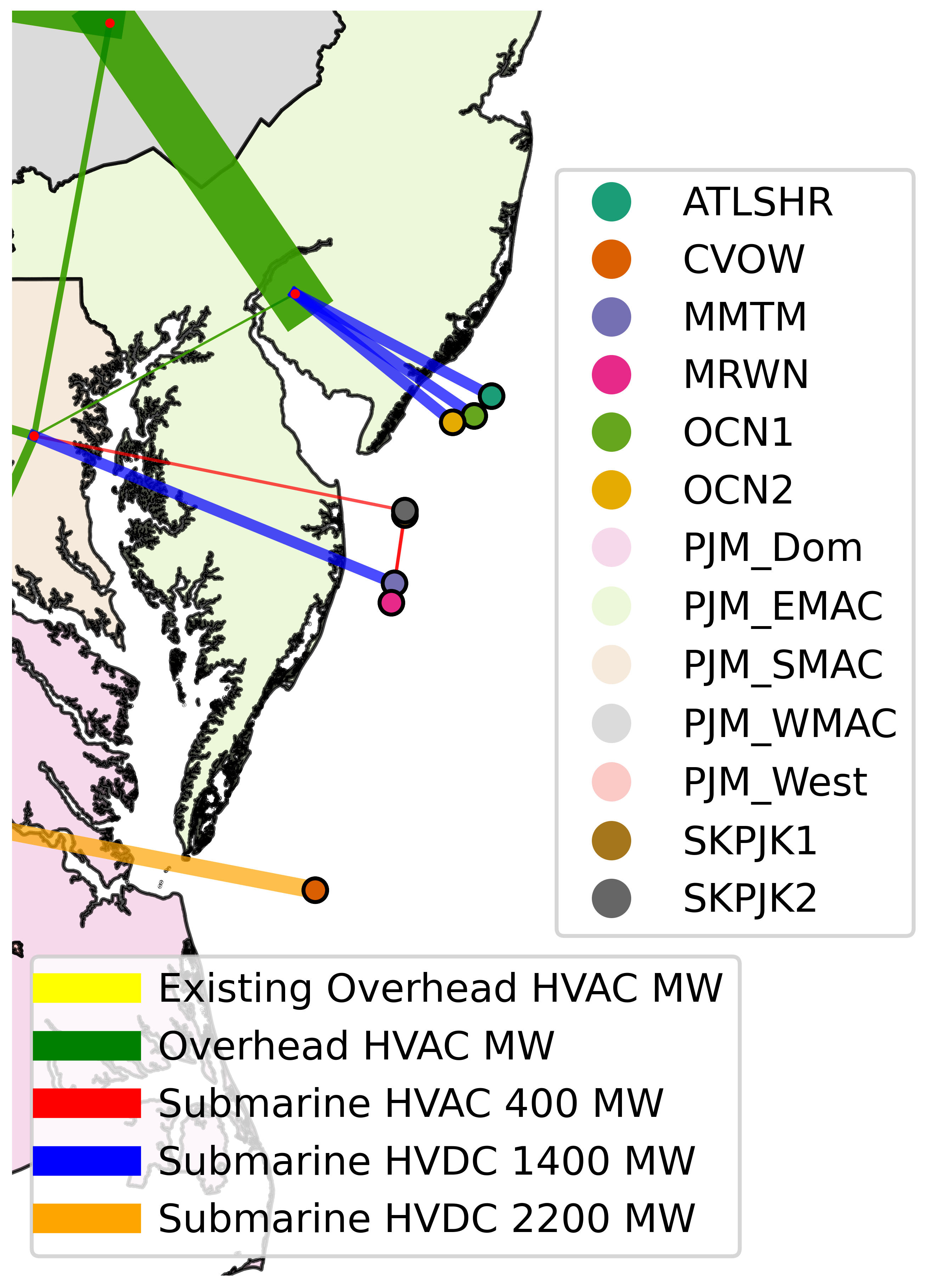}
        \caption{focused (c)}
        \label{fig:2_MO_focus}
    \end{subfigure}
    \newline
    \begin{subfigure}{0.36\textwidth}
        \includegraphics[width=\textwidth]{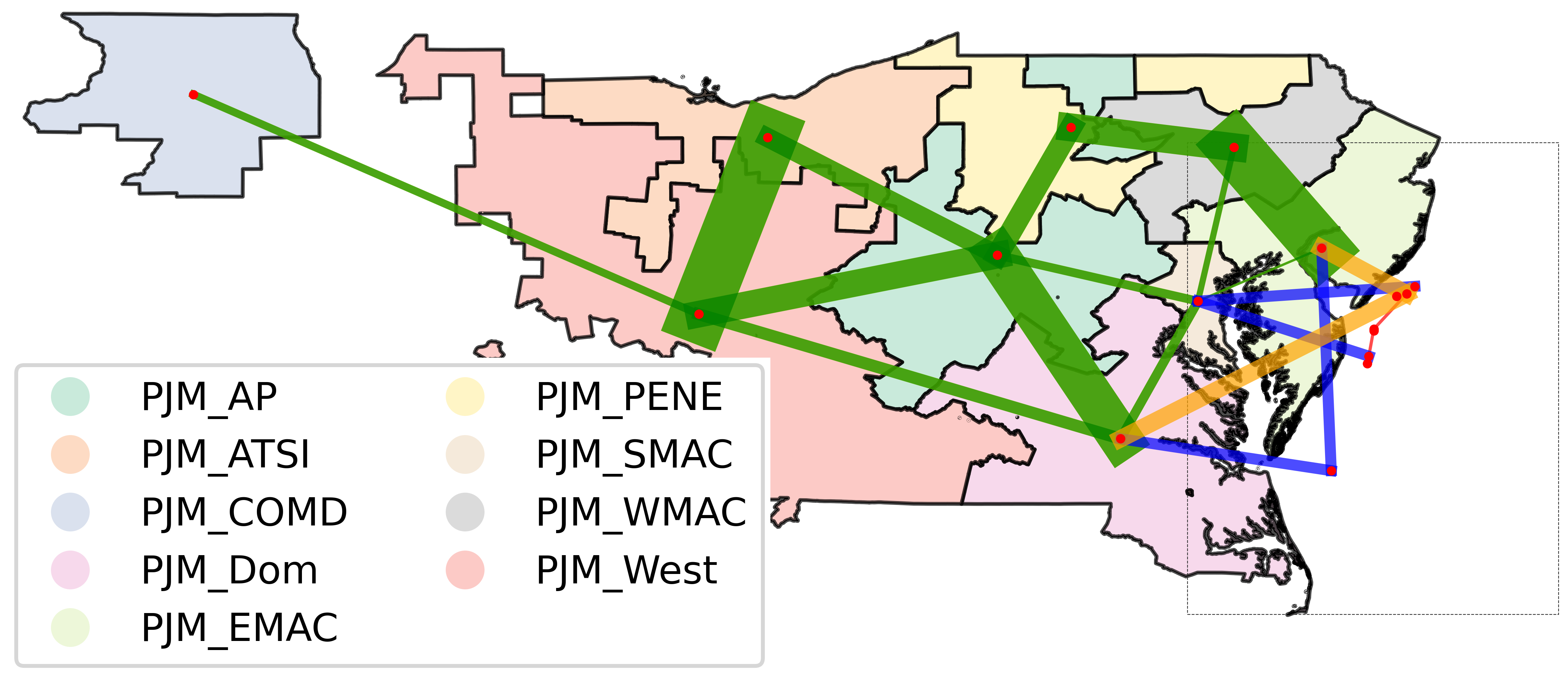}
        \caption{MO OPOI}
        \label{fig:2_MO OPOI}
    \end{subfigure}
    \hfill
    \begin{subfigure}{0.12\textwidth}
        \includegraphics[width=\textwidth]{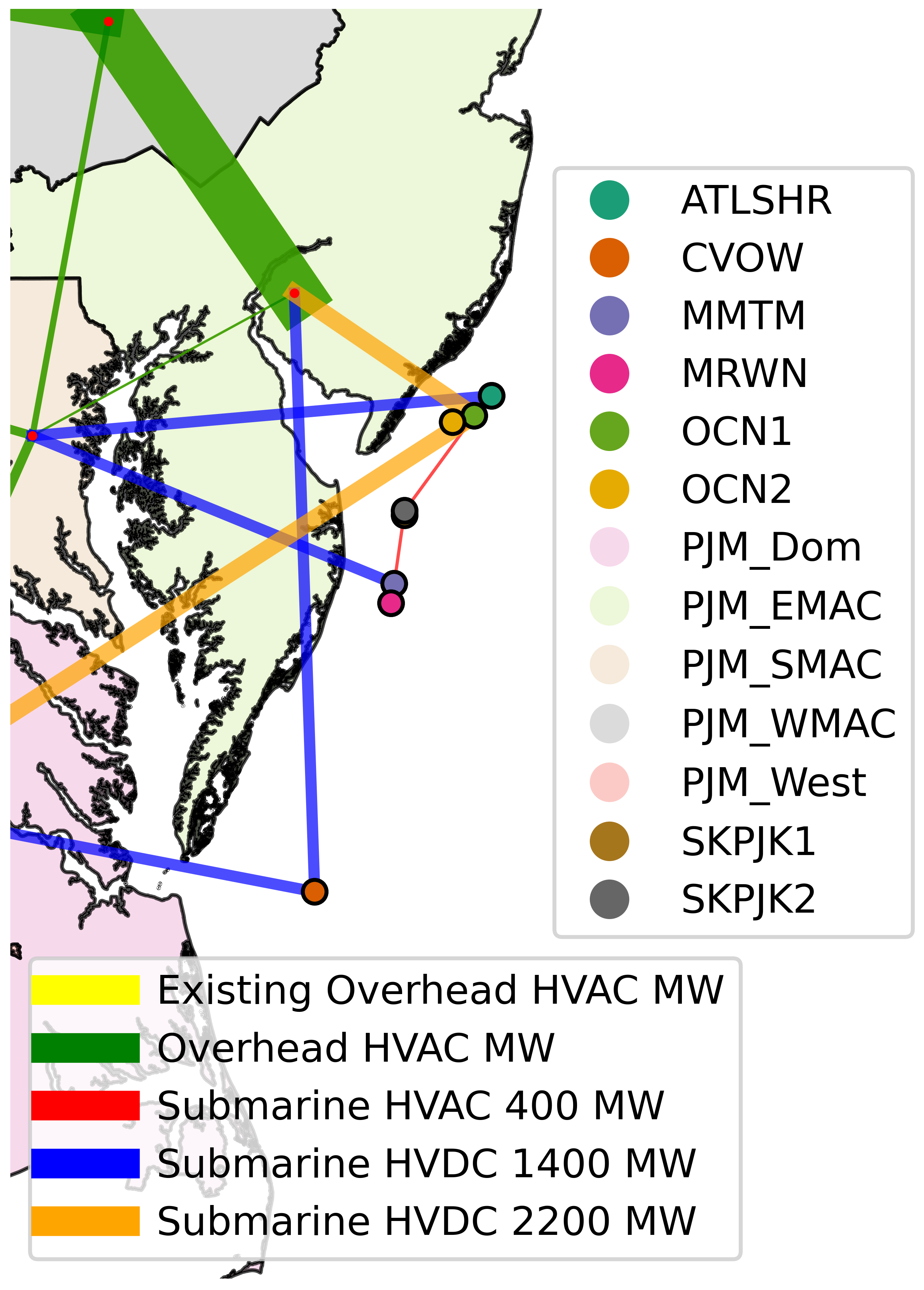}
        \caption{focused (e)}
        \label{fig:2_MO OPOI_focus}
    \end{subfigure}
    \caption{Optimal Onshore and Offshore Topology with Varying Specifications. (\textit{Notes:} MO OPOI, as well as MO OPOI X5, was solved with a MIP gap of 0.1\%.)} 
    \label{fig:2_topology}
\end{figure}

\begin{figure}
    \centering
    \includegraphics[width=\linewidth]{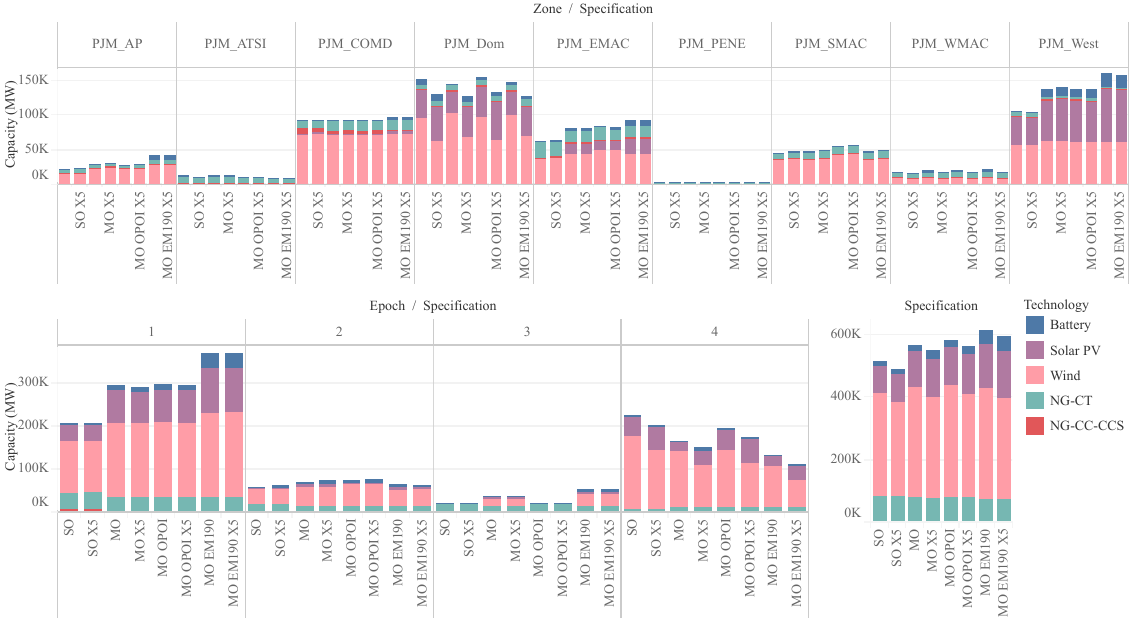}
    \caption{Optimal Generation and Storage Capacity Expansion Decisions with Varying Specifications. \emph{Top}: Across existing onshore zones and model specifications. \emph{Bottom left}: Across epochs and model specifications. \emph{Bottom right}: Total across model specifications. X5 denotes consideration of extreme scenarios.}
    \label{fig:2_capacity}
\end{figure}

\begin{figure}
    \centering
    \includegraphics[width=\linewidth]{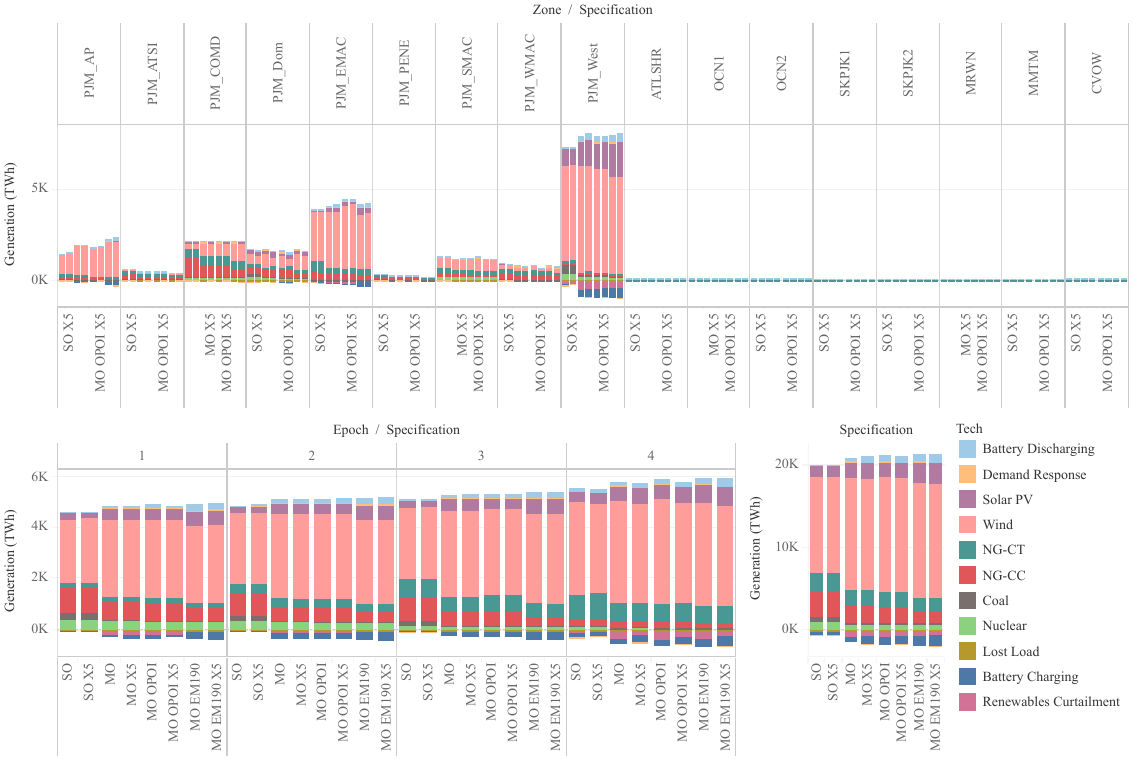}
    \caption{Optimal Generation Decisions with Varying Specifications. \emph{Top}: Across existing onshore zones and model specifications. \emph{Bottom left}: Across epochs and model specifications. \emph{Bottom right}: Total across model specifications. X5 denotes consideration of extreme scenarios.}
    \label{fig:2_generation}
\end{figure}

\begin{figure}
    \centering
    \includegraphics[width=\linewidth]{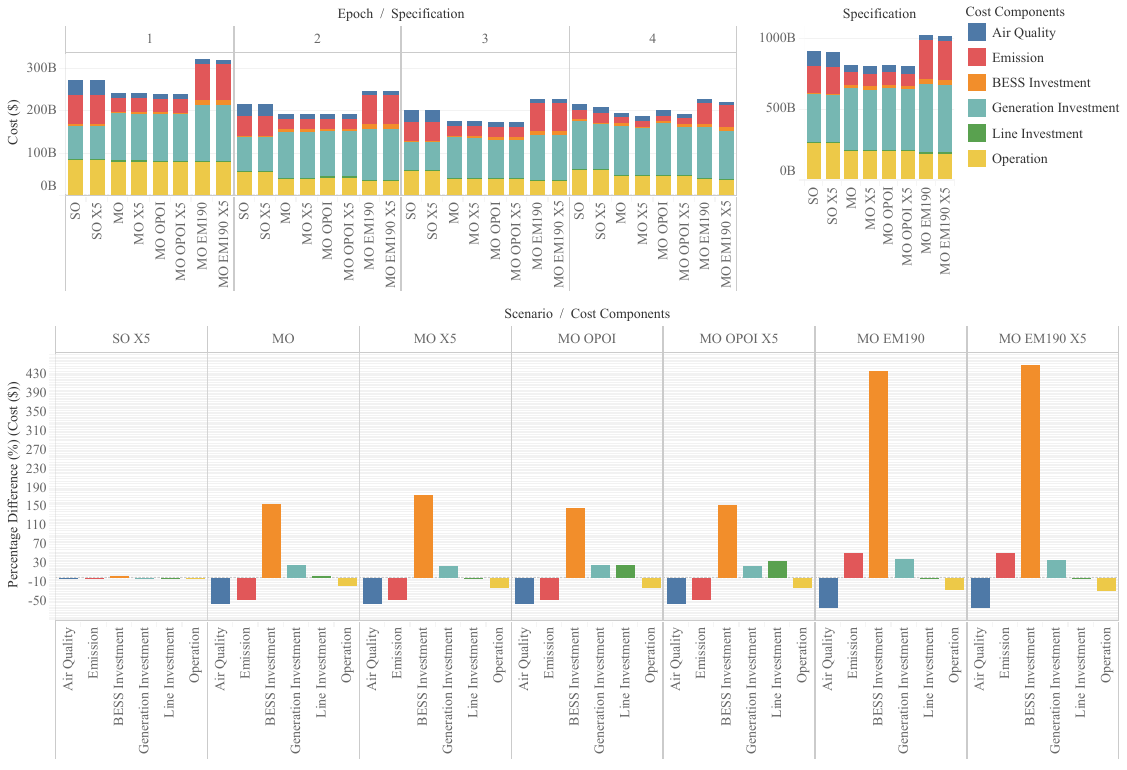}
    \caption{Optimal Costs with Varying Specifications. \emph{Top}: Costs across epochs and model specifications. \emph{Bottom left}: Cost comparisons across model specifications considering SO as the baseline. \emph{Bottom right}: Total costs across model specifications. X5 denotes consideration of extreme scenarios.}
    \label{fig:2_costs}
\end{figure}

Fig.~\ref{fig:2_topology} illustrates the optimal onshore transmission buildout and offshore topology with varying specifications. Similar to the ISO-NE system, we observe a consistent offshore topology in fixed Point of Interconnection (POI) specifications (SO and MO), characterized by primarily radial layouts with locally meshed connections. However, the onshore buildouts display distinct variations. Again, similarly to the ISO-NE case study, the offshore topology in optimized POI specifications (MO OPOI) favors meshed configurations with all candidate POIs compared to the corresponding fixed POI specifications. Fig.~\ref{fig:2_capacity} presents new capacity additions, Fig.~\ref{fig:2_generation} shows the generation profiles, and Fig.~\ref{fig:2_costs} details the associated costs. In the MO cases, there is notable investment in batteries, especially in specifications using the higher SCC estimate, coupled with a decrease in externality costs. Unlike the ISO-NE case study, significant differences are not observed when extreme scenarios are used. Similar to the ISO-NE case study, we find more investment in wind, which dominates over solar PV in terms of cleaner generation, across all specifications. However, unlike ISO-NE, there is always a need to invest in fossil-fueled generators, primarily Natural Gas Combined Cycle (NG-CC), in the PJM case study, mainly driven by flexibility needs arising from aggressive retirements of dispatchable thermal generators.

\begin{figure}
    \centering
    \begin{subfigure}{0.36\textwidth}
        \includegraphics[width=\textwidth]{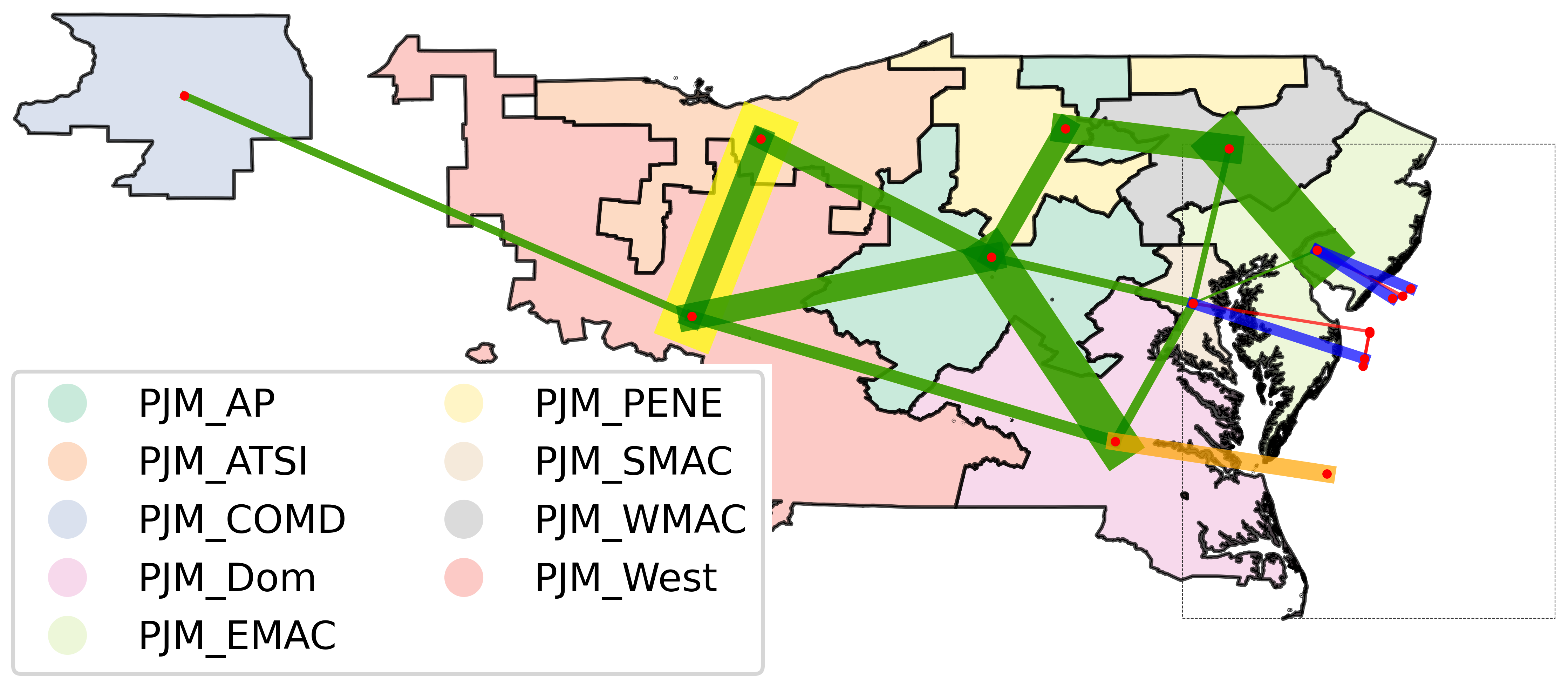}
        \caption{SO}
        \label{fig:2_NS_SO}
    \end{subfigure}
    \hfill
    \begin{subfigure}{0.12\textwidth}
        \includegraphics[width=\textwidth]{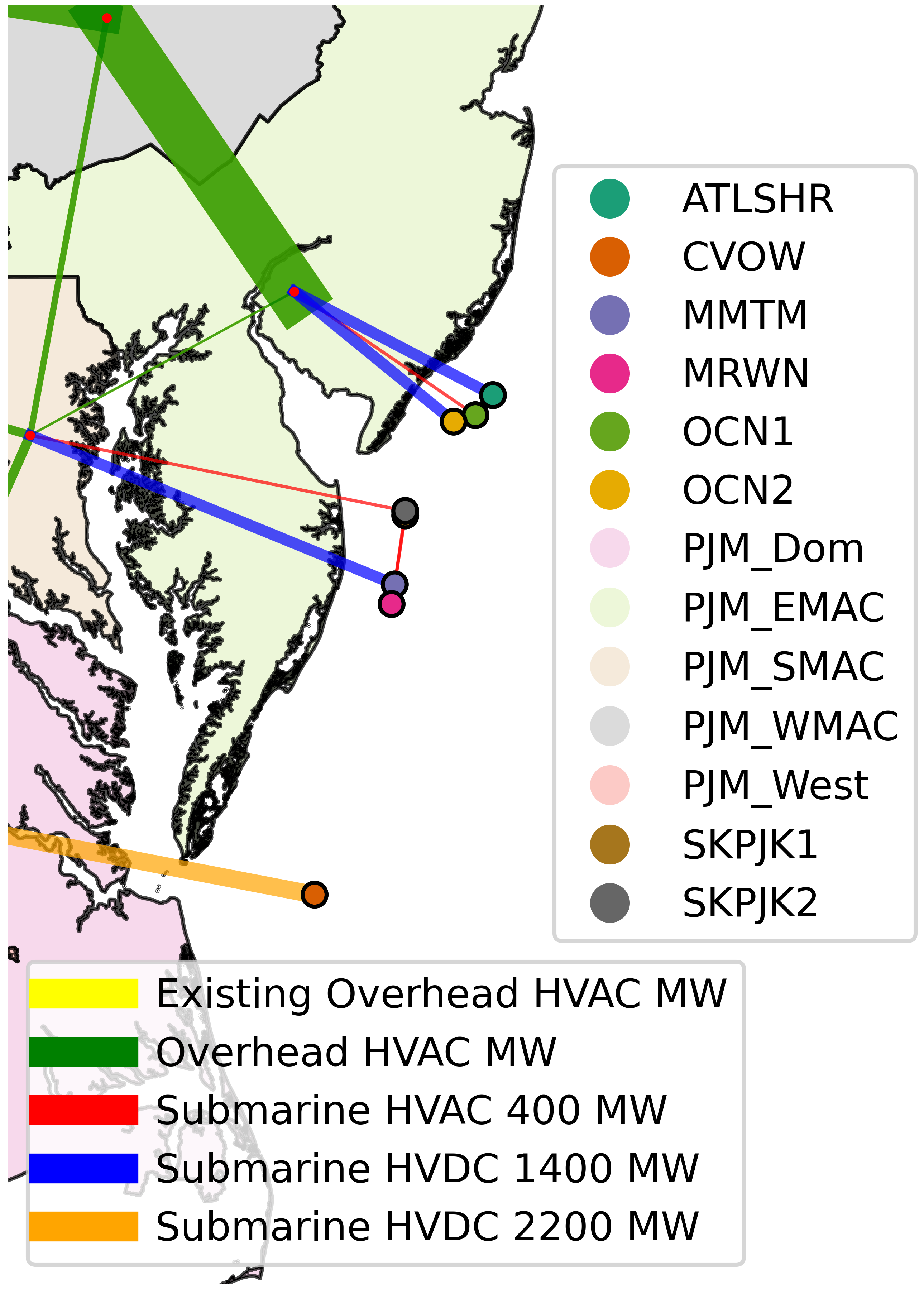}
        \caption{focused (a)}
        \label{fig:2_NS_SO_focus}
    \end{subfigure}
    \newline
    \begin{subfigure}{0.36\textwidth}
        \includegraphics[width=\textwidth]{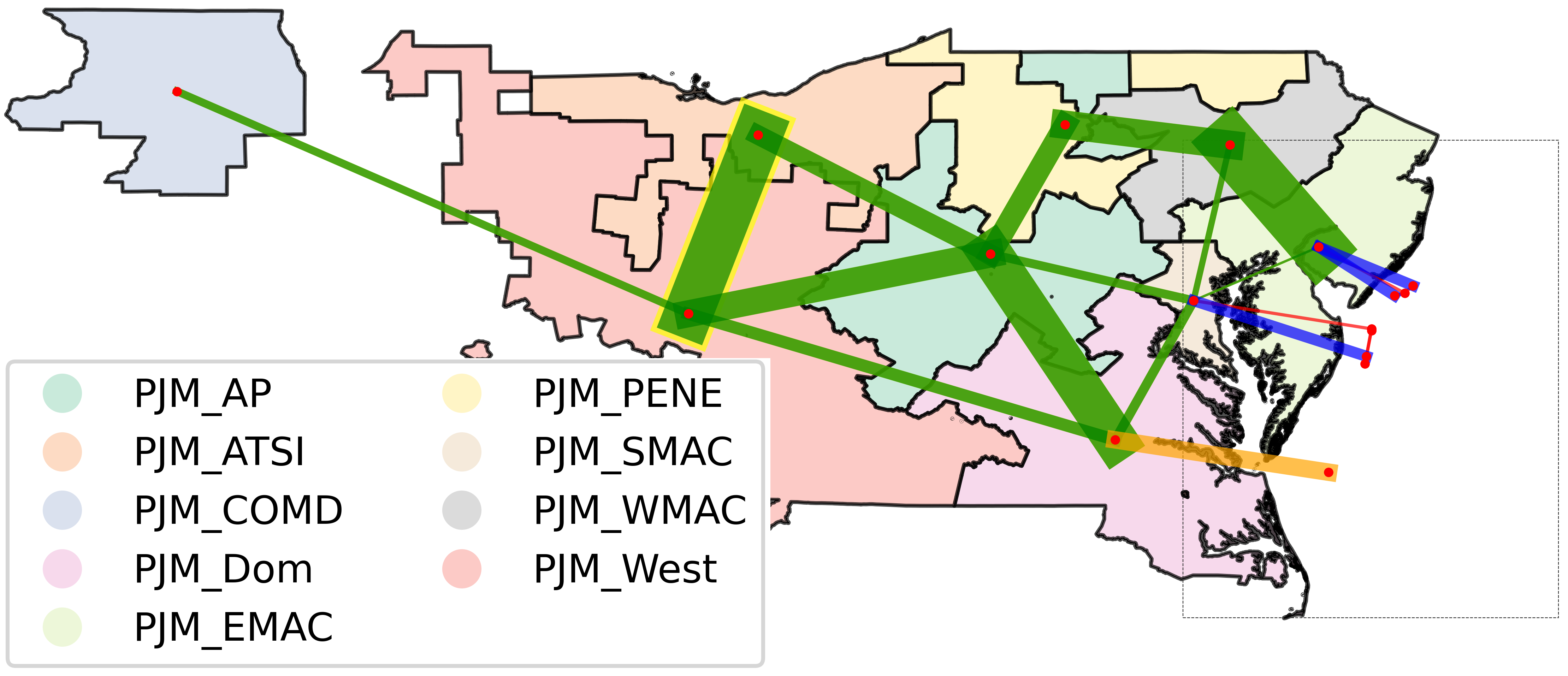}
        \caption{MO}
        \label{fig:2_NS_MO}
    \end{subfigure}
    \hfill
    \begin{subfigure}{0.12\textwidth}
        \includegraphics[width=\textwidth]{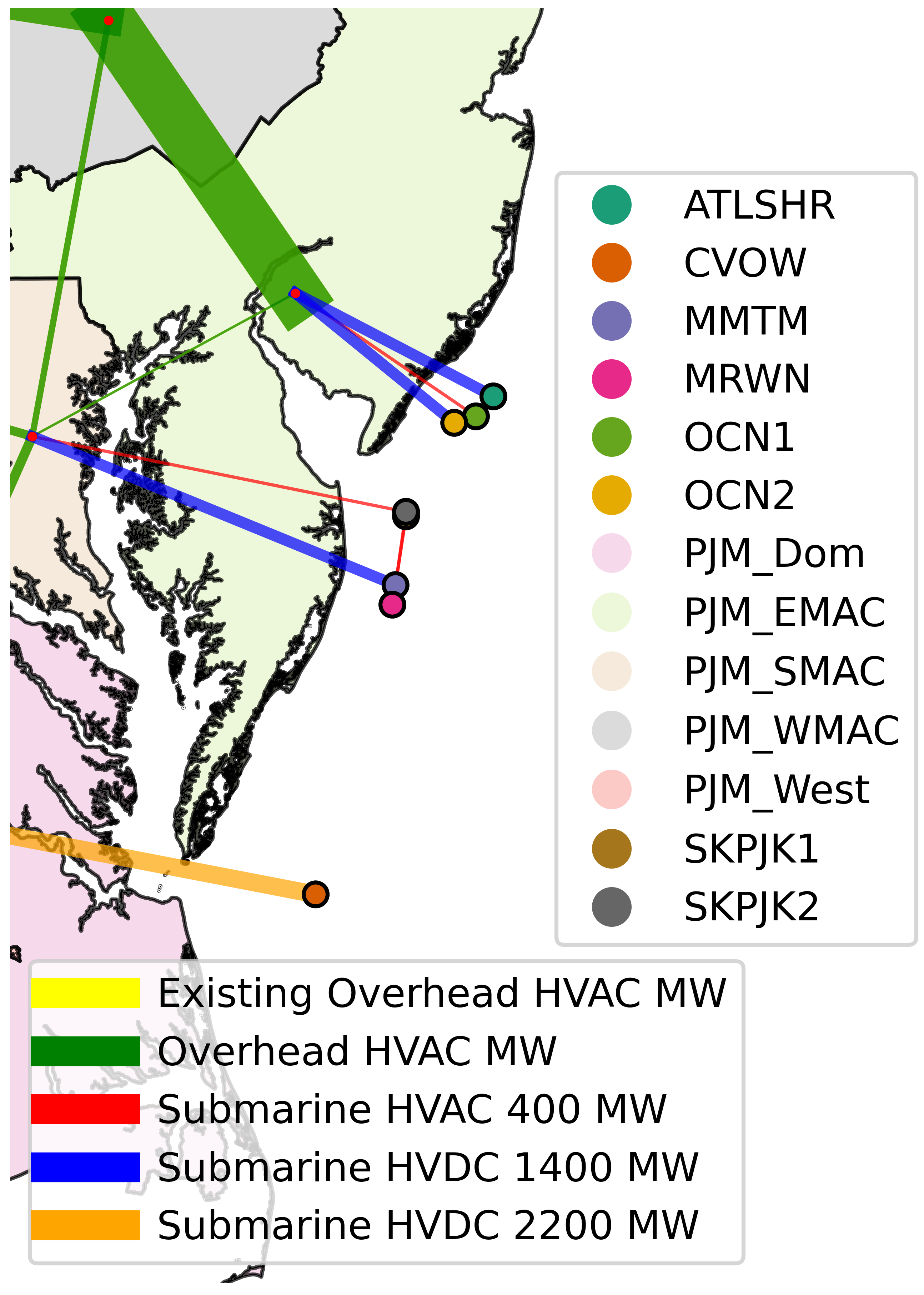}
        \caption{focused (c)}
        \label{fig:2_NS_MO_focus}
    \end{subfigure}
    \newline
    \begin{subfigure}{0.36\textwidth}
        \includegraphics[width=\textwidth]{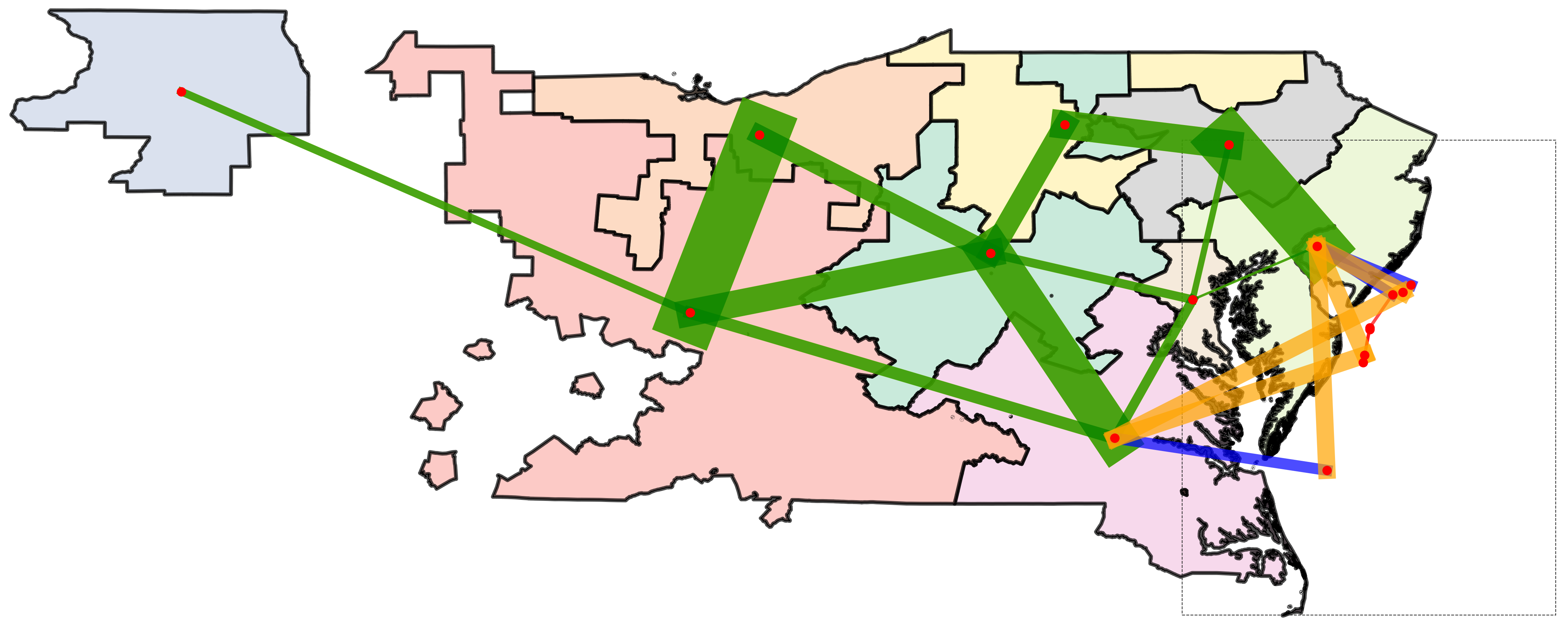}
        \caption{MO OPOI}
        \label{fig:2_NS_MO_OPOI}
    \end{subfigure}
    \hfill
    \begin{subfigure}{0.12\textwidth}
        \includegraphics[width=\textwidth]{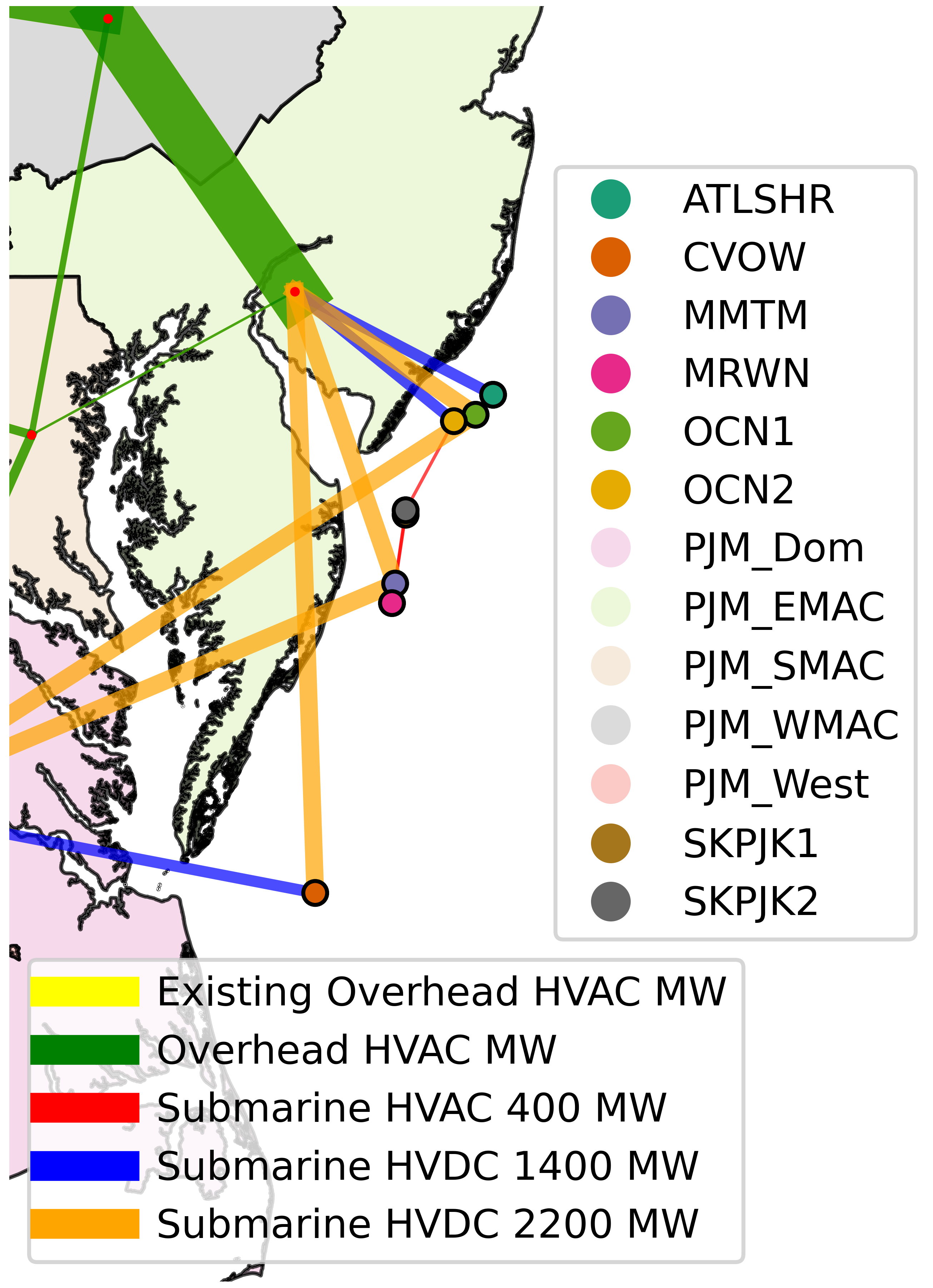}
        \caption{focused (e)}
        \label{fig:2_NS_MO_OPOI_focus}
    \end{subfigure}
    \caption{Optimal Onshore and Offshore Topology with Varying Specifications Using Operational Scenarios Derived from \cite{scott2019clustering}.} 
    \label{fig:2_ns_topology}
\end{figure}

\begin{figure}
    \centering
    \includegraphics[width=\linewidth]{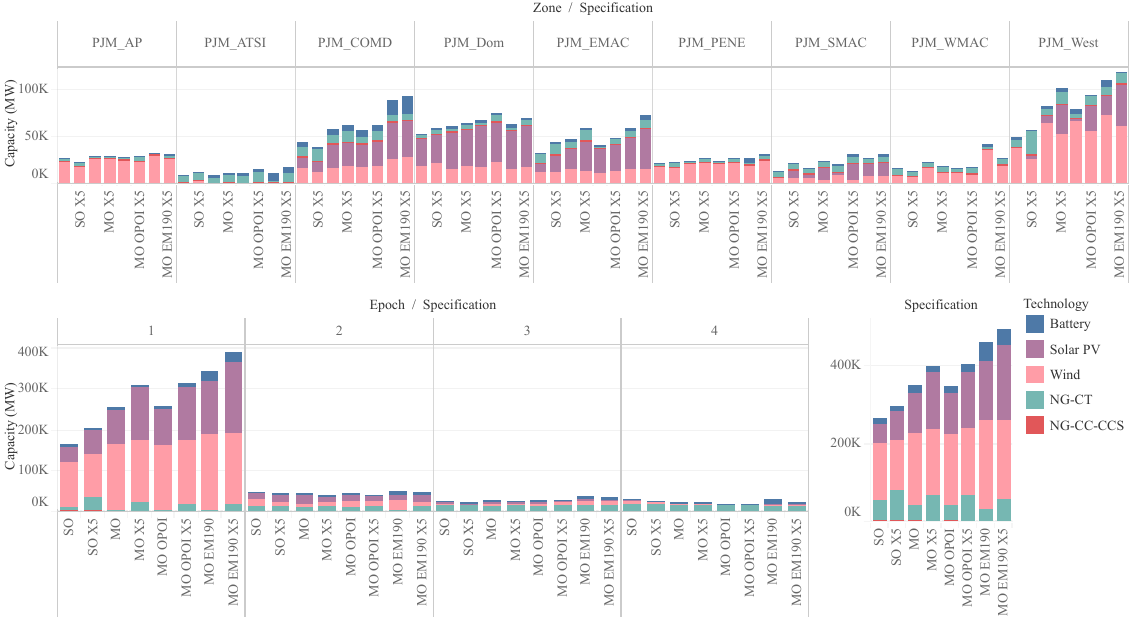}
    \caption{Optimal Generation and Storage Capacity Expansion Decisions with Varying Specifications Using Operational Scenarios Derived from \cite{scott2019clustering}. \emph{Top}: Across existing onshore zones and model specifications. \emph{Bottom left}: Across epochs and model specifications. \emph{Bottom right}: Total across model specifications. X5 denotes consideration of extreme scenarios.} 
    \label{fig:2_ns_capacity}
\end{figure}

\begin{figure}
    \centering
    \includegraphics[width=\linewidth]{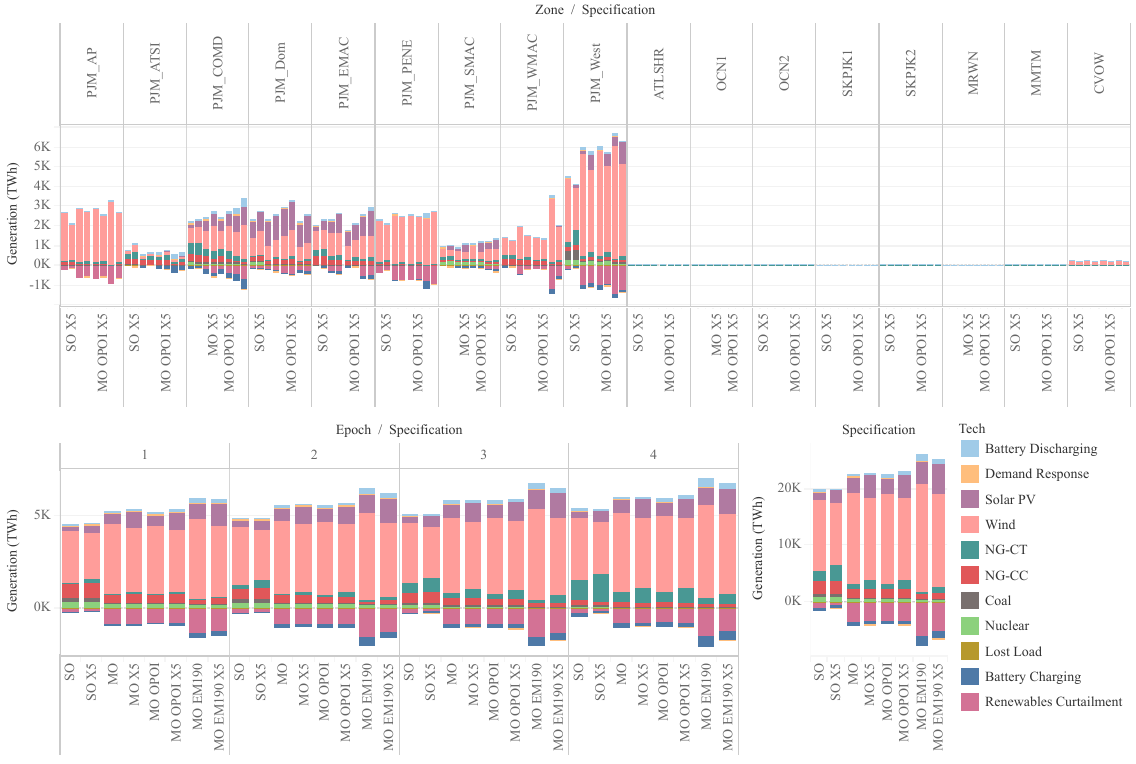}
    \caption{Optimal Generation Decisions with Varying Specifications Using Operational Scenarios Derived from \cite{scott2019clustering}. \emph{Top}: Across existing onshore zones and model specifications. \emph{Bottom left}: Across epochs and model specifications. \emph{Bottom right}: Total across model specifications. X5 denotes consideration of extreme scenarios.} 
    \label{fig:2_ns_generation}
\end{figure}

\begin{figure}
    \centering
    \includegraphics[width=\linewidth]{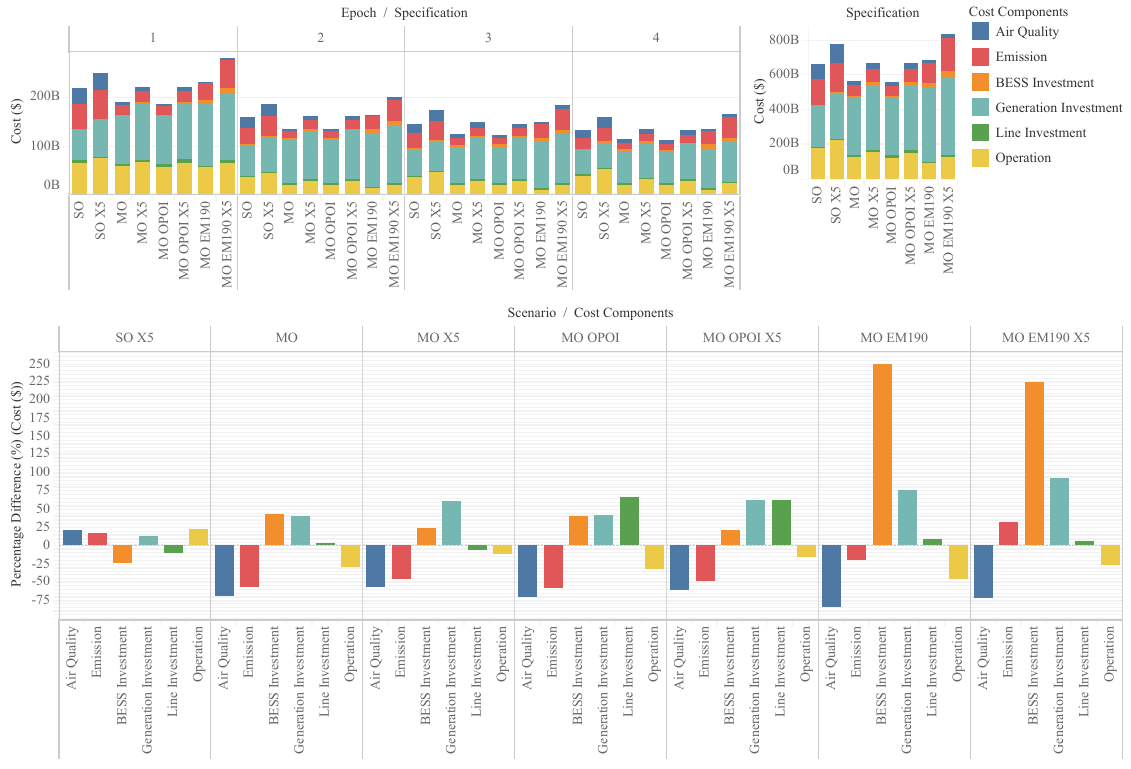}
    \caption{Optimal Costs with Varying Specifications Using Operational Scenarios Derived from \cite{scott2019clustering}. \emph{Top}: Costs across epochs and model specifications. \emph{Bottom left}: Cost comparisons across model specifications considering SO as the baseline. \emph{Bottom right}: Total costs across model specifications. X5 denotes consideration of extreme scenarios.} 
    \label{fig:2_ns_costs}
\end{figure}

Figs.~\ref{fig:2_ns_topology}, \ref{fig:2_ns_capacity} and \ref{fig:2_ns_generation} respectively show transmission decisions, capacity additions, and dispatch decisions, with a new set of operational scenarios derived from \cite{scott2019clustering}. Compared with Fig.~\ref{fig:2_topology}, the fixed POI specifications (SO and MO) do not impact the topology, while MO OPOI favors a meshed offshore network. Similar to ISO-NE, we again see higher costs for extreme scenarios, compared to cases with only normal scenarios (e.g. SO versus SO X5 costs in Figs~\ref{fig:2_costs} and \ref{fig:2_ns_costs}). Comparing Fig.~\ref{fig:2_ns_capacity} and Fig.~\ref{fig:2_capacity} shows that the initial investment in generation and storage capacity is common across two sets of operational scenarios, while the magnitude of the invested capacity ranges from 267.01 GW in SO to 495.63 GW in MO EM190 X5, unlike relatively unchanged capacity mixes from previous scenarios (ranging from 491.96 GW in SO X5 up to 617.35 GW in MO EM190). The differences are due to significantly more solar PV investment in COMD, DOM and EMAC zones. Also, resource diversity has led to less investment in gas units. The results also indicate more line investment in OPOI cases, and all MO specifications, including OPOI, favor investment in battery, significantly so when a high SCC estimate is imposed.

\section{Conclusion} \label{sec:conclusion}
We describe a multi-objective, multi-stage capacity expansion framework that encompasses generation, storage, and transmission to offer a comprehensive approach to coordinated grid planning with large-scale deployment of offshore wind power. In addition, our model accounts for unpriced or underpriced externalities, such as greenhouse gas emissions and local air pollution. In the face of large-scale offshore grid integration, this provides a deeper understanding of both economic and non-economic costs in grid expansion decisions, thereby helping reform existing planning and aiding in cost-allocation processes.

In our analysis of the ISO-NE system, we discovered that accounting for negative externalities could reduce onshore line upgrades and increase upfront investment in clean energy and storage resources, which are largely offset by lower expected operational costs. Accounting for extreme operational scenarios may sometimes alter the needs for onshore transmission upgrades and lead to meshed offshore networks. Additionally, we observed that strictly fixed POIs (adhering to predetermined offtake agreements) for offshore wind projects could reduce resilience benefits and increase overall costs. Finally, when comparing results with a temporally coupled set of operational scenarios, there was a notable shift in generation investment decisions and related costs, yet maintaining consistent overarching insights on transmission investments. For the PJM case study, optimizing POIs, considering externalities, and accounting for extreme scenarios remain important, similar to the ISO-NE case study. The cases with all fixed POIs resulted in similar offshore network topologies, and those with all optimized POIs showed little to no qualitative difference from the ISO-NE results. Furthermore, considering the extra set of operational scenarios led to differences in clean energy investment (e.g., increased investment in solar as opposed to wind-dominated investment previously) and operational costs, while the impact on optimal offshore configurations remained unchanged. In essence, these findings underscore the significance of incorporating externalities and extreme scenarios into energy policy and planning models, especially given the evolving generation mix and demand driven by clean energy and climate policies.

\bibliographystyle{IEEEtran}
{\footnotesize \bibliography{IEEEabrv,main}}

\begin{thebibliography}{10}
\providecommand{\url}[1]{#1}
\csname url@samestyle\endcsname
\providecommand{\newblock}{\relax}
\providecommand{\bibinfo}[2]{#2}
\providecommand{\BIBentrySTDinterwordspacing}{\spaceskip=0pt\relax}
\providecommand{\BIBentryALTinterwordstretchfactor}{4}
\providecommand{\BIBentryALTinterwordspacing}{\spaceskip=\fontdimen2\font plus
\BIBentryALTinterwordstretchfactor\fontdimen3\font minus \fontdimen4\font\relax}
\providecommand{\BIBforeignlanguage}[2]{{%
\expandafter\ifx\csname l@#1\endcsname\relax
\typeout{** WARNING: IEEEtran.bst: No hyphenation pattern has been}%
\typeout{** loaded for the language `#1'. Using the pattern for}%
\typeout{** the default language instead.}%
\else
\language=\csname l@#1\endcsname
\fi
#2}}
\providecommand{\BIBdecl}{\relax}
\BIBdecl

\bibitem{jenkins2021mission}
{Jenkins et al}, ``Mission net-zero america: The nation-building path to a prosperous, net-zero emissions economy,'' \emph{Joule}, vol.~5, no.~11, pp. 2755--2761, 2021.

\bibitem{brinkman2021north}
{Brinkman et al}, ``The north american renewable integration study (naris): A canadian perspective,'' NREL, Tech. Rep., 2021.

\bibitem{brown2021value}
P.~R. Brown and A.~Botterud, ``The value of inter-regional coordination and transmission in decarbonizing the us electricity system,'' \emph{Joule}, vol.~5, no.~1, pp. 115--134, 2021.

\bibitem{Morehouse2021}
\BIBentryALTinterwordspacing
C.~Morehouse, ``Cost allocation remains key challenge for ferc ahead of transmission reform, glick says,'' 7 2021. [Online]. Available: \url{https://www.utilitydive.com/news/cost-allocation-remains-key-challenge-for-ferc-ahead-of-transmission-reform/603597/}
\BIBentrySTDinterwordspacing

\bibitem{conejo2016investment}
{A. Conejo et al}, ``Investment in electricity generation and transmission,'' \emph{Switzerland: Springer}, vol. 119, 2016.

\bibitem{pfeifenbergerbenefit}
{J. Pfeifenberger et al}, ``{The Benefit and Cost of Preserving the Option to Create a Meshed Offshore Grid for New York},'' Brattle Group, Inc, Tech. Rep., Dec 2021, prepared for NYSERDA.

\bibitem{FERC2022}
\BIBentryALTinterwordspacing
{FERC}, ``Building for the future through electric regional transmission planning and cost allocation and generator interconnection,'' pp. 26\,504--26\,611, 5 2022. [Online]. Available: \url{https://www.federalregister.gov/documents/2022/05/04/2022-08973/building-for-the-future-through-electric-regional-transmission-planning-and-cost-allocation-and}
\BIBentrySTDinterwordspacing

\bibitem{nationalgrid2022}
\BIBentryALTinterwordspacing
``Pathway to 2030: A holistic network design to support offshore wind deployment for net zero,'' 2022. [Online]. Available: \url{https://www.nationalgrideso.com/document/262676/download}
\BIBentrySTDinterwordspacing

\bibitem{gacitua2018comprehensive}
L.~Gacitua, P.~Gallegos, R.~Henriquez-Auba, {\'A}.~Lorca, M.~Negrete-Pincetic, D.~Olivares, A.~Valenzuela, and G.~Wenzel, ``A comprehensive review on expansion planning: Models and tools for energy policy analysis,'' \emph{Renewable and Sustainable Energy Reviews}, vol.~98, pp. 346--360, 2018.

\bibitem{papavasiliou2011reserve}
A.~Papavasiliou, S.~S. Oren, and R.~P. O'Neill, ``Reserve requirements for wind power integration: A scenario-based stochastic programming framework,'' \emph{IEEE Transactions on Power Systems}, vol.~26, no.~4, pp. 2197--2206, 2011.

\bibitem{costa2023review}
C.~R. d.~S. Costa and P.~Ferreira, ``A review on the internalization of externalities in electricity generation expansion planning,'' \emph{Energies}, vol.~16, no.~4, p. 1840, 2023.

\bibitem{Gies2017}
E.~Gies, ``The real cost of energy: All energy production has environmental and societal effects. but calculating them - and pricing energy accordingly - is no easy task,'' \emph{Nature}, vol. 551, pp. S145--S147, 2017.

\bibitem{munoz2013engineering}
{F. Munoz et al}, ``An engineering-economic approach to transmission planning under market and regulatory uncertainties: Wecc case study,'' \emph{IEEE Trans. Pwr. Syst.}, vol.~29, no.~1, pp. 307--317, 2013.

\bibitem{qiu2016stochastic}
{T. Qiu et al}, ``Stochastic multistage coplanning of transmission expansion and storage,'' \emph{IEEE Trans. Pwr. Syst.}, vol.~32, no.~1, pp. 643--651, 2016.

\bibitem{rafaj2007internalisation}
P.~Rafaj and S.~Kypreos, ``Internalisation of external cost in the power generation sector: Analysis with global multi-regional markal model,'' \emph{Energy Policy}, vol.~35, no.~2, pp. 828--843, 2007.

\bibitem{chen2018advances}
S.~Chen, Z.~Guo, P.~Liu, and Z.~Li, ``Advances in clean and low-carbon power generation planning,'' \emph{Computers \& Chemical Engineering}, vol. 116, pp. 296--305, 2018.

\bibitem{rodgers2019assessing}
M.~Rodgers, D.~Coit, F.~Felder, and A.~Carlton, ``Assessing the effects of power grid expansion on human health externalities,'' \emph{Socio-Economic Planning Sciences}, vol.~66, pp. 92--104, 2019.

\bibitem{quiroga2019power}
D.~Quiroga, E.~Sauma, and D.~Pozo, ``Power system expansion planning under global and local emission mitigation policies,'' \emph{Applied Energy}, vol. 239, pp. 1250--1264, 2019.

\bibitem{chen2019multi}
S.~Chen, P.~Liu, and Z.~Li, ``Multi-regional power generation expansion planning with air pollutants emission constraints,'' \emph{Renewable and Sustainable Energy Reviews}, vol. 112, pp. 382--394, 2019.

\bibitem{chiu2020future}
M.-C. Chiu, H.-W. Hsu, M.-C. Wu, and M.-Y. Lee, ``Future thinking on power planning: A balanced model of regions, seasons and environment with a case of taiwan,'' \emph{Futures}, vol. 122, p. 102599, 2020.

\bibitem{lv2020generation}
T.~Lv, Q.~Yang, X.~Deng, J.~Xu, and J.~Gao, ``Generation expansion planning considering the output and flexibility requirement of renewable energy: the case of jiangsu province,'' \emph{Frontiers in Energy Research}, vol.~8, p.~39, 2020.

\bibitem{pereira2020power}
A.~Pereira and E.~Sauma, ``Power systems expansion planning with time-varying co2 tax,'' \emph{Energy Policy}, vol. 144, p. 111630, 2020.

\bibitem{gbadamosi2020multi}
S.~L. Gbadamosi and N.~I. Nwulu, ``A multi-period composite generation and transmission expansion planning model incorporating renewable energy sources and demand response,'' \emph{Sustainable Energy Technologies and Assessments}, vol.~39, p. 100726, 2020.

\bibitem{fitiwi2020enhanced}
D.~Z. Fitiwi, M.~Lynch, and V.~Bertsch, ``Enhanced network effects and stochastic modelling in generation expansion planning: Insights from an insular power system,'' \emph{Socio-economic planning sciences}, vol.~71, p. 100859, 2020.

\bibitem{sani2021decarbonization}
L.~Sani, D.~Khatiwada, F.~Harahap, and S.~Silveira, ``Decarbonization pathways for the power sector in sumatra, indonesia,'' \emph{Renewable and Sustainable Energy Reviews}, vol. 150, p. 111507, 2021.

\bibitem{verastegui2021optimization}
F.~Ver{\'a}stegui, {\'A}.~Lorca, D.~Olivares, and M.~Negrete-Pincetic, ``Optimization-based analysis of decarbonization pathways and flexibility requirements in highly renewable power systems,'' \emph{Energy}, vol. 234, p. 121242, 2021.

\bibitem{ribeiro2013evaluating}
F.~Ribeiro, P.~Ferreira, and M.~Ara{\'u}jo, ``Evaluating future scenarios for the power generation sector using a multi-criteria decision analysis (mcda) tool: The portuguese case,'' \emph{Energy}, vol.~52, pp. 126--136, 2013.

\bibitem{santos2017scenarios}
M.~Santos, P.~Ferreira, M.~Ara{\'u}jo, J.~Portugal-Pereira, A.~Lucena, and R.~Schaeffer, ``Scenarios for the future brazilian power sector based on a multi-criteria assessment,'' \emph{Journal of cleaner production}, vol. 167, pp. 938--950, 2017.

\bibitem{musial2021offshore}
{W. Musial et al}, ``Offshore wind market report: 2021 edition,'' National Renewable Energy Lab., Tech. Rep., 2021.

\bibitem{musial2022offshore}
------, ``Offshore wind market report: 2022 edition,'' National Renewable Energy Lab., Tech. Rep., 2022.

\bibitem{liu2022optimization}
Y.~Liu, Y.~Fu, L.-l. Huang, Z.-x. Ren, and F.~Jia, ``Optimization of offshore grid planning considering onshore network expansions,'' \emph{Renewable Energy}, vol. 181, pp. 91--104, 2022.

\bibitem{meng2019offshore}
{K. Meng et al}, ``Offshore transmission network planning for wind integration considering ac and dc transmission options,'' \emph{IEEE Transactions on Power Systems}, vol.~34, no.~6, pp. 4258--4268, 2019.

\bibitem{taylor2023wind}
{P. Taylor et al}, ``Wind farm array cable layout optimisation for complex offshore sites—a decomposition based heuristic approach,'' \emph{IET Renewable Power Generation}, vol.~17, no.~2, pp. 243--259, 2023.

\bibitem{mehrtash2019graph}
M.~Mehrtash, A.~Kargarian, and A.~J. Conejo, ``Graph-based second-order cone programming model for resilient feeder routing using gis data,'' \emph{IEEE Trans. Pwr. Del.}, vol.~35, no.~4, pp. 1999--2010, 2019.

\bibitem{jin2019cable}
{R. Jin et al}, ``Cable routing optimization for offshore wind power plants via wind scenarios considering power loss cost model,'' \emph{Applied Energy}, vol. 254, p. 113719, 2019.

\bibitem{nguyen2016quantifying}
T.~L.~T. Nguyen, B.~Laratte, B.~Guillaume, and A.~Hua, ``Quantifying environmental externalities with a view to internalizing them in the price of products, using different monetization models,'' \emph{Resources, Conservation and Recycling}, vol. 109, pp. 13--23, 2016.

\bibitem{EPA2022finalized}
\BIBentryALTinterwordspacing
{US EPA}, ``Report on the social cost of greenhouse gases: Estimates incorporating recent scientific advances,'' National Center for Environmental Economics, 2023. [Online]. Available: \url{https://www.epa.gov/system/files/documents/2023-12/epa_scghg_2023_report_final.pdf}
\BIBentrySTDinterwordspacing

\bibitem{TeHi17}
C.~W. Tessum, J.~D. Hill, and J.~D. Marshall, ``{InMAP: A model for air pollution interventions},'' \emph{PLOS ONE}, vol.~12, no.~4, pp. 1--26, 2017.

\bibitem{EPA2023MortalityRisk}
\BIBentryALTinterwordspacing
{US EPA}, ``Mortality risk valuation,'' 2023, accessed on: Oct 25, 2023. [Online]. Available: \url{https://www.epa.gov/environmental-economics/mortality-risk-valuation}
\BIBentrySTDinterwordspacing

\bibitem{dvorkin2017co}
Y.~Dvorkin, R.~Fernandez-Blanco, Y.~Wang, B.~Xu, D.~S. Kirschen, H.~Pand{\v{z}}i{\'c}, J.-P. Watson, and C.~A. Silva-Monroy, ``Co-planning of investments in transmission and merchant energy storage,'' \emph{IEEE Transactions on Power Systems}, vol.~33, no.~1, pp. 245--256, 2017.

\bibitem{arroyo2020use}
J.~M. Arroyo, L.~Baringo, A.~Baringo, R.~Bola{\~n}os, N.~Alguacil, and N.~G. Cobos, ``On the use of a convex model for bulk storage in mip-based power system operation and planning,'' \emph{IEEE Transactions on Power Systems}, vol.~35, no.~6, pp. 4964--4967, 2020.

\bibitem{hobbs2008improved}
{B. Hobbs et al}, ``Improved transmission representations in oligopolistic market models: quadratic losses, phase shifters, and dc lines,'' \emph{IEEE Trans. Pwr. Syst.}, vol.~23, no.~3, pp. 1018--1029, 2008.

\bibitem{krishnamurthy20158}
D.~Krishnamurthy, W.~Li, and L.~Tesfatsion, ``An 8-zone test system based on iso new england data: Development and application,'' \emph{IEEE Transactions on Power Systems}, vol.~31, no.~1, pp. 234--246, 2015.

\bibitem{us3form}
{US EIA}, ``Form eia-860 detailed data with previous form data (eia-860a/860b), 2021 form eia-860 data.''

\bibitem{IRENA2019Flexibility}
{International Renewable Energy Agency (IRENA)}, ``Innovation landscape brief: Flexibility in conventional power plants,'' \url{https://www.irena.org/-/media/Files/IRENA/Agency/Publication/2019/Sep/IRENA_Flexibility_in_CPPs_2019.pdf}, Abu Dhabi, 2019, accessed: Nov 12, 2023.

\bibitem{ISO_NE2023}
``{ISO New England Website},'' \url{https://www.iso-ne.com/}, 2023, accessed: Nov 1, 2023.

\bibitem{DOEOEDI2023}
{Department of {E}nergy}, ``{O}pen {E}nergy {D}ata {I}nitiative ({OEDI}),'' \url{https://data.openei.org/s3_viewer?bucket=nrel-pds-wtk&prefix=wtk-techno-economic%2Fpywtk-data%2F}, accessed: Nov 1, 2023.

\bibitem{scott2019clustering}
I.~J. Scott, P.~M. Carvalho, A.~Botterud, and C.~A. Silva, ``Clustering representative days for power systems generation expansion planning: Capturing the effects of variable renewables and energy storage,'' \emph{Applied Energy}, vol. 253, p. 113603, 2019.

\bibitem{merrick2016representation}
J.~H. Merrick, ``On representation of temporal variability in electricity capacity planning models,'' \emph{Energy Economics}, vol.~59, pp. 261--274, 2016.

\bibitem{HuTa21}
\BIBentryALTinterwordspacing
{J. Huetteman et al}, \emph{{EPA-EIA Power Sector Data Crosswalk}}, ver 0.3, U.S. EPA, 2021. [Online]. Available: \url{https://www.epa.gov/power-sector/power-sector-data-crosswalk}
\BIBentrySTDinterwordspacing

\bibitem{LeLa12}
J.~L. et~al, ``{Chronic Exposure to Fine Particles and Mortality: An Extended Follow-up of the Harvard Six Cities Study from 1974 to 2009},'' \emph{Env. Health Pers.}, vol. 120, no.~7, pp. 965--970, 2012.

\bibitem{USEPA2023}
\BIBentryALTinterwordspacing
{US EPA}, ``Power sector emissions data,'' 2023. [Online]. Available: \url{https://www.epa.gov/power-sector/power-sector-emissions-data}
\BIBentrySTDinterwordspacing

\bibitem{spglobal_renewable}
\BIBentryALTinterwordspacing
{S\&P Global Market Intelligence}, ``New england renewable policies to drive 12,500 mw of renewable capacity by 2030,'' 2023, accessed on: Oct. 24, 2023. [Online]. Available: \url{https://www.spglobal.com/marketintelligence/en/news-insights/research/new-england-renewable-policies-to-drive-12500-mw-of-renewable-capacity-by-2030}
\BIBentrySTDinterwordspacing

\bibitem{vimmerstedt2022annual}
L.~Vimmerstedt, S.~Akar, B.~Mirletz, A.~Sekar, D.~Stright, C.~Augustine, P.~Beiter, P.~Bhaskar, N.~Blair, S.~Cohen \emph{et~al.}, ``Annual technology baseline: The 2022 electricity update,'' National Renewable Energy Lab.(NREL), Golden, CO (United States), Tech. Rep., 2022.

\bibitem{ho2021regional}
{J. Ho et al}, ``Regional energy deployment system (reeds) model documentation (2020),'' National Renewable Energy Lab, Tech. Rep., 2021.

\bibitem{xiang2021comparison}
{X. Xiang et al}, ``Comparison of cost-effective distances for lfac with hvac and hvdc in their connections for offshore and remote onshore wind,'' \emph{CSEE J. of Pwr. \& En. Syst.}, vol.~7, no.~5, pp. 954--975, 2021.

\bibitem{xiang2016cost}
X.~Xiang, M.~M. Merlin, and T.~C. Green, ``Cost analysis and comparison of hvac, lfac and hvdc for offshore wind power connection,'' 2016.

\bibitem{Catapult2024WindFarmCosts}
{Catapult Offshore Renewable Energy}, ``Wind farm costs,'' \url{https://guidetoanoffshorewindfarm.com/wind-farm-costs}, 2024, accessed: 2024-02-26.

\bibitem{EPA2024PowerSectorModeling}
{U.S. Environmental Protection Agency}, ``Epa's power sector modeling platform v6 using ipm,'' \url{https://www.epa.gov/power-sector-modeling}, U.S. Environmental Protection Agency, U.S. EPA, 2024, accessed: 2024-03-01.

\bibitem{grubert2020fossil}
E.~Grubert, ``Fossil electricity retirement deadlines for a just transition,'' \emph{Science}, vol. 370, no. 6521, pp. 1171--1173, 2020.

\bibitem{pjm2023retirements}
{PJM}, ``Energy transition in pjm: Resource retirements, replacements \& risks,'' \url{https://www.pjm.com/-/media/library/reports-notices/special-reports/2023/energy-transition-in-pjm-resource-retirements-replacements-and-risks.ashx}, Feb 2023, accessed March 1, 2024.

\end{thebibliography}

\end{document}